\PassOptionsToPackage{prologue, cmyk}{xcolor}
\PassOptionsToPackage{pdfa, colorlinks}{hyperref}
\documentclass[acmsmall]{acmart}
\usepackage{colorprofiles}

\usepackage{preamble}
\usepackage{iris}
\usepackage{prob}
\usepackage{examples}

\newcommand{\appref}[1]{the appendix}

\setcopyright{rightsretained}
\acmDOI{10.1145/3674635}
\acmYear{2024}
\acmJournal{PACMPL}
\acmVolume{8}
\acmNumber{ICFP}
\acmArticle{246}
\acmMonth{8}
\acmSubmissionID{icfp24main-p44-p}
\received{2024-02-28}
\received[accepted]{2024-06-18}



\usepackage[disable]{todonotes}

\newcommand\philipp[1]{\todo[author=Philipp, size=\small, color=purple!20, inline]{#1}}

\title{Error Credits: Resourceful Reasoning about Error Bounds for Higher-Order Probabilistic Programs}

\author[A. Aguirre]{Alejandro Aguirre}
\orcid{0000-0001-6746-2734}
\affiliation{
  \institution{Aarhus University}
  \country{Denmark}
}
\email{alejandro@cs.au.dk}

\author[P. G. Haselwarter]{Philipp~G. Haselwarter}
\orcid{0000-0003-0198-7751}
\affiliation{
  \institution{Aarhus University}
  \country{Denmark}
}
\email{pgh@cs.au.dk}

\author[M. de Medeiros]{Markus de Medeiros}
\orcid{0009-0005-3285-5032}
\affiliation{
  \institution{New York University}
  \country{USA}
}
\email{mjd9606@nyu.edu}

\author[K. H. Li]{Kwing Hei Li}
\orcid{0000-0002-4124-5720}
\affiliation{
  \institution{Aarhus University}
  \country{Denmark}
}
\email{hei.li@cs.au.dk}

\author[S. O. Gregersen]{Simon Oddershede Gregersen}
\authornote{The majority of this work was carried out while the author was affiliated with Aarhus University.}
\orcid{0000-0001-6045-5232}
\affiliation{
  \institution{New York University}
  \country{USA}
}
\email{s.gregersen@nyu.edu}

\author[J. Tassarotti]{Joseph Tassarotti}
\orcid{0000-0001-5692-3347}
\affiliation{
  \institution{New York University}
  \country{USA}
}
\email{jt4767@nyu.edu}

\author[L. Birkedal]{Lars Birkedal}
\orcid{0000-0003-1320-0098}
\affiliation{
  \institution{Aarhus University}
  \country{Denmark}
}
\email{birkedal@cs.au.dk}

\begin{document}

\begin{abstract}
  Probabilistic programs often trade accuracy for efficiency, and thus
  may, with a small probability, return an incorrect result.
  It is important to obtain precise bounds for
  the probability of these errors, but existing verification approaches have limitations that
  lead to error probability bounds that are excessively coarse, or only apply to first-order programs.
  In this paper we present \theaplog, a higher-order separation logic for proving
  error probability bounds for probabilistic programs written in an expressive higher-order language.

  Our key novelty is the introduction of \emph{error credits}, a separation logic resource
  that tracks an upper bound on the probability that a program returns an erroneous result.
  By representing error bounds as a resource,
  we recover the benefits of separation logic, including compositionality,
  modularity, and dependency between errors and program terms, allowing
  for more precise specifications. Moreover, we enable novel reasoning principles
  such as expectation-preserving error composition, amortized error reasoning,
  and error induction.

  We illustrate the advantages of our approach by proving amortized error
  bounds on a range of examples, including collision probabilities
  in hash functions, which allow us to write
  more modular specifications for data structures that use them as clients. We
  also use our logic to prove correctness and almost-sure termination of
  rejection sampling algorithms. All of our results have been mechanized
  in the Coq proof assistant using the Iris separation logic framework and
  the Coquelicot real analysis library.
\end{abstract}

\maketitle

\section{Introduction}
\label{sec:intro}

Randomness is an important tool in the design of efficient algorithms and data structures and is widely used in many application domains, including cryptography and machine learning.
In many cases, probabilistic programs are only approximately
correct because they admit unwanted behaviors with low probability, usually as a
trade-off for better performance.

%

One major class of randomized algorithms are so-called Monte Carlo algorithms; they admit wrong results with small probability, but are usually much faster than their
deterministic counterparts.
For example, primality tests such as Miller-Rabin \cite{miller,rabin} or Solovay-Strassen \cite{solovay-strassen} are Monte Carlo algorithms: they check the divisibility of a potential prime by a number of randomly selected candidates and answer in polynomial time with either ``probably prime'' (which may happen for composite inputs with low probability) or with ``certainly composite''. The other major class are Las Vegas algorithms. These never return wrong results, but their running time is a random variable which may take high values but generally with low probability. For example, a rejection sampler searching for a ``good'' sample in a large sample space might give up after a bounded number of iterations if no good sample can be found in time (Monte Carlo) or continue searching indefinitely (Las Vegas).

The trade-off between efficiency and accuracy/termination is typically justified by observing that the unwanted behavior occurs only with small probability.
Establishing bounds on the probability of these errors (from now on, \emph{``error bounds''}) is therefore an important prerequisite for the use of Monte Carlo and Las Vegas algorithms.
However, probabilistic reasoning is often counterintuitive and when combined with reasoning about the complex state space of probabilistic programs, it quickly becomes infeasible to manually establish error bounds for even moderately complicated programs.
To address this problem, probabilistic program logics offer a rigorous way to establish trust in the correctness of randomized programs.
For randomized first-order \texttt{WHILE} programs, the approximate Hoare logic~(aHL) of \citet{ub} provides a convenient way to over-approximate the error behavior of an algorithm.
Formally, aHL judgments are annotated with an ``error budget'' $\err$.
The judgment $\aHL{\err}{P}{c}{Q}$ means that when the precondition~$P$ holds, the probability that the postcondition~$Q$ is \emph{violated} after executing~$c$ is at most~$\err$.
The logic supports local reasoning through \emph{union bounds}:\footnote{The principle behind union bounds is also known as ``Boole's inequality''.} the error of a sequence of commands $c_1 ; c_2$ is bounded by $\err_1+\err_2$ if $c_1$ (resp.\ $c_2$) has error $\err_1$ (resp.\ $\err_2$) when considered in isolation. This principle is formalized in the rule for sequential composition:
\begin{equation*}
  \infrule[Right]{aHL seq}{\aHL{\err_1}{P}{c_1}{Q}
    \and
    \aHL{\err_2}{Q}{c_2}{R}
  }{\aHL{\err_1+\err_2}{P}{c_1; c_2}{R}}
\end{equation*}
Subsequent work~\cite{sato,adversarial} develops a higher-order union bound logic (HO-UBL) for a monadic presentation of a probabilistic $\uplambda$-calculus without recursion.

However, by baking the error bounds into the judgmental structure of the logic rather than treating them as ordinary propositions, these works on approximate correctness forego some of the ability to reason about errors in a modular way.
For instance, an error bound in aHL~\cite{ub} cannot depend on a program term, and HO-UBL~\cite[\S4.1]{adversarial} cannot prove the expected approximate higher-order specifications for simple functions such as \iterexpr because the error is a part of the judgement, and not a first class proposition which may itself occur in, \eg{}, a pre- or postcondition.

Furthermore, reasoning about composition via union bounds over-approximates error and can produce excessively coarse bounds when errors are not independent.
Specifically, the union bound for two events $A$ and $B$ bounds the probability $\pr{A \lor B} = \pr{A} + \pr{B} - \pr{A \land B}$ by $\pr{A} + \pr{B}$, thereby losing precision when $\pr{A \land B}$ is large.

%



In this paper, we present \theaplog: a higher-order separation logic with support for advanced reasoning principles for proving bounds on the probabilities of errors for programs written in \thelang, an expressive ML-like language with random sampling, full recursion, higher-order functions, and higher-order store.
%
Inspired by time credits~\citep{atkey,unionfind, tcred-iris} to reason about cost as a resource, we introduce and develop \emph{error credits}.
Ownership of $\err$ error credits is a first-class proposition in \theaplog, written $\upto \err$ (read: ``up to $\err$''), and proving an \theaplog{} specification $\vdash \hoare{P \ast \upto \err} e Q$ intuitively means: ``if $P$ holds then the probability of $e$ crashing or returning a result that violates $Q$ is at most $\err$''.

The resource approach affords \theaplog great flexibility: if we own an error credit $\upto \err$, we can choose to spend it however suits our proof.
We can ``pay'' for an operation which fails with probability $\err$, store it in an invariant describing a probabilistic data structure, frame it away during a function call to keep it for later, or split it into any number of credits $\upto{\err_1}, \ldots, \upto{\err_n}$ so long as we stay below the initial error budget, \ie $\sum_{i = 1}^n\err_i \leq \err$.




We now proceed to outline how \theaplog{} addresses the two limitations of prior work mentioned above (lack of modularity and conservative error bounds), and afterwards we explain how the error credits of \theaplog{} also support a novel form of reasoning about \emph{amortized error bounds} for randomized data structures.
Finally, we give an overview of how error credits in a
\emph{total correctness} version of \theaplog{} can be used to prove \emph{almost-sure termination} of Las Vegas algorithms. 

\paragraph{Modular specifications of higher-order programs}
\Theaplog can be used to give modular specifications to higher-order functions.
For a concrete example, consider the specification below for a (parallel) iterator function $\iterexpr$,
which is just the standard specification one would prove in a non-probabilistic setting.
In the precondition, the first line states that the argument $l$ is program-level representation for the mathematical sequence of values $xs$.
Then, the second line assumes a specification for the function $\iterf$ to be iterated over the list:
for each argument $\var$, $\iterf$ takes $P(\var)$ as a precondition and returns $Q(\var)$ as a postcondition.
The last line states that the precondition $P$ holds for each $\var$ in the list $xs$.
Finally, the postcondition states that $Q$ holds for each $\var$ in $xs$.
\begin{mathpar}
  \hoare {
    \begin{array}{c}
      \islist\ xs\ l
      \sep \\
      \All \var .
      \hoare{P (\var)}
      {\iterf\ \var}
      {Q(\var)}
      \sep \\
      \textstyle\Sep_{\var\in xs} P(\var)
    \end{array}}
  {\iterexpr\ \iterf\ l}
  {\Sep_{\var \in xs} Q(\var)}
\end{mathpar}
By instantiating the higher-order specification, we can reason in a setting in which the specification of $e$ only holds up to some error bound as shown below.
Now, the second line states that assuming that the precondition $R(\var)$ holds, the postcondition $Q(\var)$ will hold except with some error probability $\Err(\var)$.
Notably, the error can depend on the value of the argument $\var$.
The third line requires that the user of the \iterexpr has have enough error credits to ``pay'' for each call to $\iterf$.
\begin{mathpar}
  \hoare
  {
    \begin{array}{c}
      \islist\ xs\ l
      \sep \\
      \All \var .
      \hoare{R (\var) \sep \mhl{\upto{\Err(\var)}}}
      {\iterf\ \var}
      {Q(\var)}
      \sep \\
      \textstyle\Sep_{\var\in xs} \left(R(\var)  \sep \mhl{\upto{\Err(\var)}}\right)
    \end{array}}
  {\iterexpr\ \iterf\ l}
  {\Sep_{\var \in xs} Q(\var)}
\end{mathpar}
It is noteworthy that by treating error as a resource, the version \emph{with} error credits can be \emph{derived} from the standard version by instantiating $P(\var)$ to be $R(\var) \ast \upto{\Err(\var)}$.
The specification of $\iterexpr$ does not need to be reproven and the probabilistic version just becomes a special case.

\paragraph{More precise error bounds via expectation-preserving composition}
%

In addition to bringing union-bound reasoning in the style of \citet{ub,sato,adversarial} to a richer programming language, \theaplog supports a novel form of reasoning about errors \emph{in expectation}, which leads to more precise error bounds.
This feature hinges on two observations.

The first observation is a simple but useful consequence of treating errors bounds as separation logic resources:
as mentioned above, error bounds can depend on values of computations.
Concretely, consider the following instantiation of the so-called \emph{bind} rule:
\[
    \infrule<ht-bind-exp-intro>[Right]{ht-bind-exp}{
    \vdash \hoare{P \sep \upto {\err_1}}{\expr_1}{x \ldotp \upto {\Err_2(x)} \sep Q}
    \and
    \vdash \All x . \hoare{Q \sep \upto {\Err_2(x)}}{\expr_2}{R}
  }
  { \vdash \hoare{P \sep \upto {\err_1}}{\Let x = \expr_1 in \expr_2}{R}}
\]
The rule expresses that starting with $\err_1$ credits to begin with, if all evaluations of $\expr_1$ leave enough credits $\Err_{2}(x)$ to verify the continuation $R$, then $\err_1$ credits suffice to verify the entire $\langkw{let}$ expression.
Recall that the $x.$ in the postcondition of $\expr_1$ acts as a binder that captures the return value
of the computation.
By this rule, \theaplog supports \emph{value-dependent error composition}: in different branches of
$\expr_2$ we can spend different amounts of credits depending on $x$, giving us a more precise error
bound than the maximum error across all cases.
We present a concrete example of this phenomenon in \cref{sec:logic}.

The second observation is that, whenever the program takes a step and we initially have $\err_1$ credits, then we can split our $\err_1$ error credits across all possible branches as a \emph{weighted sum} according to the probability of each branch.
For example, if we sample $i$ uniformly from the set $\{0,\ldots,N\}$, and moreover we know that any continuation needs $\Err_2(i)$ credits, then it suffices to have $\expect{\Err_2} = \sum_{i=0}^N \Err_2(i)/(N+1)$ credits at the start.
This is captured formally in the proof rule for the random sampling operation shown below:
\[
  \infrule<ht-rand-exp-intro>[Right]{ht-rand-exp}
  { \textstyle\sum_{i=0}^{N}\dfrac{\Err_2(i)}{N+1} = \err_1 }
  { \vdash\hoare{\upto{\err_1}}{\Rand N}{n \ldotp \upto {\Err_2(n)}}}
\]
%
To put the two observations to use, let us consider a concrete example of a composite computation for which we get more precise error bounds than would be possible in previous work:
\begin{mathpar}
  \hoare{
  \begin{array}{c}
        \All \var .
        \hoare{\err}
        {\iterf\ 0}
        {\TRUE}
        \sep \\
        \upto{\err \cdot \frac K 2}
        \end{array}
  }{
  \begin{array}{l}
  \Let n = \Rand K in \\ \Let l = \langv{List.make}\ 0\ n in \iterexpr\ e\ l
  \end{array}}
  {\TRUE}
\end{mathpar}
Here we intend $\langv{List.make}\ 0\ n$ to construct a list of zeros of length $n$, where $n$ is sampled uniformly from the set $\{0,\dots,K\}$, and that $e$ is a function to be iterated, such that it with the argument $0$ requires $\err$ credits to execute safely (i.e.
without crashing).
Using the \rref[ht-bind-exp-intro]{ht-bind-exp} rule, we can then prove that the composed program executes safely if we have $\err \cdot \frac K 2$ credits in our precondition and set $\Err_2(n) = n \cdot \err$ in the postcondition of $\Rand K$.
By the specification of $\iterexpr$ we will need $\upto{\err}$ for every element of the list we iterate over, that is $\upto{\err \cdot n}$ for a list of length $n$.
Note that $\err \cdot \frac K 2$ is precisely the expected value of $\Err_2$, and in combination, the two observations mean that we obtain a form of \emph{expectation-preserving composition}.
We further discuss these rules in \cref{sec:logic}.

\paragraph{Amortized error bounds}
By representing error bounds as a resource, \theaplog not only addresses the limitations of prior
work mentioned above, but also supports reasoning about \emph{amortized error bounds} for operations of randomized data structures.
Inspired by the work on type-based resource analysis of~\citet{DBLP:conf/popl/HofmannJ03}, \Citet{atkey} pioneered the use of ``time credits'' to reason about amortized time complexity in separation logic, and the idea was subsequently extended and formalized in different separation logics \cite{unionfind,tcred-iris}.
Our use of error credits in turn allows us to give modular, amortized specifications of randomized data structures which hide implementation details such as the timing of ``costly'' (\ie error-prone) internal operations.
In \cref{sec:case-studies} we present several case studies that demonstrate how \theaplog{} supports modular reasoning about amortized error bounds.

\paragraph{Almost-sure termination using error credits}
%
So far, we have implicitly considered a partial correctness interpretation of the \theaplog{} Hoare triples.
In particular, this means that a divergent program trivially satisfies any Hoare triple;
 in \theaplog{} this can be proven using so-called guarded recursion/L{\"o}b induction.
This proof principle is sound in \theaplog{} because the semantics of Hoare triples is defined using a guarded fixed point.
Now, an interesting observation is that we can easily make a ``total-correctness'' version of \theaplog{}
called \theaplogtotal{} by instead defining the semantics of Hoare triples as a \emph{least fixed point}.
Because of the approximate up-to-error interpretation of Hoare triples this
yields an ``approximate total-correctness'' interpretation: a Hoare triple
with $\err$ credits in the precondition bounds the probability of never reaching a value satisfying the postcondition, which includes both the possibility of not satisfying the postcondition and the possibility of diverging. In particular, the probability of termination is \emph{at least} $1-\err$. In turn, if we can show a total Hoare triple for \emph{an arbitrary} error bound $\err$,
then we can conclude that it holds when $\err$ becomes vanishingly small and thus that the program almost-surely terminates, i.e. it
terminates with probability 1.
We state and prove these properties more formally in \cref{sec:ubtwp} and \cref{sec:soundness}; the soundness of this approach relies on a continuity argument for the semantics.
To the best of our knowledge, this argument and the approach of showing almost-sure termination via error credits is novel.
We demonstrate in \cref{sec:total-examples} how it can be used to prove correctness of several Las Vegas algorithms, including rejection samplers.

\paragraph{Contributions} To summarize, we provide:
\begin{itemize}[topsep=0pt]
  \item The first probabilistic higher-order separation logic, \theaplog, for \emph{modular} approximate reasoning (up-to-errors) about probabilistic programs written in \thelang, a randomized higher-order language with higher-order references.
  \item A resourceful account of errors which allows for more precise accounting of error bounds via value-dependent and expectation-preserving composition, and for reasoning about a richer class of properties, in particular amortized error bounds.
  \item A total correctness version of \theaplog{}, which can be used to establish \emph{lower} bounds on probabilities of program behaviors and thus to prove almost-sure termination.
  \item A substantial collection of case studies which demonstrate how the proof principles mentioned work in practice.
  \item All of the results in this paper have been mechanized (see~\cite{aguirre_2024_11489778}) in the \rocq proof assistant, building on the Iris separation logic framework \cite{irisho, irismonoid, irisjournal, irisinteractive} and the Coquelicot real analysis library \cite{coquelicot}.
\end{itemize}

\paragraph{Outline} In \cref{sec:lang} we recall some preliminaries and define the operational semantics of \thelang.
We then introduce \theaplog in \cref{sec:logic}.
We demonstrate how to use \theaplog on a range of case studies, focusing on amortized error bounds, in \cref{sec:case-studies}.
Afterwards, in \cref{sec:ubtwp} we describe how a total version \theaplogtotal of \theaplog can be used to reason about almost-sure termination via error credits; the section includes a number of case studies (and more can be found in the long version of the paper\footnote{A full version of the paper with appendix can be found at \url{https://arxiv.org/pdf/2404.14223}}).
Then we present the model of \theaplog in \cref{sec:soundness} and sketch how the model is used to prove soundness and adequacy of the logics.
Finally, we discuss related work in \cref{sec:related} and conclude and discuss future work in \cref{sec:conclusions}.


\section[Preliminaries and the Language λ\_(μ,ref,rand)]{Preliminaries and the Language \thelang{}}
\label{sec:lang}

In \cref{sec:prelim-proba} we first recall elements of discrete probability theory required to define the semantics of our probabilistic language \thelang and introduce the definitions we use to express approximate reasoning.
We subsequently define the syntax and operational semantics of \thelang{} in~\cref{sec:prelim-opsem}.

\subsection{Probabilities and Programs}
\label{sec:prelim-proba}

As a first approximation, one might expect the execution of a randomized program $\expr$ to produce a (discrete) probability distribution on values.
However, since programs may not terminate, programs might not induce \emph{proper} distributions, but rather \emph{subdistributions}, whose total mass is upper-bounded by~1, but may be lower.
\begin{definition}[Mass]
  Let $\nnreal$ denote the non-negative real numbers. For a countable set $X$, the \defemph{mass} of a function $f : X \rightarrow \nnreal$ is given by $\mass f = \sum_{x \in X} f(x)$ if this sum is finite.
\end{definition}
\begin{definition}[Subdistribution]
  A \defemph{(discrete) probability subdistribution} on a countable set $X$ is a function $\distr : X \to [0,1]$ such that $\mass \distr \leq 1$. We say that $\distr$ is a \defemph{proper} probability distribution if $\mass \distr = 1$.
  We write $\Distr{X}$ for the set of all subdistributions on $X$.
\end{definition}
We simply write \emph{distribution} to mean ``discrete probability subdistribution'' in the remainder of the paper. Unless otherwise specified, the variable $\distr$ denotes a distribution, and $X$ or $Y$ a countable set, typically the set of values, expressions, or configurations of \thelang.
\begin{lemma}[Probability Monad]
  Let $\mu \in \Distr{X}$, $x \in X$, and $f : X \to \Distr{Y}$.
  Then
  \begin{enumerate}
    \item $\mbind(f,\mu)(y) \eqdef{} \sum_{x \in X} \mu(x) \cdot f(x)(y)$
    \item $\mret(x)(x') \eqdef{}
            \begin{cases}
              1 & \text{if } x = x' \\
              0 & \text{otherwise}
            \end{cases}$
  \end{enumerate}
  gives a monadic structure to $\DDistr$.
  We write $\mu \mbindi f$ for $\mbind(f, \mu)$.
\end{lemma}
\begin{definition}[Restriction]
  Let $P$ be a predicate on $X$. The \defemph{restriction of $\distr$ to $P$} is given by:
  \vspace*{-1ex}
  \[ \restr \distr P(x) =
    \begin{cases}
      \distr(x) & \text{if $P(x)$ holds,} \\
      0         & \text{otherwise.}
    \end{cases}
  \]
\end{definition}
\begin{definition}[Probability of a Predicate]
  The \defemph{probability of a predicate $P$ with respect to $\distr$}, written as $\pr[\distr]P$, is the total probability mass of $\distr$ satisfying $P$, \ie
  $\pr[\mu]{P} = \mass{\restr \distr P}$.
\end{definition}


\subsection{Language Definition and Operational Semantics}
\label{sec:prelim-opsem}

The syntax of $\thelang{}$, the language we consider in this paper, is defined by the grammar below.
\begin{align*}
  \val, \valB \in \Val \bnfdef{}
                             & z \in \integer \ALT
  b \in \bool \ALT
  \TT \ALT
  \loc \in \Loc \ALT
  \Rec \lvarF \lvar = \expr \ALT
  (\val,\valB) \ALT
  \Inl \val  \ALT
  \Inr \val 
  \\
  \expr \in \Expr \bnfdef{}  &
  \val \ALT
  \lvar \ALT
  \Rec \lvarF \lvar= \expr \ALT
  \expr_1~\expr_2 \ALT
  \expr_1 + \expr_2 \ALT
  \expr_1 - \expr_2 \ALT
  \ldots \ALT
  \If \expr then \expr_1 \Else \expr_2 \ALT \\
  &(\expr_1,\expr_2) \ALT
  \Fst \expr \ALT
  \Snd \expr \ALT                                                                                       
  \Inl(\expr) \ALT
  \Inr(\expr) \ALT 
  \Match \expr with \Inl \val~ => \expr_1 | \Inr \valB => \expr_2 end \ALT  \\
  & \AllocN \expr_1~\expr_2 \ALT
  \deref \expr \ALT
  \expr_1 \gets \expr_2 \ALT
  \Rand \expr
  \\
  \lctx \in \Ectx \bnfdef{}  &
  -
  \ALT \expr\,\lctx
  \ALT \lctx\,\val
  \ALT \AllocN\lctx
  \ALT \deref \lctx
  \ALT \expr \gets \lctx
  \ALT \lctx \gets \val
  \ALT \Rand \lctx
  \ALT \ldots
  \\[.5ex]
  \state \in \State \eqdef{}& (\Loc \fpfn \Val)
                              \qquad\qquad
                              \cfg \in \Conf \eqdef{} \Expr \times \State
\end{align*}
The term language is mostly standard:
$\AllocN \expr_1~\expr_2$ allocates a new array of length $\expr_1$ with each cell containing the value returned by $\expr_2$, $\deref \expr$ dereferences the location $\expr$ evaluates to, and $\expr_{1} \gets \expr_{2}$ assigns the result of evaluating $\expr_{2}$ to the location that $\expr_{1}$ evaluates to. 
We introduce syntactic sugar for lambda abstractions $\Lam \var . \expr$ defined as $\Rec {\_} \var = \expr$,
let-bindings $\Let \var = \expr_{1} in \expr_{2}$ defined as $(\Lam \var . \expr_{2})~\expr_{1}$,  sequencing $\expr_{1} ; \expr_{2}$ defined as $\Let \_ = \expr_{1} in \expr_{2}$, and references $\Alloc \expr$ defined as $\AllocN~\expr~1$. We write $l[b]$ as sugar for offsetting location $l$ by $b$, defined as $(l + b)$.

Our language matches that of Clutch~\cite{clutch}, modulo the minor difference that we add arrays.\footnote{As in Clutch, we support presampling tapes, but since they are not used until \cref{sec:ubtwp}, we relegate this discussion to \cref{sec:rules-tapes}.}
States in \thelang{} are finite maps from memory locations to values.

To define full program execution, we define $\stepdistr(\cfg) \in \Distr{\Conf}$, the distribution induced by the single step reduction of configuration $\cfg \in \Conf$. The semantics of $\stepdistr$ is standard: all non-probabilistic constructs reduce deterministically as usual, \eg{}, $\stepdistr(\If \True then \expr_1 \Else \expr_2, \sigma) = \mret(\expr_{1}, \sigma)$, and the probabilistic choice operator $\Rand \tapebound$ reduces uniformly at random:
\begin{align*}
   & \stepdistr(\Rand \tapebound, \sigma)(n, \sigma) =
  \begin{cases}
    \frac{1}{\tapebound + 1} & \text{for } n \in \{ 0, 1, \ldots, N \}, \\
    0                        & \text{otherwise}.
  \end{cases}
\end{align*}
The Boolean operation $\Flip$ is syntactic sugar for $(\Rand 1 == 1)$.

With the single step reduction $\stepdistr$ defined, we now define a stratified execution probability $\exec_{n}\colon \Conf \to \Distr{\Val}$ by induction on $n$:
\begin{align*}
 \exec_{n}(\expr, \state) \eqdef{}
 \begin{cases}
   \mathbf{0}                                           & \text{if}~\expr \not\in\Val~\text{and}~n = 0, \\
   \mret(\expr)                                         & \text{if}~\expr \in \Val, \\
   \stepdistr(\expr, \state) \mbindi \exec_{(n - 1)} & \text{otherwise.}
 \end{cases}
\end{align*}
where $\mathbf{0}$ denotes the everywhere-zero distribution.
The probability that a full execution, starting from configuration $\cfg$, reaches a value $\val$ is taken as the limit of its stratified approximations, which exists by monotonicity and boundedness:
\begin{align*}
 \exec(\cfg)(\val) \eqdef{} \lim_{n \to \infty} \exec_{n}(\cfg)(\val)
\end{align*}
We simply write $\exec \expr$ as notation for $\exec (\expr, \state)$ if $\exec(\expr, \state)$ is the same for all states $\state$.


As an example, consider the program $\expr \eqdef \If \Flip \And \Flip then 42 \Else \Omega$, where $\Omega$ is a diverging term and $\And$ denotes logical conjunction. If we execute $\expr$, we either obtain the value $42$ in a few steps (both $\Flip$s return $\True$ with probability $0.5 \times 0.5 = 0.25$), or we do not obtain a value at all otherwise. In other words, $\exec \expr$ induces the subdistribution
$\set{42 \mapsto 0.25,\; \_~\mapsto 0 } :\  \Val \rightarrow [0,1]. $


\section{The \theaplog Logic}
\label{sec:logic}

In this section we introduce the \theaplog{} logic.
We first present the propositions of \theaplog{}, and then to provide some intuition for the program logic proof rules we present the adequacy theorem, which expresses what one can conclude by proving a Hoare triple in \theaplog{}.
The adequacy theorem itself is only proved later (\cref{sec:soundness-proof}) when we introduce the semantic model of \theaplog{}.
After the adequacy theorem, we then present a selection of the program logic rules of \theaplog.


\Theaplog is based on the Iris separation logic framework~\cite{irisjournal} and inherits all of the basic propositions and their associated proof rules.
An excerpt of \theaplog propositions is shown below, including the later modality $\later$, the persistence modality $\always$ and the points-to
connective $\progheap{\loc}{\val}$, which asserts ownership of the location $\loc$ and its content $\val$:
\begin{align*}
  \prop,\propB \in \iProp \bnfdef{}
  & \TRUE \ALT \FALSE \ALT \prop \land \propB \ALT \prop \lor \propB \ALT \prop \Ra \propB \ALT
   \All \var . \prop \ALT \Exists \var . \prop \ALT \prop \sep \propB \ALT \prop \wand \propB \ALT \\
  & \always \prop \ALT
    \later \prop \ALT
    \progheap{\loc}{\val} \ALT
    \upto{\err} \ALT
     \hoare{\prop}{e}{\propB} \ALT
    \ldots
\end{align*}

The main novelty of \theaplog is the program logic $\hoare{\prop}{\expr}{\propB}$ and the new $\upto \err$ assertion which denotes ownership of $\err$ error credits.
Error credits satisfy the following rules:
\begin{mathpar}
	\upto{\err_1} \ast \upto{\err_2} \dashv \vdash  \upto{\err_1 + \err_2}
	\and
	\upto{\err_1} \ast (\err_2 < \err_1) \vdash  \upto{\err_2}
	\and
	\upto{1} \vdash \FALSE
\end{mathpar}
From the point of view of a user of the logic, this interface (and the rule \rref{ht-rand-exp} below) is all they need to know about credits.
The first rule expresses that ownership of $\err_1 + \err_2$ error credits is the same as ownership of $\err_1$ credits and ownership of $\err_2$ credits.
The second rule says that it is sound to throw away credits that we own.
Finally, the last rule says that if we own $1$ full error credit, then we can immediately conclude a contradiction.
Intuitively, ownership of $\upto{1}$ corresponds to proving a statement holds with probability at least $0$, which is trivially true.




We define the semantics of the program logic in \cref{sec:soundness}.
The adequacy theorem shown below captures what a specification with error credits means in terms of probabilities and the operational semantics.
\begin{theorem}[Adequacy]\label{thm:adequacy}
	If $\vdash \hoare{\upto \err}{\expr}{\pprop}$ then $\prex{\expr}{\neg \pprop} \leq \err$. Moreover, the probability of $\expr$ getting stuck is at most $\err$.
\end{theorem}
The adequacy theorem states that if we prove $\hoare{\upto \err}{\expr}{\pprop}$ in \theaplog, for any meta-logic postcondition $\pprop$, then the final distribution obtained from running $\expr$ does \emph{not} satisfy $\pprop$ with at most probability $\err$.
It also states that $\expr$ is safe with probability at least $1-\err$.

As mentioned in \cref{sec:intro}, \theaplog is a partial correctness logic, which means that a diverging program satisfies any specification, and diverging traces of $\expr$ will also be considered correct.
Hence  $\err$ is an \emph{upper bound} on the probability of terminating and not satisfying $\pprop$.
Later on, we will present \theaplogtotal, a total correctness version of \theaplog, which allows us to establish lower bounds on the probability of terminating and satisfying $\pprop$.

Although our definition of Hoare triples is new,
the program logic rules for the deterministic fragment of \thelang are essentially standard, \eg{},
  \begin{mathpar}
	  \infrule[lab]{ht-frame}
	  {\vdash \hoare{\prop}{\expr}{\propB}}
	  {\vdash \hoare{\prop \ast \propC}{\expr}{\propB \ast \propC }}
    \and
	  \infrule[lab]{ht-bind}
	  {\vdash \hoare{\prop}{\expr}{\val . \propB} \\ \vdash \All \val. \hoare{\propB}{\fillctx\lctx[\val]}{\propC}}
	  {\vdash \hoare{\prop}{\fillctx\lctx[\expr]}{\propC}}
    \and
	  \infrule[lab]{ht-load}
    { \progheap{\loc}{\valB} \vdash \prop(\valB)}{\proves \hoare{\progheap{\loc}{\valB}}{\deref\loc}{\val \ldotp \prop(v) }}
    \and
	  \infrule[lab]{ht-rec}
    {\All \valB . \hoare{\prop}{(\Rec f x = e)\ \valB}{\propB} \vdash \hoare{\prop}{\subst{\subst{e}{x}{\val}}{f}{(\Rec f x = e)}}{\propB} }
    {\vdash \hoare{\prop}{(\Rec f x = e)\ \val}{\propB}}
  \end{mathpar}
Note in particular that \theaplog includes the standard rule \rref{ht-rec} for reasoning about recursive functions.

We do not have a specialized rule for composing errors for composite computations as in previous works.
The error is just a resource which, as mentioned in \cref{sec:intro}, allows us to derive an aHL-style composition rule as well as the value-dependent composition rule shown in \cref{sec:intro} from the resource rules of credits and the \rref{ht-bind} and \rref{ht-frame} rules.
These derived rules are shown below.
\begin{mathpar}
	\infrule[Right]{ht-bind-simpl}
	  {\vdash \hoare{\prop \ast \upto{\err_1}}{\expr}{\val . \propB} \\ \vdash \All \val. \hoare{\propB \ast \upto{\err_2}}{\fillctx\lctx[\val]}{\propC}}
	  {\vdash \hoare{\prop \ast \upto{\err_1 + \err_2}}{\fillctx\lctx[\expr]}{\propC}}
   \and
  \infrule[Right]{ht-bind-exp}{
    \vdash \hoare{P \sep \upto {\err_1}}{\expr_1}{x. \upto {\Err_2(x)} \sep Q}
    \and
    \vdash \All x . \hoare{Q \sep \upto {\Err_2(x)}}{\expr_2}{R}
  }
  { \vdash \hoare{P \sep \upto {\err_1}}{\Let x = \expr_1 in \expr_2}{R}}
\end{mathpar}

\paragraph{Proof rules for sampling}
The only rules that make direct use of error credits are our novel rules involving sampling. 
As mentioned in \cref{sec:intro}, \theaplog{} includes a general rule for sampling, which takes the \emph{expected} number of error credits into account.
When taking a probabilistic step, this allows us to make the error depend on the result of the step.
Suppose that we sample $i$ uniformly from the set $\{0, \dots, N \}$ and, moreover, that we know that any continuation needs $\Err_2(i)$ credits, where $\Err_2 : \{0, \dots, N \} \to [0,1]$.
Then it suffices to have $\expect{\Err_2}$ at the start, since it is the expected number of error credits that our proof will need.
We can also interpret this principle forwards: whenever we take a step, we can split our $\err$ error credits across all possible branches as a weighted sum according to the probability of each branch.
Formally, this is captured in the following rule:
\newcommand{\RandAdvComp}{
  \infrule[right]{ht-rand-exp}
  { \textstyle\sum_{i=0}^{N}\dfrac{\Err_2(i)}{N+1} = \err_1 }
  { \vdash\hoare{\upto{\err_1}}{\Rand N}{n \mathrel{.} \upto {\Err_2(n)}} }
}
\begin{mathpar}
  \RandAdvComp
\end{mathpar}


In practice, we often want to avoid the generality of this rule as it requires explicit accounting of error credits for every possible outcome.
We derive rules for applications that only require a more coarse-grained analysis.
For instance, when sampling from $\{0, \dots, N \}$, we can guarantee that the result of our sampling will not be in a list of error values $xs$ by using the following derived rule: \vspace*{-.7\baselineskip}
\[
  \infrule[Right]{ht-rand-err-list}
  {  }
  { \vdash \hoare{\upto{\mathsf{length}(xs) / (N+1)}}{\Rand N}{n \mathrel{.} n \not\in xs} }
\]
In order to derive the rule, it suffices to apply \rref{ht-rand-exp} using
\[ 
  \Err_2(n) \triangleq \If (n \in xs) then 1 \Else 0
\]
and immediately discard all outcomes in $xs$ using $\upto{1} \proves \bot$.




%

\paragraph{Example}


\begin{figure}[t]
  \centering
  \small
  \phantom{MMM}\begin{minipage}[c]{.15\linewidth}\small
    \begin{align*}
       & \tikzmark{top}\Let n = \Rand 3 in                              \\
       & \If n \leq 1 then                                              \\
       & \quad \True \tikzmark{trueN}                                   \\
       & \Else                                                          \\
       & \quad\tikzmark{bottop}\Let k = \Rand 1 in                      \\
       & \quad\If (n+k \leq 2) then \                                   \\
       & \quad\quad \True \tikzmark{trueK}                              \\
       & \quad\ElIf (n+k = 3) then                                      \\
       & \quad\quad \False \tikzmark{false}                             \\
       & \quad\Else                                                     \\
       & \tikzmark{topbot}\quad\tikzmark{bot}\quad\Omega \tikzmark{div}
    \end{align*}
    \DrawVerticalBrace[gray,thick][node[black, midway, left]{$\expr \eqdef$\;}]{top}{topbot}
    \DrawVerticalBrace[gray,thick][node[black, midway, left]{$\expr' \eqdef$\hspace*{.5pt}}]{bottop}{bot}
  \end{minipage}
  \qquad
  \begin{minipage}[c]{.65\linewidth}
    \begin{tikzpicture}
      [edge from parent path={[->](\tikzparentnode\tikzparentanchor) -- (\tikzchildnode\tikzchildanchor)},
      level distance=18mm,
      level 1/.style={sibling distance=25mm},
      level 2/.style={sibling distance=12mm,level distance=13mm}]
      \node[draw, rounded rectangle] {\parbox[b][9ex][c]{10ex}{\center $\expr$\\$\err_p{=}\frac 2 8, \err_t{=}\frac 3 8$}}
      child {
        node[draw, rounded rectangle] {\tikzmark{phiN} {\parbox[b][3ex][c]{14ex}{\center $\frac{4}{8}:\phi$}}}
        edge from parent node[left=2mm, pos=0.3]{$\frac 1 2$}
      }
      child {
        node[draw, rounded rectangle] { {\parbox[b][3ex][c]{8ex}{\tiny \center $\expr'[n{:=}2]$ \\ $\err_p = \err_t {=} \frac 1 2$ }}}
        child {
          node[draw, rounded rectangle]
          {\parbox[b][3ex][c]{5ex}{$\frac{1}{8}:\tikzmark{phiK}\phi$}}
          edge from parent node[left,pos=0.4]{$\frac 1 2$}}
        child {
          node[draw, rounded rectangle]
          {\parbox[b][3ex][c]{5ex}{$\frac{1}{8}{:\tikzmark{negphiOne}}\,\neg \phi$}}
          edge from parent node[right,pos=0.4]{$\frac 1 2$}}
        edge from parent node[right,pos=0.45]{$\frac 1 4$}
      }
      child {
        node[draw, rounded rectangle] { {\parbox[b][3ex][c]{8ex}{\tiny \center $\expr'[n{:=}3]$ \\ $\err_p{=}\frac 1 2, \err_t{=}1$ }}}
        child {
          node[draw, rounded rectangle]
          {\parbox[b][3ex][c]{5ex}{$\frac{1}{8}{:\tikzmark{negphiTwo}}\,\neg \phi$}}
          edge from parent node[left,pos=0.4]{$\frac 1 2$}}
        child {
          node[draw, rounded rectangle]
          {\parbox[b][3ex][c]{5ex}{$\frac{1}{8}:\tikzmark{Uparrow}{\Uparrow}$}}
          edge from parent node[right,pos=0.4]{$\frac 1 2$}}
        edge from parent node[right=2mm, pos=0.3]{$\frac 1 4$}
      };
    \end{tikzpicture}
    \vspace*{2cm}
  \end{minipage}
  \begin{tikzpicture}[overlay,remember picture]
    \draw[->,thick] ($(trueN)+(.5ex,.5ex)$) .. controls +(right:1cm) and +(left:1cm) .. ($(phiN)+(-1.85ex,1.25ex)$);%
    \draw[->] ($(trueK)+(.5ex,.5ex)$) .. controls +(right:1.5cm) and +(left:1cm) .. ($(phiK)+(-5.25ex,0.5ex)$);%
    \draw[->] ($(false)+(.5ex,.5ex)$) .. controls +(right:3cm) and +(down:0.5cm) .. ($(negphiOne)+(0ex,-2.25ex)$);%
    \draw[->] ($(false)+(.5ex,.5ex)$) .. controls +(right:3cm) and +(down:2cm) .. ($(negphiTwo)+(0ex,-2.25ex)$);%
    \draw[->] ($(div)+(.5ex,.5ex)$) .. controls +(6cm,0cm) and +(down:2.25cm) .. ($(Uparrow)+(-.5ex,-2.25ex)$);%
  \end{tikzpicture}%
  \caption{Expected error analysis, partial and total.}
  \label{fig:expected-error}
\end{figure}

Consider the program $\expr$ and its probabilistic execution tree shown in \cref{fig:expected-error}.
The program first samples a natural $n$ at random from $0$ to $3$, returning true if it chooses $0$ or $1$.
Otherwise the sub-program $\expr'$ samples a random bit $k$ and branches on the sum of $n + k$: when
$n + k \leq 2$  the program returns $\True$, when $n + k = 3$ the program returns $\False$, and
otherwise the program diverges (denoted by $\Omega$).

%
We will now show how to use \theaplog{} to show that $\expr$ returns $\True$ up to some
error bound.
Before we present the proof, let us take a step back and ask ourselves: What specification can we hope to show for $\expr$? 
If we consider all the possible executions of $\expr$, we can determine that it returns a value that satisfies $\phi$ with probability $\frac{5}{8}$, returns a value that does \emph{not} satisfy $\phi$ with probability $\frac{1}{4}$, and loops forever with probability $\frac{1}{8}$. Therefore, we should be able to prove the Hoare triple:
\begin{mathpar}
  \hoare{
    \upto{\textstyle{\frac{1}{4}}}
  }{\expr}
  {\phi}
\end{mathpar}
To prove this triple, we first prove a triple for the subexpression $\expr'$:
\begin{mathpar}
  \hoare{
    (n = 3 \lor n = 2) \ast \upto{\textstyle{\frac{1}{2}}}
  }{\expr'}
  {\phi}
\end{mathpar}
This Hoare triple is not difficult to prove; during the $\Rand 1$ step, we use the \rref{ht-rand-err-list} rule to spend $\upto{\frac{1}{2}}$ and ``avoid'' values that eventually lead to undesirable outcomes, e.g.\ returning $\False$.
To be more specific, for the case where $n=2$, we avoid sampling $1$ as that branch eventually reduces to $\False$. Similarly, for the case where $n=3$, we avoid sampling $0$.
After verifying the sub-program $\expr'$, we can now turn to verifying the overall program $\expr$. Notice that after assigning a random value to $n$, the number of error credits that we need for the continuation depends on $n$. This dependency is captured by the $\Err_2$ function defined below:
\[ \Err_2(n) = \begin{cases}
    0           & x<2          \\
    \frac{1}{2} & x=2 \lor x=3
  \end{cases}
\]
Since $\sum \Err_2(i)/4 = 1/4$ we apply the \rref{ht-rand-exp} rule to conclude $\hoare{\upto{\frac{1}{4}}}{\Rand 3}{n \ldotp \Err_2(n)}$ and by using \rref{ht-bind-exp} we complete the proof.

To summarize, this example demonstrates how we can use the proof rules of \theaplog to distribute error credits across many branches in a fine-grained manner and establish a strict expected error bound.
Moreover, the error analysis in the proof is modular in the sense that we prove the properties of the sub-program $\expr'$ without taking into account the context in which it is used.


\section{Case Studies}
\label{sec:case-studies}

In this section we present a series of case studies that showcase the features
and reasoning principles introduced by \theaplog. Due to the representation of
errors as a resource, \theaplog allows us to do precise, value-dependent
reasoning about error bounds for operations on randomized data structures.
However, when using these randomized data structures as a client, value
dependent specifications often reveal too many details about the internal state
of the data structure and its implementation. Here we also show how to do
\emph{amortized error} reasoning which, analogously to bounds on amortized running time,
assigns a uniform error cost to every operation on the data structure despite
the real error cost varying over time.



\subsection{Dynamic Vectors under a Faulty Allocator}
\label{sec:faulty-vect-alloc}

A quintessential example of amortized \emph{time complexity} reasoning is
that of a vector with dynamic resizing, see~\cite{clrs}. We assume that on
initialization we will allocate a memory block of size $N$, which allows us
to do $N$ insertions (each of cost $1$) into the vector. For the $(N+1)$-th
insertion however, we allocate a new block of memory of size $2N$ and
copy the contents of the vector into the new block. This operation incurs a
cost of $N+1$, paying $N$ to copy the initial $N$ elements into
the new memory block, and $1$ for the actual insertion. Using amortized cost
reasoning we can argue that each insertion has amortized cost $3$: $1$
for the insertion itself, $1$ to pay for the first time it gets copied, and $1$
to pay to copy another element that was inserted and moved previously.

We will use a similar intuition to introduce amortized \emph{error} reasoning.
Here we consider a faulty memory allocator which has a small probability
$\err$ of failing on each write operation. Specifically, we assume that the allocator
offers two methods $\langv{extend}$ and $\langv{store}$ with specifications:
\begin{align*}
  &\hoare {\upto{n \cdot \err} \sep l \mapsto^\ast vs}{\langv{extend}~n~l}
  {l' \ldotp l' \mapsto^* (vs \dplus \langv{replicate}(n, ()) ) } \\
  &\hoare {\upto{\err} \sep l \mapsto^* vs \sep n < \langv{length}(vs)}
  {\langv{store}~l~n~v}
  {l \mapsto^* (vs[ n := v])}
\end{align*}
Here, $\mapsto^* vs$ denotes ownership of a points-to connective for each
element of the list $vs$. Provided we own $l \mapsto^* vs$, we can get a
new, $\langv{extend}$ed memory block starting at a new location $l'$
containing the old array with $n$ new empty locations (containing $()~$)
appended. This incurs an error cost of $\upto{n \cdot \err}$.
We can also use the $\langv{store}$ operation to write to any position $n$ within $vs$ and
update its value, again by incurring an error cost of $\upto{\err}$.
Consider now the following code for the $\langv{pushback}$ method which adds an element $v$ to the end of the vector $\langv{vec}$. It is parametrized by two methods $\langv{ext}$ and $\langv{str}$ for extending and
storing:
\begin{align*}
  \langv{pushback}\ \langv{ext}\ \langv{str}\ \langv{vec}\ v\ \triangleq \quad
  &\Let (l, s, r) = \langv{vec} in\\
  &\langv{str}\ l\ s\ v ; \\
  &\If s + 1 == r then (\langv{ext}\ r\ l, s+1, 2 \cdot r)\\
	&\hspace{4.9em}\Else (l, s+1, r)
\end{align*}
A vector is a tuple $(l, s, r)$ of a location $l$ pointing to the start of the vector and two integers $s, r$ denoting the current size of the vector and the current size of the allocated block, respectively.
On insertion we store value $v$ at position $s$, and if we reach the end of the current allocated space we resize it so that the new block has size $2 \cdot r$.

The representation predicate for the vector looks as follows:
\begin{align*}
  \logv{vec\_spec}\ \langv{vec}\ \langv{vs} \triangleq
  \Exists l, s, r, xs, p . & vs = (l, s , r) \sep \upto{p} \sep l \mapsto^* (vs \dplus xs) \sep{} \\
                           &  s < r \sep s = \langv{length}(vs) \sep{}
                             r = \langv{length}(vs) + \langv{length}(xs) \sep{} \\
                           & p + 2 \cdot \err \cdot \langv{length}(xs) = r \cdot \err
\end{align*}
Here, $\logv{vec\_spec}\ \langv{vec}\ \langv{vs}$ should be read as ``$\langv{vec}$ is a vector containing
the values $\langv{vs}$''. Internally, the representation predicate contains the starting location
of the vector $l$, its current size $s$, the size of the allocated space $r$ and  a list of dummy values $xs$.
Crucially, it also stores a reserve of $p$ error credits. We also know that there are $\langv{length}(xs)$
insertions remaining until resizing, and on each we will leave $2\err$ credits to spare.
Altogether, this suffices until the next resizing, which has cost $r \cdot \err$.

With this representation predicate, we can prove the following specification for $\langv{pushback}$.
\begin{align*}
  \hoare{ \logv{vec\_spec}\ \langv{vec}\ \langv{vs} \sep \upto{3\cdot \err} }
	{ \langv{pushback}\ \langv{extend}\ \langv{store}\ \langv{vec}\ \langv{v}}
	{ \langv{vec'}, \logv{vec\_spec}\ \langv{vec'}\ \langv{vs}\dplus[\langv{v}] }
\end{align*}
Ignoring the error credits for the moment, this is a natural specification: if we have a
vector containing $\langv{vs}$ and we append $\langv{v}$, we get a vector
containing $\langv{vs}\dplus[\langv{v}]$. We just give a quick sketch of the
proof, focusing on the accounting of credits, as the rest is
standard separation logic reasoning. First, we split $\upto{3\cdot \err}$
into $\upto{\err} \sep \upto{2\cdot\err}$, using the first
$\err$ credits to pay for the call to $\langv{store}$. From the definition of
the representation predicate we get $\upto p$, and we can split the proof
into two cases depending on whether $s + 1 < r$ or $s + 1 = r$. In the first
case, which steps into the $\langkw{else}$ branch of the conditional, we store back
$\upto{2\cdot\err + p}$ into the representation predicate. It is easy to see that
\[
  p + 2 \cdot \err + 2 \cdot \err \cdot (\langv{length}(xs) - 1) = r \cdot \err
\]
since we will overwrite the first dummy location of $xs$.
In the second case, we step into the $\langkw{then}$ branch to resize. Here we know from the
representation predicate that $\langv{length}(xs) = 1$ since there is only one
dummy location left to be overwritten. Therefore, the representation predicate implies
\[
  p + 2 \cdot \err = r \cdot \err
\]
We own exactly $\upto{p} \sep \upto{2 \cdot \err}$, which we use to
pay for the extend operation with cost $\upto{r \cdot \err}$. At the
end, we store 0 error credits into the representation predicate. This completes
the proof.

\subsection{Amortized Error for Collision-Free Hash Functions}
\newcommand{\cfhf}{\logv{cf\_hashfun}}
\newcommand{\acfhf}{\logv{amort\_cf\_hashfun}}
\newcommand{\totalsize}{n+1}
\newcommand{\maxsize}{\text{\textsf{MAX}}}
\newcommand{\amortizederr}{\err_{\maxsize}}
\newcommand{\queryf}{\logv{hfun}}

We now implement a model of an idealized hash function under the \textit{uniform hash assumption}~\cite{uniform-hash-assumption}, \ie, a hash function $h$ from a set of keys $K$ to values $V$ that behaves as if, for each key $k$, the hash $h(K)$ is randomly sampled from a uniform distribution over $V$ independently of all other keys.
We implement the model using a mutable map $lm$, which serves as a cache of hashes computed so far.
If the key $k$ has already been hashed we return the value stored in $lm(k)$.
Otherwise, we sample a fresh value uniformly from $V=\{0, \dots, n\}$, store it in $lm(k)$, and return it.
\begin{align*}
  \computehash\ lm\ v \eqdef{}
   & \MatchML \mapget\ lm\ v with
  \Some(b) => b
  | \None =>
  {\begin{array}[t]{l}
     \Let b = \Rand n  in \\
     \mapset\ lm\ v\ b;    \\
     b
   \end{array}}
  end {}
\end{align*}

To reason about the correctness of many data structures, we often assume that a
hash function is \textit{collision-free} in the sense that for the finite
number of times we query the hash function, different input keys will return
different hash values. In reality collisions may occur, but when the size of $V$ is
magnitudes larger than the number of times we use the hash function it is common
to postulate that the hash function will remain collision-free, up
to some small error. 

To be precise, suppose we have queried $\queryf\eqdef{}\computehash\ lm$ a total of $s$
times, each with a distinct input, and that the map is still collision-free
(that is, we have observed $s$ different values). If we apply the hash
function to a completely new input, in order to maintain the collision-free
property the hash function needs to ``avoid'' sampling any of the
previous $s$ hash outputs. We can reason about this by means of the \rref{ht-rand-err-list}
rule, meaning we would need to pay $\upto{\frac{s}{\totalsize}}$ when choosing the new hash.
We can encode this as a specification for our hash function in \theaplog{}:
\begin{mathpar}
  \hoare
  {n\notin \dom\ m \sep 
    \cfhf\ lm\ m\ V \sep     
    \upto{\tfrac{\text{size}(m)}{\totalsize}}
  }
  {\queryf\ n}
  {v \ldotp \cfhf\ lm\ (\mapinsert{n}{v}{m})\ V}
\end{mathpar}

The predicate $\cfhf\ lm\ m\ V$ states that the mutable map $lm$ tracks the finite partial function $m : \nat \rightharpoonup \{0, \dots, V \}$ represented as a finite map, and furthermore states that $m$ is injective (\ie there are no collisions).
After querying the hash function for an unhashed key $n$, it will return a value $v$ and update the mutable map to track the finite map $\mapinsert{n}{v}{m}$, which is again injective.

One limitation of the above specification is that the error requirement for each
hash operation is proportional to the size of the map. This leads to worse
modularity, since a client of this data structure needs to know how many
queries have been performed before, which may be challenging \eg, in the presence of concurrency
where multiple clients may share the same hash function.

One possible solution is to fix a maximum global number of hash queries $\maxsize$
and amortize the error over all those queries, so that for each
query, the error one needs to pay is a fixed constant that is not dependent on
the inner map. As with the previous example, we will implement this using error credits.

Starting from an empty map, if we bound the number of queries to be $\maxsize$, the total number of error credits used for the $\maxsize$ queries is $\sum_{i=0}^{\maxsize-1} \frac{i}{\totalsize} = \frac{(\maxsize-1)*\maxsize}{2(\totalsize)}$. We will require that the client always incurs the mean error $\frac{(\maxsize-1)*\maxsize}{2(\totalsize)*\maxsize}=\frac{(\maxsize-1)}{2(\totalsize)}$.
Let $\amortizederr= \frac{(\maxsize-1)}{2(\totalsize)}$. Updating our specification, 
\begin{mathpar}
  \hoare
  {\begin{array}{c}
     \text{size}(m)<\maxsize \sep
      n\notin \dom\ m \sep \\
      \acfhf\ lm\ m\ V \sep
      \upto{\amortizederr}
    \end{array}
  }{\queryf\ n}
  {v \ldotp \acfhf\ lm\ (\mapinsert{n}{v}{m})\ V}
\end{mathpar}

In \theaplog{}, this new specification is \emph{derivable} from the original non-amortized specification. We accomplish this by defining the abstract predicate $\acfhf$ to not only contain the $\cfhf$ resource, but also a reserve of extra error credits which the clients paid in excess for the first half of the hash operations (similar to the dynamic vector example). For the second half of the hash operations, when the mean error $\amortizederr$ is insufficient to apply the original specification, we draw the additional error credits from the reserve in $\acfhf$.
By using error credits, we provide a simpler interface to our initial specification which alleviates the error accounting burden from clients of $\acfhf$.

\subsection{Collision-Free Resizing Hash Functions}
\label{sec:cf-hash}

\newcommand{\computehashrs}{\langv{hash\_rs}}
\newcommand{\acfhfrs}{\logv{cf\_hash\_rs}}

We can go one step further and implement a collision-free hash function with constant amortized insertion error, but without imposing any \emph{a priori} limit on the number of insertions.
Of course with a fixed set $V$ of possible hash values (as in the implementation above), collisions are eventually unavoidable.
Instead, we will keep the probability of collision low by resizing the \textit{sample space} once a threshold of inserted elements is reached.
One way to think of this model is to assume that the hash function gets values over a much larger sample space, but initially we only look at the first $n$, and every time we resize we look at the $(n+1)$-th  bit.

As in the previous example, the hash is sampled lazily. In addition to a
mutable map $m$ the hash function will keep track of three quantities $V$, $S$, and $R$.
Here $V$ represents the size of the value space of our hash function, which are nonnegative integers
over $\{0,\dots,V\}$. The value $S$ represents the current size of the domain of the hash function, and $R$
represents a threshold on the amount of stored values after which the hash will resize.
That is, once $S$ reaches $R$, we will update the hash so that  $R$ becomes $2\cdot R$ and $V$ becomes $2\cdot V$.
Initially $V,S,R$ are set to some default values $V_0, 0, R_0$. We will prove that
overall, the hash will remain collision-free with an amortized error of $(3
\cdot R_0) / (4 \cdot V_0)$ per insertion, \emph{no matter the number of insertions}.\footnote{Of course,
even with this constant error cost, if we execute a large enough number of insertions we will eventually have
consumed over $1$ error credit. The advantage of this specification is
that it will enables us to do more modular proofs, since the cost will be constant
independently of the internal state of the hash function.}
For instance, to keep the amortized error below $2^{10}$ we can initially set $V_0 = 2^{10} \cdot R_0$.

The code for querying the hash function is shown below:
\begin{align*}
  \computehashrs\ hf\ w \eqdef{}
   & \Let{(lm, v, s, r)} = hf in \\
   & \MatchML \mapget\ lm\ w with
  \Some(b) => (b, hf)
  | \None =>
  {\begin{array}[t]{l}
     \Let b = \Rand~(v-1)  in \\
     \mapset\ lm\ w\ b;    \\
     \If s + 1 = r then (b, (lm, 2 \cdot v, s+1, 2 \cdot r)) \\
     \Else (b, (lm, v, s+1, r))
   \end{array}}
  end {}
\end{align*}
Note that, the code is analogous to the non-resizing hash besides tracking the size: in the case where $s + 1 = r$ we double the value of both $v$ and $r$. The
specification uses the following predicate:
\begin{align}
  \acfhfrs\ \mathit{hf}\ m\ v\ s\ r \triangleq
  \Exists lm, p . & hf = (lm, v, s, r) \sep \upto{p}\\
                  & p + (rval - sval)\cdot((3 \cdot R_0) / (4 \cdot V_0)) \geq \textstyle\sum_{i=s}^{r-1} i/v \sep \label{eqn:ec-cond}\\
                  & \cfhf\ lm\ m\ v\ \sep (\dots)
\end{align}
We explain the representation predicate line by line. The first line
contains the internal representation of the hash function as a tuple, and
a reserve of $p$ error credits. The second line imposes a condition
on $p$, namely that the current number of credits in the reserve plus the
credits we will get until the next resizing is enough to pay for all of the error of the insertions until
the next resizing. Note that when there are $s$ elements in the image of the
hash function and we sample uniformly over $\{0, \dots, v-1\}$, the error we
will have to pay is $\upto{s/v}$. The third line states that $lm$ points to a
list that represents the partial map $m$, and that there are no collisions.
Finally, the rest of the predicate contains some constraints on the sizes
$v,s,r$ that we omit for brevity. In this specification we have decided to
expose $v$, $s$ and $r$ to the client as we will use those values in the next section,
however it is also possible to hide these values from the client when those details are not needed.

We prove three specifications for $\computehashrs$, depending on the
initial conditions. If we query an element that was already in the domain of
the hash function, we just get back its hash value, without the need for spending
error credits:
\begin{align*}
  &\hoare{ \begin{array}{c} m[w] = \mathsf{Some}\ b\ \sep \\   \acfhfrs\ f\ m\ v\ s\ r  \end{array} }
    { \computehashrs\ f\ w }
    { (b', f') \ldotp \begin{array}{l} b' = b \sep \\ \acfhfrs\ f'\ m \ v\ s\ r \end{array} }
\end{align*}
If we query for an element that is not in the domain we will have to sample it in
a collision-free manner, at a cost of $\upto{s/v}$.
From the precondition we have $\upto{(3 \cdot r_0) / (4 \cdot v_0)}$,
and from the representation predicate we can get the reserve $\upto{p}$.
We can derive that this is enough to pay for $\upto{s/v}$ from
condition \eqref{eqn:ec-cond} by an uninteresting, albeit nontrivial, calculation.
Since the code branches depending on whether $s+1$ is equal to
$r$, we have the following specification:
\begin{align*}
  \hoareV{
  m[w] = \mathsf{None}\ \sep (s + 1 \leq r )\ \sep \acfhfrs\ f\ m\ v\ s\ r\ \sep \upto{\tfrac{3 r_0}{4 v_0}}}
  { \computehashrs\ f\ w  }
  { (b, f') \ldotp
  \begin{array}{l}
    (s+1 < r \sep \acfhfrs\ f'\ m[w := b] \ v\ (s+1)\ r) \lor{} \\
    (s+1 = r \sep \acfhfrs\ f'\ m[w := b] \ (2\cdot v)\ (s+1)\ (2\cdot r))
  \end{array}}
\end{align*}
In both cases we will have to reestablish the representation predicate. In particular,
we will have to store the remaining credits back into the reserve, and
prove that condition \eqref{eqn:ec-cond} is still valid (\ie, that we will have enough
credits to pay for future insertions). This follows again by arithmetic calculations.

\subsection{Collision-Free Resizing Hash Map}
\label{sec:cf-hash-map}

Hash maps (or hash tables) are one of the most ubiquitous data structures in programming, since they
can represent large sets with efficient insertion, deletion, and lookup.
Their efficiency relies on having a low number of collisions, so that each location
on the table contains a small number of values. As the number of collisions increases,
and thus the performance of the hash map worsens, it is often beneficial to resize the
table, redistributing the hashed values and freeing up space for new insertions.

In order to be able to reason about the efficiency of hash maps, we need to compute
the probability of a hash collision. Computing this probability over a
sequence of insertions can be cumbersome, as it depends on the current size of
the hash map and the number of elements it contains. As a consequence, it can lead to less
modular specifications for programs that use hash maps inside their components.


We will use the dynamically-resizing hash function defined above to implement a
collision-free dynamically-resizing hash map, and specify it with an amortized cost for insertion.
Namely we will use an array of size $v$, in which $s$ entries are filled with a
hashed value and the rest are uninitialized. Once we fill in $r$ elements, we
resize the table to have size $2v$ and we set $r$ to $2r$. New hash elements
are sampled in a collision-free manner following the specification shown in the previous
sections, thus ensuring that the hash map is also collision-free. We can then prove
the following specification\footnote{The code for this example and the definition of the representation predicate can be found in \cref{app:cf-hash-map}}
for inserting a value $w$ into a hash map $hm$:
\[
  \hoare{ \mathsf{isHashmap}\ hm\ ns\ \sep \upto{(3 \cdot R_0) / (4 \cdot V_0)} }
	{  \mathit{insert}\ hm\ w }     
 { hm' \ldotp \mathsf{isHashmap}\ hm' (ns \cup \{ w \})  }
\]
Recall here that $V_0$ is the initial capacity of the hash table, and $R_0$ is the
threshold on which we will first resize.
The representation predicate $\mathsf{isHashmap}\ hm\ ns$ should be understood as
``$hm$ is a collision-free hash map representing the set (of natural numbers) $ns$''.
Crucially, this predicate does not keep track of error credits as all of the error accounting is done
through the $\acfhfrs$ predicate, which is used as a client within $\mathsf{isHashmap}$.
This specification states that an insertion of an element $w$ fail with probability
at most $(3 \cdot R_0) / (4 \cdot V_0)$. There are two cases
in proving this specification: either $w$ was already in the hash map (and
therefore $ns = ns \cup \{ w \}$) or it is a new element.
The former case is immediate; if it is a new
element, we can use $\upto{(3 \cdot R_0) / (4 \cdot V_0)}$ to sample a
fresh value from the hash function using the specifications proven in the
previous section. This ensures that the location in the table corresponding to
that index is uninitialized. Since the hash map resizes
at the same time as the hash function does, this establishes our specification
no matter how many insertions have been performed before.


\subsection{Amortized Hash Functions and Merkle Trees}
\label{sec:hashfun-merkle}

\newcommand{\mtree}{\mathit{tree}}
\newcommand{\mtreeheight}{\mathit{height}(\mtree)}
\newcommand{\checker}{\langv{checker}}
\newcommand{\treevalid}{\mathit{tree\_valid}}
\newcommand{\correctproof}{\mathit{correct\_proof}}
\newcommand{\incorrectproof}{\mathit{incorrect\_proof}}
\newcommand{\treeleafproofmatch}{\mathit{tree\_leaf\_proof\_match}}
\newcommand{\nottreeleafproofmatch}{\mathit{not\_tree\_leaf\_proof\_match}}

A \emph{Merkle tree}~\cite{merkle-tree} is a data structure that relies on a hash function. It is used to ensure the authenticity and validity of data received from a potentially unreliable and malicious source and used widely in, \eg, distributed file systems~\cite{ipfs} and databases~\cite{dynamo}.

A Merkle tree is a binary tree whose nodes contain pairs consisting of a value
and a label. For leaves, the label is the hash of the value stored in the
leaf. In the case of inner nodes, the label corresponds to the hash of the
concatenation of the labels of its children. We call the label of the root of a
Merkle tree a \emph{root hash}.
Merkle trees are interesting because they support constructing a cryptographic \emph{proof certificate} that a value is in a leaf of the tree.
These proofs can be validated by a client who only knows the root hash of the tree.

To construct a proof that a value $v$ is in the tree, we start from the leaf $l$ containing the value $v$,
The proof starts with the hash of the \emph{sibling} of $l$.
We then traverse up from the leaf $l$ to the root along the ancestors of $l$, appending to the proof the hash annotations of the \emph{siblings} of each ancestor we traverse.
A client who has the root hash $h$ can check the proof by effectively computing a list fold over the proof, successively hashing each element of the proof against an accumulated hash.
The client then checks whether the result of the fold matches the root hash $h$; if it matches, the proof is deemed valid, and otherwise it is rejected as invalid.

Why is this proof checking procedure sound?
For an invalid proof to be (incorrectly) validated by a checker, the invalid proof must contain values that cause a colliding hash value to be computed during the checker's list fold.
Thus, if an adversary cannot find a collision, they cannot maliciously convince a checker with an invalid proof.
In particular, if the hash is treated as a uniform random function, and the total number of distinct hashes ever computed is relatively small (\eg, because of constraints on the adversary's computational power), the probability that the proof checking procedure will accept an invalid proof is very small.
In this example, we prove such an error bound under the assumption of a bound on the total number of hashes ever computed.

%

\newcommand{\hashf}{\langv{f}}
\newcommand{\lproof}{\langv{lproof}}
\newcommand{\lleaf}{\langv{lleaf}}
\newcommand{\lroot}{\langv{v}}
\newcommand{\listhead}{\langv{List.head}}
\newcommand{\listtail}{\langv{List.tail}}
\newcommand{\hash}{\langv{hash}}
\newcommand{\hashpath}{\langv{hash\_path}}
\newcommand{\valbitsize}{\langv{val\_bit\_size}}
\newcommand{\roothash}{\langv{root\_hash}}
\newcommand{\fun}{\langkw{fun}}
\newcommand{\helper}{\langv{helper}}

We use the fixed-size amortized hash with values in $0,\dots, 2^{V-1}$ to implement a library for Merkle trees.
Given a possible leaf value and a purported proof, together with an error credit of
$\upto {\amortizederr \cdot \mtreeheight}$ in the precondition, the specification for the proof checker will ensure that when a proof is invalid the checker will return false (\ie, the checker is sound up to this probability of error). 
The amortization of the hash simplifies the specification of the checker since it incurs a constant amount of error credits which only depends on the amortized error $\amortizederr$  and the tree but \emph{not} the size of the map in the hash.

The checker function as shown in \cref{fig:merkle-code} is implemented using the $\hashpath$ helper function, which recursively computes the potential root hash from the input proof and leaf value.
\begin{figure*}[t]
  \centering
  {\newcommand{\hd}{\langv{hd}}
    \newcommand{\tl}{\langv{tl}}
    \newcommand{\bb}{\langv{b}}
    \newcommand{\hp}{\langv{hp}}
    \begin{align*}
      &\hashpath\ \hashf\ \lproof\ \lleaf \eqdef{}\\
      & \hspace{1em} \MatchML \lproof with
        (\hd :: \tl) =>
        {\begin{array}[t]{l}
           \Let (\bb, \hash) = \hd  in \\
           \If \bb then \hashf\ ((\hashpath\ \hashf\ \tl\ \lleaf)*2^V+ \hash) \\
           \Else \hashf\ (\hash * 2^V+(\hashpath\ \hashf\ \tl\ \lleaf))
         \end{array}}
        | \langv{nil} => \hashf\ \lleaf
        end{}\\\\
      &\checker\ \roothash\ \hashf \eqdef{}\\
      & \hspace{1em} \Lam \lproof, \lleaf .\\
      & \hspace{2em} \Let \hp = \hashpath\ \hashf\ \lproof\ \lleaf in \\
      & \hspace{2em} \roothash == \hp
    \end{align*}
  }
  \caption{A proof checker for Merkle trees.}
  \label{fig:merkle-code}
\end{figure*}
We represent a proof as a list of tuples following the path from the leaf to the root:
each tuple consists of a boolean flag to determine which child of the current node is on the path to the leaf,
and the hash of the child node that is not on the path.
In the base case where the proof is an empty list, we arrive at the leaf of the Merkle tree, and we return the hash of our input leaf value.
In the intermediate step, where we arrive at a branch of the tree, we recursively compute the potential hash value of the branch containing the leaf node and bit-wise concatenate it with the hash found in the head element of our proof.
We then return the hash of this concatenated number.

Our simplified specification for the checker is displayed below:
%
  \begin{mathpar}
    \hoareVH {
      \begin{array}{c}
        \islist\ l\ \lproof \sep                       \\
        \treevalid\ \mtree\ m \sep                     \\
        \acfhf\ \hashf\ m \sep                              \\
        \mathit{size}(m)+\mtreeheight <= \maxsize \sep \\
        \upto{\amortizederr \cdot \mtreeheight}
      \end{array}
    }
    {\qquad\checker\ \roothash\ \hashf\  \lproof\ \lroot\quad}
    {b .
      \begin{array}{l}
        \text{if}\ b                                                    \\
        \text{then}\ \treeleafproofmatch\ \mtree\ \lroot\ \lproof \sep \dots \\
        \text{else}\ \nottreeleafproofmatch\ \mtree\ \lroot\ \lproof  \sep \dots
      \end{array}
    }
  \end{mathpar}
%
Line-by-line, the precondition here says that:
\begin{enumerate}
  \item the value $\lproof$ is represented by the abstract mathematical list $l$,
  \item the Merkle tree $\mtree$ is built correctly according to the hash map $m$,
  \item the function $\hashf$ encodes the amortized hash function under the map $m$,
  \item the size of $m$ plus the height of the tree is smaller or equal to $\maxsize$,
  \item we have credits equal to amortized error multiplied by the height of the tree.
\end{enumerate}
The postcondition states that the Boolean returned by $\checker$ soundly represents the inclusion of $v$ in the tree.
We impose the inequality of the size of the map $m$ in the beginning since $\checker$ runs the hash function exactly $\mtreeheight$ times (recall that there is a total limit on the number of distinct hashes that can be computed).

How are the error credits used to derive this specification?
As long as the hash function remains collision-free throughout the checking procedure then any corrupted
data will modify all hashes above it in the tree---in particular, it will change the root hash.
Therefore, we will spend error credits at each of the $\mtreeheight$ hashes computed by $\checker$ to
preserve collision-freedom throughout the checking process.
We remark that if we chose to use a non-amortized hash for the implementation of the Merkle tree library, the amount of error credits paid as one traverses the tree may change if a new value is ever encountered, leading to a more convoluted specification.

\subsection{Further Case Studies}

For reasons of space, we omit other case studies, which can be found in the long version of the paper. In particular, we include an example
that uses the Merkle tree as a client to store data into an unreliable storage system and prove that, with high probability,
it can be used to detect data corruption.



\section{Almost-Sure Termination via Error Credits}
\label{sec:ubtwp}

In this section, we introduce \theaplogtotal, an \emph{approximate total-correctness} version of \theaplog, and show how it can be used to prove almost-sure termination via reasoning about error credits.
Before we embark on the technical development, we consider for a second time the example from \cref{fig:expected-error}, which illustrates the distinction between partial and total correctness interpretations of approximate reasoning up to some error.
A summary of the example is depicted in \cref{fig:partial-total}.

\begin{figure}[t]
  \centering
  \Forest{
    for tree={rounded rectangle, draw, edge={->}, s sep=2em, l sep=2.0em}
      [{\parbox[b][7ex][c]{14ex}{\center $\expr$}}
          [{\parbox[b][3ex][c]{18ex}{\center $\phi$}},
            edge label={node[pos=0.35, left=1.5ex] {$\frac{5}{8}$}}
          ],
        [{\parbox[b][3ex][c]{10ex}{\center \tikzmark{totalbeg} $\neg \phi$}},name=Ftotalbeg,
            edge label={node[pos=0.45, right] {$\frac{2}{8}$}}
          ],
        [{\parbox[b][3ex][c]{4ex}{\center ${\Uparrow}$ \tikzmark{totalend}}},
            name=Ftotalend,
            edge label={node[pos=0.35, right=1.5ex] {$\frac{1}{8}$}}
          ]
      ]
    \draw[decorate,decoration={brace,amplitude=.85ex},black!50] ($(Ftotalbeg.south east)+(2mm,-.7mm)$) -- ($(Ftotalbeg.south west)+(-2mm,-.7mm)$) node[black, midway, below=.5mm]{\small partial error};
    \draw[decorate,decoration={brace,amplitude=.85ex},black!50, thick] ($(Ftotalend.south east)+(3mm,-6mm)$) -- ($(Ftotalbeg.south west)+(-3mm,-6mm)$) node[black, midway, below=.5mm]{\small total error};
  }
  \vspace*{-1.5ex}
  \caption{Partial and total approximate correctness contrasted.}
  \label{fig:partial-total}
  \vspace*{-1.5ex}
\end{figure}
In \cref{sec:logic} we showed that for the program $\expr$ in \cref{fig:expected-error}, we can show the specification Hoare triple $\hoare{\upto {\err_p}}{\expr}{x.\, x = \True}$ for $\err_p = \frac{2}{8} = \frac{1}{4}$ in \theaplog, intuitively because the program terminates with a result not satisfying the postcondition ($\False$) with probability $\frac{1}{4}$.
Note that the probability of non-termination ($\frac{1}{8}$) is not included in the error, since non-termination is considered acceptable by the partial-correctness interpretation of \theaplog.

In \theaplogtotal on the other hand, one cannot satisfy a Hoare triple by not terminating, and thus $\err_t = \frac{3}{8}$ is needed to show an approximate \textit{total} Hoare triple $\thoare{\upto {\err_t}}{\expr}{x.\, x = \True}$ (we use angle brackets when we want to emphasize that we are stating a Hoare triple in \theaplogtotal).
The proof of this Hoare triple in \theaplogtotal is very similar to the \theaplog proof for $\expr$ in \cref{sec:logic}, only changing the distribution of error credits at each sample so that our additional starting credit can discharge the nonterminating branch (specifically, $\err_t = 1$ in the rightmost branch of \cref{fig:expected-error}).



Stepping back from the example, we can make a simple but crucial observation:
if the total error necessary for showing a Hoare triple
$\thoare{\upto \err}{\expr}{\prop}$ (for any postcondition $\prop$)
can be proven for any arbitrary (but positive) $\err$, then we can
make the probability of divergence vanishingly small, and $\expr$ must
be almost-surely terminating.

In this section, we show how to make this argument precise and then demonstrate how it yields an approach for showing almost-sure termination via error credits, which allows us to prove correctness of Las Vegas algorithms.

\subsection[The Erisₜ Logic,  Adequacy, and Almost-Sure Termination]{The \theaplogtotal Logic,  Adequacy, and Almost-Sure Termination}
\label{sec:twp-adequacy}\label{sec:cont-ast}

It is important to note that all the proof rules of \theaplog shown earlier are still sound for
\theaplogtotal, with the exception of \rref{ht-rec}. Note in particular that \rref{ht-rec} could
be applied to the diverging function $\Rec f x = f~x$, which is unsound in the total-correctness setting.
To reason about recursive programs in \theaplogtotal we can instead use a novel technique we call
\emph{induction on the error amplification}.
We will see examples below of how this principle works.


The meaning of a Hoare triple in \theaplogtotal is given by its adequacy theorem, which
states that given a program $\expr$, if
from the assumption $\upto \err$ we can prove a metalogical postcondition $\phi$, then
the program will terminate and satisfy $\phi$ with at least probability
$1-\err$:
\begin{theorem}[Total Adequacy]\label{thm:twp-adequacy}
  If $\vdash \thoare{\upto \err}{\expr}{\phi}$
  then $\prex{(\expr,\state)}{\phi} \geq 1 - \err$
  for any state $\state$.
\end{theorem}
\noindent Since the logic is total this also implies that the probability of the program crashing is
bounded from above by $\err$. By a continuity argument in the meta logic (outlined in
\cref{sec:soundness}), we then obtain the following theorem.
\begin{theorem}[Almost-Sure Termination]\label{thm:twp-ast}
  Let $\err \geq 0$. If for all $\err'>\err$, $\vdash \thoare{\upto{\err'}}{\expr}{\phi}$ 
  then $\prex{(\expr,\state)}{\phi} \geq 1 - \err$
  for any state $\state$.
\end{theorem}
\noindent If we pick $\err=0$, this theorem allows us to conclude almost-sure termination of $\expr$ by
proving $\vdash \thoare{\upto {\err'}}{\expr}{\phi}$ for any $\err'>0$.
Note that we here quantify over $\err'$ in the meta-logic; the \theaplogtotal portion of
this argument freely assumes ownership over some arbitrary positive $\upto{\err'}$.

\subsection{Case Studies}
\label{sec:total-examples}

We now present case studies of how we can prove almost-sure termination via error credits.
In the first case study we demonstrate how we can prove
a Hoare triple in \theaplogtotal by a form of induction on error credits.
The second case study presents a technique to upper-bound the probability
of excessively long runs of a program.
In the third case study we present a novel \textit{planner} proof rule, which
can separate credit arithmetic from concrete program steps,
and we show how to use it to prove correctness of general rejection samplers.
Finally, we revisit an example due to~\citet{newRuleAST}, and show how error
credits can be used to provide a simple proof that it is almost surely terminating.

\subsubsection{Induction by Error Amplification}
\label{sec:induction-amplification}

\newcommand{\ucheck}[1]{\langv{uc}_{#1}}
\newcommand{\usample}[1]{\langv{us}_{#1}}
\newcommand{\bcheck}[1]{\langv{cp}_{#1}}
\newcommand{\bsample}[1]{\langv{cs}_{#1}}


A variety of \textit{Las Vegas} algorithms employ a ``sample and retry'' approach, whereby a \textit{sampler} program produces a possibly undesirable value and a \textit{checker} program forces the sampler to retry until it produces an acceptable value.
In a partial-correctness logic, it is trivial to prove that such algorithms return acceptable values when they terminate (using \rref{ht-rec} in \theaplog).
However, it is more challenging to bound the probability that they fail to terminate.

%

\textit{Rejection samplers} are one common example of this kind of algorithm.
Rejection samplers simulate complex probability distributions using sequences of
samples from simpler distributions, by strategically rejecting sequences which do not correspond to values in
the target distribution.
%
Consider the implementation of a typical
rejection sampling scheme $({\sf RSamp}\ s\ c)$ with sampler program $s$ and checker
program $c$ below.
\[
  {\sf RSamp} \eqdef
  \;\Lam \langv{s} . \Lam \langv{c} . \Let \langv{try} = (\Rec {\langv{try}}\ \_ =
  \;\Let \langv{v} = \langv{s}\ () in
  \;\If (\langv{c}\ \langv{v}) then \langv{v} \Else \langv{try}\ ()) in \langv{try} \; ()
\]
As an archetypical example, we can emulate samples from $\{0, 1, \ldots, N\}$ using a uniform
sampler of size $M > N \ge 1$ by providing the sampler $\usample{M} \eqdef (\Lam \_. \Rand M)$
and checker $\ucheck{N} \eqdef (\Lam \langv{v} . \langv{v} \le N)$.

Let us show that this uniform rejection sampler $({\sf RSamp}\ \usample{M}\ \ucheck{N})$ terminates
almost surely.
Using \cref{thm:twp-ast}, it suffices to show, for arbitrary nonnegative $\err$,
\begin{equation}
  \thoare{\upto \err }{{\sf RSamp}\ \usample{M}\ \ucheck{N}}{\TRUE}.
  \label{eqn:sampler-goal}
\end{equation}

In proving this, we need to reason about the recursion in ${\sf RSamp}$.
Since we no longer have \rref{ht-rec}, we will have to use some form of induction, yet there appears to be no argument to induct on.
The solution is a technique we call \textit{induction by error amplification}.
We first show how the principle works in detail, then derive rules that makes its use
more practical.
Note that using expectation-preserving composition, for any $\err'$ we are able to prove
\begin{gather*}
  \thoare{\upto{\err'}}{\usample{M}}{v \ldotp \upto{E(v) \cdot\err'}}
  \quad\quad\text{where}\quad
  E(v) \eqdef
  \begin{cases}
      0 & 0 \leq v \le N \\
      \frac{M+1}{M-N} & N < v
  \end{cases}
\end{gather*}
since $\frac{1}{M+1}\,\sum_{i = 0}^{M}E(i) = 1$.
In other words, each sampling attempt either produces a value which the checker will certainly
accept, or scales our error credit by a factor of $\frac{M+1}{M-N} > 1$.
This means that we can grow our error credit geometrically in the cases where a sample does
not immediately terminate, and  we need only repeat this procedure $d(\err) = \lceil - \log_{(M+1)/(M-N)}(\err) \rceil$ times:
either some sampling attempt will succeed, or they all fail and we will have amplified the original error to $\upto{1}$, from which we obtain a proof of $\FALSE$.
Starting with any $\err > 0$, induction over $d(\err)$ allows us to prove \eqref{eqn:sampler-goal},
completing the proof.
While this proof technique appears to be novel among total correctness logics, under the hood
it closely mirrors a standard analysis in probability theory where one
shows that longer and longer traces are increasingly unlikely, and concludes by taking a limit.

This induction principle is abstractly captured by the rule below:
\[
    \inferH{ind-err-amp}
    {\err > 0 \\  k > 1 \\ \All \err'. (\upto{k\cdot \err'} \wand \prop) \sep \upto{\err'} \vdash \prop }
    {\upto \err \vdash \prop}
\]

The rule requires us to (1) own a positive amount of error credits $\err$ and (2) choose an amplification factor $k > 1$.
Using the rule gives us an arbitrary initial amount of error
credits $\upto{\err'}$ and an inductive hypothesis for which we need to pay $\upto{k \cdot \err'}$. Soundness of the rule
is proven by induction on the number of times we need to scale $\err$ by $k$ until we accumulate $\upto{1}$,
i.e. $\lceil - \log_{k}(\err) \rceil$.

Note that
while this principle holds for an arbitrary proposition $P$, it only ever makes sense to use it when reasoning
about programs, because otherwise we have no way of amplifying our error credits. To reason about recursive functions
in the total correctness logic, we can derive the following rule as a consequence:
\[
    \inferH{ht-rec-err}
    {\err > 0 \\  k > 1 \\ \All \err'. (\All \valB .\ \thoare{\prop \sep \upto{k\cdot \err'}}{(\Rec f x = e)\ \valB}{\propB})~\vdash~\thoare{\prop \sep \upto{\err'} }{\subst{\subst{e}{x}{\val}}{f}{(\Rec f x = e)}}{\propB} }
    {\vdash \thoare{\prop \ast \upto \err }{(\Rec f x = e)\ \val}{\propB}}
\]

Turning again to our example, we can prove \cref{eqn:sampler-goal} in an arguably simpler manner
by applying \rref{ht-rec-err}
and setting $k = \frac{M+1}{M-N}$. This reduces to proving, for an arbitrary $\err'$,
\begin{align*}
\thoare{\upto{k \cdot \err'}}{\langv{try}\ ()}{\TRUE}~ \vdash~
\thoare{ \upto{\err'} }{\Let \langv{v} = \usample{M} in
\;\If (\ucheck{N}\ \langv{v}) then \langv{v} \Else \langv{try}\ ()}{\TRUE}
\end{align*}
Following the credit splitting strategy above, after sampling from $\usample{M}$ we will either have a
value $\langv{v}$ that passes the check $\ucheck{N}$, or we will amplify our error credits to $\upto{k\cdot\err'}$
and go to the recursive case, in which case we can conclude immediately by instantiating the inductive
hypothesis.

%
%

\subsubsection{Reasoning about Tail Bounds}

Another property one may want to prove about a rejection sampler is that
long runs only happen with low probability. The explicit proof
of termination in the example above
using induction on $d(\varepsilon)$ is implicitly proving an upper bound on
the probability of a run taking longer than $d(\varepsilon)$ steps. Indeed,
we can
make this concrete by instrumenting the rejection sampler with a counter,
and make it fail once a limit count is reached.
\begin{align*}
  {\sf RSamp}_{\sf bd} \eqdef \Lam \langv{s} . \Lam \langv{c} . \Lam \langv{m} .
  &\DumbLet \Rec{} \langv{try}~n = \\
  &\quad
      \If n \leq 0 then {\sf None} \\
  & \quad \Else
    (\Let \langv{v} = \langv{s}\ () in
    \If (\langv{c}\ \langv{v}) then {\sf Some}~\langv{v} \Else \langv{try}\ (n-1))) in \\
  & \langv{try} \; m
\end{align*}
We can now prove the triple below:
\begin{align*}
  \thoare{\upto{{\left(\tfrac{M-N}{M+1}\right)}^n}}{{\sf RSamp}_{\sf bd}\ \usample{M}\ \ucheck{N}\ n}{v \ldotp \Exists w . v = {\sf Some}~w }.
\end{align*}

Since the sampler will only return something of the shape ${\sf Some}~w$ if it
succeeds in $n$ tries or fewer, {$\left(\frac{M-N}{M+1}\right)^n$} is an upper bound on the probability of
the sampler of taking more than $n$ tries to produce a valid sample. We can prove
this specification by a simple (Coq-level) induction on $n$. The case $n=0$ is trivial,
because we begin the proof with $\upto{1}$. In the inductive case, after the probabilistic
sampling, we will either terminate or amplify our error by $\frac{M+1}{M-N}$, as in the
previous case study, and then we can apply the inductive hypothesis.

The caveat of this approach is that the reasoning about the runtime is entirely
intrinsic.  Our adequacy theorem does not allow us to derive from the proof of
a Hoare triple a statement about the concrete runtime of our programs, but we
believe that we can achieve this via an integration of (deterministic) time credits~\cite{tcred-iris}
into this setting.

\subsubsection{Presampling Tapes}
\label{sec:rules-tapes}

In order to extract a more general proof rule from our credit amplification argument,
it is useful to separate the credit accounting steps from the symbolic execution
of a program.
Because the amount of error credits after an expectation-preserving composition can depend on the value sampled,
this necessitates some way to express the outcome of random sampling events ahead of time.
Luckily, the presampling tapes by \citet{clutch} provide this mechanism exactly.
In this section we first briefly recall the semantics of \thelang with tapes as well as the proof rules for a $\progtape{\lbl}{\tapebound}{\tape}$ proposition, which expresses ownership of a presampling tape. In the next section we show how to use them to get a general proof rule for credit amplification.
For more details on presampling tapes, we refer the reader to \loccit, where tapes were originally introduced and used to sample in an asynchronous manner.

To introduce presampling tapes in \thelang, we extend the syntax of the language as follows.
\begin{align*}
  \expr \in \Expr \bnfdef{}&  \ldots \ALT \AllocTape \expr \ALT \Rand \expr_1\, \expr_2 \\
  \val, \valB \in \Val \bnfdef{}& \ldots \ALT \lbl \in \Lbl \\
  \state \in \State \eqdef{}& (\Loc \fpfn \Val) \times (\Lbl \fpfn \Tape) \\
  t \in \Tape \eqdef{}& \{ (\tapebound, \tape) \mid \tapebound \in \mathbb{N} \wedge \tape \in \mathbb{N}_{\leq \tapebound}^{\ast} \}
\end{align*}
In addition to the heap, a state in \thelang with presampling also contains a map from tape labels to presampling tapes. Tapes are formally pairs $(\tapebound, \tape)$ of an upper bound $\tapebound \in \mathbb{N}$  and a finite sequence $\tape$ of natural numbers less than or equal to $\tapebound$.
Tape allocation via $\AllocTape \expr$ returns a fresh label $\lbl$ and extends the state with a new empty tape $\epsilon$ with bound $\tapebound$ if $\expr$ evaluates to $\tapebound \in \nat$. Sampling from a tape with label $\lbl$ via $\Rand \tapebound\, \lbl$ either \emph{deterministically} pops the first element from the list $\tape$ or uniformly samples a new integer between $0$ and $\tapebound$.

From the point of view of the \theaplogtotal logic, a presampling tape behaves somewhat similarly to
standard heap location.
Like in Clutch, the proposition $\progtape{\lbl}{\tapebound}{\tape}$ asserts ownership of a tape labelled $\lbl$ with bound $\tapebound$ and contents $\tape$.
We can allocate an empty tape $\epsilon$ with a specified bound.
\vspace{-1.5ex}
\begin{mathpar}
  \infrule[Right]{alloc-tape}
  {}
  {\proves \thoare{\TRUE}{\AllocTape(\tapebound)}{ \lbl \ldotp \progtape{\lbl}{\tapebound}{\epsilon}}
  }
\end{mathpar}
Sampling from a non-empty tape consumes the first value of the list.
\vspace{-1.5ex}
\begin{mathpar}
  \infrule[Right]{load-tape}{ }
  {\proves
    \thoare
    {\progtape{\lbl}{\tapebound}{n\cdot\tape}}
    {\Rand \tapebound \,\lbl}
    {x \ldotp x=n \sep \progtape{\lbl}{\tapebound}{\tape}}
  }
\end{mathpar}
Note that there are no primitives in $\thelang$ for directly writing to or adding values to tapes and values are only added to tapes via \emph{ghost operations} that appear purely at the logical level of an \theaplog proof.
\begin{mathpar}
  \infrule[Right]{presample}
  {\proves \thoare{\progtape{\lbl}{\tapebound}{\tape\cdot n}}{\expr}{P}
    \\
    0 \leq n \leq N
  }
  {\proves \thoare{\progtape{\lbl}{\tapebound}{\tape}}{\expr}{\prop}
  }
\end{mathpar}
Crucially, \theaplog extends the above proof rules for tapes taken from \cite{clutch} with the following additional rule, connecting error credits in expectation to presampling via tapes.
\newcommand{\PresampleAdvComp}{
  \infrule[Right]{presample-exp}
  {
    \textstyle\sum_{i=0}^{N}\tfrac{\Err_2(i)}{N+1} = \err_1
    \\
    \All n . \proves \thoare{\progtape{\lbl}{\tapebound}{\tape \cons n} \sep \upto{\Err_2(n)}}{ e }{\prop}
  }
  { \proves \thoare{\progtape{\lbl}{\tapebound}{\tape} \sep \upto{\err_1}}{ e }{\prop }}
}
\begin{mathpar}
  \PresampleAdvComp
\end{mathpar}
Similarly to \rref{ht-rand-exp}, this rule allows us to sample a random integer $n$, write it to
a tape, and get $\Err_2(n)$ error credits, assuming that our initial amount of error credits $\err_1$
is the expected value of $\Err_2(n)$. Since no execution step is taken, this allows us to disentangle
credit arithmetic from the operational semantics of the program.
Analogous rules also hold in the partial version of the
logic, but the interaction between tapes and error credits is particularly useful for total correctness.

\subsubsection{The Planner Rule}
\label{sec:planner}

%


Equipped with the ability to perform credit reasoning and symbolic execution
separately, we can now derive an induction principle for \theaplogtotal which
eliminates the need to perform fine-grained credit arithmetic as in
\cref{sec:induction-amplification}.
Our proof rule is directly inspired by the ``planner'' method by \citet{planner}.
\citeauthor{planner} establish a
proof system wherein to prove termination, one can reason as if a sequence of randomized choices will yield some prover-selected sequence of outputs infinitely often (as justified by the Borel-Cantelli lemma from classical probability theory).
%
Expressed in \theaplogtotal, we have the following \textit{planner} rule:
\[
  \infrule[Right]{presample-planner}
  {
    0 < \err
    \\
    \All {s} . \vert z({s}) \vert \le L
    \\\\ \rule{0pt}{1\baselineskip}
    \proves \thoare{\Exists {ys}. \progtape{\lbl}{\tapebound}{{xs}\,\dplus\,{ys}\,\dplus\,z({xs}\,\dplus\,{ys})}}{e}{\phi}
  }
  { \proves \thoare{\progtape{\lbl}{\tapebound}{{xs}} \sep \upto{\err}}{e}{\phi}}
\]
%
The rule states that if we have a tape with contents ${xs}$ and any positive amount
of error credits $\err$, then we can update our tape with some unknown sequence of ``garbage'' samples
${ys}$ in order to ensure it includes a prover-selected target sequence $z({xs}\dplus{ys})$
at the end.
Generalizing \citeauthor{planner}'s planner rule, our version allows the target sequence to be a function $z$ of
the current state of the tape, provided that the length of the target word has a fixed upper bound $L$ (and consequently, cannot become arbitrarily unlikely).
After invoking the planner rule, a prover can consume the samples in ${xs}\dplus{ys}$ using a
regular induction on lists, eventually ending up in their desired tape state $\progtape{\lbl}{\tapebound}{z({xs}\,\dplus\,{ys})}$.

The planner rule is derivable entirely within \theaplogtotal using the rules we have already seen.
The proof proceeds by induction by error amplification, and
most of the proof is directly analogous to the argument in \cref{sec:induction-amplification}.
The key step lies in proving the following \textit{amplification lemma} which, for a particular
constant $K_{\tapebound,w} > 1$ (dependent on $\tapebound$ and $|w|$),
allows us to either sample a target word ${w}$ onto a tape, or sample a garbage string $j$
and scale our error credit by a factor of $K_{\tapebound,w}$:
\begin{equation*}
  \infrule[Right]{}
  {
    \proves \thoare{\Exists j . \progtape{\lbl}{\tapebound}{\tape \dplus j} \sep \upto{K_{\tapebound,w} \cdot \err}}{e}{\phi}
    \\
    \proves \thoare{\progtape{\lbl}{\tapebound}{\tape \dplus {w}}}{e}{\phi}
  }
  { \proves \thoare{\progtape{\lbl}{\tapebound}{\tape} \sep \upto \err}{e}{\phi} }
\end{equation*}

Proving this involves $|z(\tape)|$ applications of \rref{presample-exp}.
We provide the details, including the amplification constant $K_{\tapebound,w}$, in \cref{appendix:iter-adv-comp}.

\paragraph{Example}
\label{sec:planner-example}

To demonstrate how the planner rule can eliminate the credit accounting in
\textit{induction by error amplification}, we will prove that a Poisson trial
almost surely terminates.
A Poisson trial is a random process, whose value comes from counting how many independent attempts
a random variable takes before meeting some criteria.
We can implement a Poisson trial that flips pairs of fair coins until they are both heads
as an instantiation of a rejection sampler $S$:
\begin{gather*}
  \bsample{l} \eqdef \Lam \_. l \gets (\deref l + 1); (\Rand 1,\ \Rand 1) \qquad
  \bcheck{l} \eqdef \Lam v. v == (1,\ 1)
\end{gather*}
Note that the sampler here both maintains internal state (counting the number of trials) and uses multiple calls to $\Rand$.
We seek to show
\begin{equation*}
  \thoare{\upto \err \sep 0 < \err \sep l \mapsto 0}{S\ \bsample{l}\ \bcheck{l}}{\TRUE}.
\end{equation*}

Starting with any tape $\progtape{\lbl}{\tapebound}{{xs}}$, we invoke the planner rule with $\upto{\err}$ and the target function
\begin{equation}
  z(s) =
  \begin{cases}
    [1, 1] & |s|~\textrm{is even,} \\
    [0, 1, 1] & \textrm{otherwise}.
  \end{cases}
\end{equation}
This results in a tape of the form
$\progtape{\lbl}{1}{{xs}\,\dplus\,{j}\,\dplus\,[1, 1]}$ where ${j}$ has even length.
By induction over ${j}$, we consume the entire garbage section of the tape,
with each invocation of $\bsample{l}$ pulling off the two samples.
Either the garbage section will happen to contain some spurious $[1, 1]$ sample, in which case the program terminates,
or we step through the entire garbage section and end up with tape $\progtape{\lbl}{1}{[1, 1]}$ which will also cause termination.
%
%
We obtain our almost-sure termination result using \cref{thm:twp-ast}, as
our initial error credit was arbitrarily small.

\subsubsection{The Escaping Spline}
\label{appendix:spline}

In this case study, we revisit Example 5.4 from~\cite{newRuleAST}, which presents a so-called \emph{escaping spline}. This
consists in a random walk over the non-negative integers. An agent is at a position indicated by an integer $n$ and on every step
it chooses probabilistically between stopping, or moving to $n+1$.
However, the bias changes with the current position $n$,
the probability it chooses to stop is $\frac{1}{n+1}$. Our goal is to show that, despite the probability of
choosing to stop decreasing as time goes on, this walk is almost surely terminating, no matter
the initial position. We can implement
the walk by the following program:
\begin{align*}
	\Rec {\sf spline}~n~=\quad &\Let x = \Rand n in \\
		&{}\If x = 0 then () \Else {\sf spline}~(n+1)
\end{align*}

Using our total correctness logic, and~\cref{thm:twp-ast}, it suffices to show, for any arbitrary positive $\err$,
and any initial position $n$,
\begin{equation}
\label{eqn:spline}
	\vdash \thoare{\upto{\err}}{{\sf spline}~n}{\TRUE}
\end{equation}
The crucial part of the proof is the following auxiliary result, proven by induction on $k$
\[
  \vdash \thoare{\upto{\frac{n}{n+k+1}}}{{\sf spline}~n}{\TRUE}
\]
for all $n$.
In the case $k=0$ we own $\upto{\frac{n}{n+1}}$, which is precisely what we need in position $n$ to ensure we
choose to stop. For the successor case, we own $\upto{\frac{n}{n+(k+1)+1}}$. Here we can apply \rref{ht-rand-exp}
to ensure we either jump to $0$ or we jump to $n+1$ and scale our credits by $\frac{n+1}{n}$, which leaves
us with $\upto{\frac{n+1}{n+1+k+1}}$. This is what we require to apply our inductive hypothesis at position $n+1$,
and finish the proof of this lemma.

Going back to proving \cref{eqn:spline}, since our initial budget $\upto{\err}$ is strictly positive, and the sequence $\frac{n}{n+k+1}$
over $k$ gets arbitrarily small, we can always pick a $K$ such that $\frac{n}{n+K+1} < \err$, and then
weaken our assumption to $\upto{\frac{n}{n+K+1}}$ and conclude by applying the auxiliary lemma above.

To contrast with our proof, \citeauthor{newRuleAST} introduce a specialized
rule to prove that this and other complex examples terminate almost surely.
Within our setting, error credits and (meta-logic) induction can be
used directly, with no need for additional rules. It is not clear if their
other case studies can also be proven AST using error credits, or if their
termination rule can be encoded in terms of error credits.

\subsubsection{Additional Case Studies}
\label{sec:twp:additional-case-studies}

While the planner rule is a versatile technique for proving almost-sure termination, it is not the only way to abstract \textit{induction by error amplification}.

In particular, when the ``target sample'' has complex dependencies on program state
it may be cumbersome to explicitly produce a target sample function $z$.
%
In the long version we outline an alternative approach for proving
almost-sure termination, which leverages a higher-order specification to directly
express a relationship between the behavior of a sampler, checker, and error
credit values.
We then apply this specification to show that \textit{WalkSAT}, a randomized SAT solver
whose behavior is highly dependent on state, almost surely recognizes
satisfiable 3SAT formulas.



\section{Semantic Model and Soundness}
\label{sec:soundness}

We now turn our attention to the semantic model of \theaplog, which we use to prove soundness of the proof rules for \theaplog and to prove the adequacy theorem presented in \cref{sec:logic}.

Following standard practice~\cite{irisjournal}, we define \theaplog Hoare triples in terms of a \emph{weakest precondition} predicate
\[
  \hoare P e Q \eqdef \always (P \wand \wpre e Q)
\]
where the persistence modality $\always$ ensures that the predicate can be duplicated.
However, our definition of the weakest precondition predicate $\wpre e Q$ is novel.
The definition is shown below.
We omit from the definition the parts pertaining to how the Iris logic handles modifications to resources via ``update modalities'', since these details would distract from the definition and are completely standard.
The full definition can be found in \cref{sec:wp-full}.
\begin{align*}
  \wpre{\expr_1}{\Phi}
  \eqdef{}
  &(\expr_1 \in \Val \mathrel{\land} \Phi(\expr_1))\\
  \lor{}\,
  &(\expr_1 \not\in \Val \mathrel{\land} \All \state_1, \err_1 . \stateinterp(\state_1) \sep \uptoauth{\err_1} \wand\\
  & \quad
    \execub
    (\expr_1, \state_1, \err_1,
    \ (\Lam \, \expr_2, \state_2, \err_2
    \ . \later (\stateinterp(\state_2) \sep \uptoauth{\err_2} \sep \wpre{\expr_2}{\Phi})))
\end{align*}
The overall structure of this definition is similar to the weakest precondition for a non-probabilistic language \cite[\S6.3]{irisjournal}.
In particular, $\wpre {\expr_1} \Phi$ is defined by guarded recursion.
The first clause of the disjunction indicates that the weakest precondition for a value simply means that the postcondition $\Phi(\expr_1)$ must be satisfied.
The second clause of the disjunction deals with the non-value case.
It requires that the \defemph{state interpretation} $\stateinterp(\state)$ is valid, which connects the logical points-to connectives to the physical state of the program.
Both the heap and the presampling tapes are handled in this way, using the standard interpretation of state as partial finite maps from locations (resp.\ labels) to values (resp.\ presampled values) \cite{irisjournal, clutch}.

The weakest precondition gives meaning to ownership of the error resource $\upto \err$ through the \defemph{error interpretation} $\uptoauth{\err_1}$.
Just like for the state interpretation, the error interpretation $\uptoauth{\err_1}$ connects the logical connective for error credits $\upto{\err}$ to the errors during program execution.
Error credits are defined using the authoritative resource algebra~\cite{iris, irisjournal} over the positive real numbers with addition and the natural order, whose valid elements are the numbers in the half-open interval $[0,1)$.
The definition of the error credit resource is thus similar to that of later credits~\cite{DBLP:journals/pacmpl/SpiesGTJKBD22}, but instead of $\Auth(\nat,+)$ we use $\Auth(\real_0^+,+)$ with validity restricted to elements strictly smaller than 1.
The proposition $\upto{\err}$ asserts ownership of a fragmental element of the resource algebra, while $\uptoauth \err$ stands for the authoritative view.
The error rules from \cref{sec:logic} then follow directly from the definition of the error credit resource together with the rules for the authoritative resource algebra.

The novel part of our definition of weakest precondition (besides the addition of the error interpretation) is that the recursive appearance of the weakest precondition is wrapped in the \emph{graded lifting modality} $\execub$.
The exact way in which $\execub$ connects the operational semantics to errors will be explained in the next section.
For now, we focus on its intuitive use in the weakest precondition. Think of $(\expr_1,\state_1)$ as the starting configuration and of $\err_1$ as the current error budget.
Through $\execub$, we quantify over the configurations we may step to according to the operational semantics as $(\expr_2,\state_2)$, and $\err_2$ stands for the left-over error budget.
The final part of the definition then indicates that the state and error interpretations with respect to $\state_2$ and $\err_2$ have to be satisfied, and that the weakest precondition has to hold recursively for $\err_2$ and $\Phi$.
Crucially, this recursive appeal to the weakest precondition occurs under the later modality $\later$.
This is what allows us to take the guarded fixed point of $\wpname$ (which, in turn, allows us to prove soundness of the recursion rule \rref{ht-rec}).

For \theaplogtotal, the only difference is that we omit the $\later$ modality in the definition of weakest precondition and instead define the predicate by the least fixed point (this is well-defined since $\wpname$ only occurs positively inside its own definition).

\subsection{The Graded Lifting Modality}
\label{sec:modality}


We now turn our attention to the graded lifting modality $\execub(\expr_1,\state_1,\err,Z)$.
\Theaplog uses the graded lifting modality to construct approximate predicate liftings of the graded predicate on configurations $Z$ with respect to the distributions induced by the execution of $(\expr_1,\state_1)$.
As we shall see, to prove the modality, the initial error budget $\err$ may be shared between the modality and $Z$.
Our use of the graded lifting modality in the weakest precondition bears similarity with the coupling modality of Clutch \cite{clutch}, which was used to construct couplings between the execution of a specification program and its refinement, but the definition of our modality itself is rather different.

To focus the discussion on the most interesting aspects of the modality, we first present a simplified version \rref{step-simple} that only supports reasoning about uniform error bounds. We then show how to modify the definition to enable expected error bound reasoning in \rref{step-exp}.
The full definition, which additionally supports expected error reasoning for presampling tapes, can be found in \cref{sec:modality-full}.


The simplified version is specified by the following rule (which should be read as a definition, expressing that the $\execub (\expr_1, \state_1, \err, Z)$ predicate in the conclusion holds if the separating conjunction of the premises above the line holds):
\begin{mathpar}
  \infrule{step-simple}{
    \red (\expr_1, \state_1)
    \quad
    \err_1 + \err_2 \leq \err
    \quad
    \pgl[\stepdistr(\expr_1, \state_1)] {\err_1} R
    \quad
    \All \expr_2, \state_2 .  R(\expr_2,\state_2) \wand Z(\expr_2, \state_2, \err_2)
  }{
    \execub (\expr_1, \state_1, \err, Z)
  }
\end{mathpar}
%

The intuitive meaning of \rref{step-simple} is that we can split the starting error budget $\err$ into $\err_1 + \err_2$ and then likewise split the reasoning about the behaviour of the program into reasoning about the first step and about the rest of the execution separately.
The error bounds can then be composed to yield a bound on the execution of the whole program.

On a technical level, the first premise of \rref{step-simple} ensures that the program does not get stuck ($\red$ is short for reducible).
The second premise states that the error budget can be split into two parts $\err_1$ and $\err_2$ provided their sum does not exceed $\err$.
The inequality gives some flexibility in error accounting by allowing one to ``weaken'' the error bound: it is always sound to leave error budget unused.
The user of the rule then picks an auxiliary intermediate predicate on configurations~$R$.
The premise $\pgl[\stepdistr(\expr_1, \state_1)] {\err_1} R$ states that the configurations $(\expr_2, \state_2)$ which $(\expr_1, \state_1)$ can reduce to in one step do not violate $R$ with error more than $\err_1$, \ie, $\prst{(\expr_1, \state_1)}{\neg R} \leq \err_1$.
Finally, the last premise requires a proof of $Z$ for configurations $(\expr_2, \state_2)$ with error budget $\err_2$, but in that proof we may now assume that $R(\expr_2, \state_2)$ is satisfied, since we ``paid'' for this assumption with $\err_1$.

\paragraph{Error in expectation}
The rule \rref{step-simple} imposes a constant bound on the error credit $\err_2$ that is left available for the correctness proof of the remainder of the program $(\expr_2,\state_2)$.
However, as we saw in the examples on expected error analysis, some expressions $\expr_2$ may need more or less error credit than others.
This intuition is realized via the next rule.
\begin{mathpar}
  \infrule{step-exp}{
    \red (\cfg_1)
    \and
    \pgl[\stepdistr(\cfg_1)] {\err_1} R
    \and
    \Exists r . \All \cfg_2 . \Err_2(\cfg_2) \leq r
    \\\\ \rule{0pt}{1.3\baselineskip}
    \err_1 + \textstyle\sum_{\cfg_2 \in \Conf} \stepdistr(\cfg_1)(\cfg_2) \cdot \Err_2(\cfg_2) \leq \err
    \\ \rule{0pt}{1.1\baselineskip}
    \All \cfg_2 .
    R(\expr_2,\state_2) \wand
    \Err_2(\cfg_2) \geq 1
    \lor
    Z(\cfg_2, (\Err_2(\cfg_2)))
  }{ \execub (\cfg_1, \err, Z)}
\vspace{-2ex}
\end{mathpar}
The first two premises serve the same purpose as in \rref{step-simple}, and the third premise is a purely technical side-condition that guarantees that the sum in premise four exists.
The novelty in \rref{step-exp} is that instead of a fixed error for the ``rest of the program'', we have a configuration-indexed family of errors $\Err_2$.
Premise four states that the error budget $\err$ can be split into $\err_1$ and, for each $\cfg_2$ the starting configuration $\cfg_1$ can step to, $\Err_2(\cfg_2)$ error credits, so long as the weighted sum of the errors multiplied by the probability of attaining each $\cfg_2$ is below $\err$.
This weighted sum is, of course, nothing other than the expectation of the random variable $\Err_2$ over the distribution $\stepdistr(\cfg_1)$.
%
%
The last premise is similar to that of \rref{step-simple}, except that \rref{step-exp} of course uses the rescaled error $\Err_2(\cfg_2)$.
Another detail that was omitted from the simple rule is that we include a clause that allows us to conclude immediately if the remaining error budget exceeds~$1$.

\subsection{Soundness, Adequacy, and Almost-Sure Termination}
\label{sec:soundness-proof}

Using our model of weakest preconditions and Hoare triples, we can prove soundness of
the program logic proof rules. For reasons of space, we refer the reader to the accompanying
Coq formalization for details. 
The adequacy theorem for Hoare triples follows directly from the corresponding theorem for the weakest precondition:
\vspace*{-0.3ex}
\begin{theorem}[Limit WP adequacy]
  If $\upto \err \vdash \wpre \expr { \phi }$
  then $\All \state . \pgl[\exec(\expr,\state)]{\err}{ \phi }$
  \vspace*{-0.5ex}
\end{theorem}
Since $\exec$ is continuous in the sense that
$(\All n . \prex[n] {(\expr,\state)} \phi \leq x) \implies \prex  {(\expr,\state)} \phi \leq x$,
it suffices to prove the corresponding statement about finite executions of arbitrary length. By applying the standard soundness theorem of the Iris base logic, we can thus restrict our attention to showing that $\vdash \later^n\;  \pgl[\exec_n(\expr,\state)]{\err}{\phi} $ holds.
The proof then proceeds by induction on the step index~$n$.
The inductive step for this argument hinges on the following lemma:
\vspace*{-0.3ex}
\begin{gather*}
  \execub(\cfg_1, \err_1, (\lambda (\cfg_2, \err_2),\
  \later{}^{n+1} (\pgl[\exec_n(\cfg_2)] {\err_2} \phi)))
  \vdash \later{}^{n+1}\ (\pgl[\exec_{n+1}(\cfg_1)] {\err_1} \phi)
  \vspace*{-0.3ex}
\end{gather*}
%
Intuitively, this says that the graded lifting modality can be composed with an error bound for an $n$-step execution of the program~$\cfg_1$ to obtain an error bound on the execution of $\cfg_1$ for $(n+1)$ steps.
This should come as no surprise, since $\execub$ requires the existence of an error bound on a single execution step. The key lemma that allows this composition is then the corresponding lemma
for composing error bounds along monadic composition:
\vspace*{-0.4ex}
\begin{lemma}
  Let $\mu \in \Distr A$, and let $f$ be an $A$-indexed family of distributions, and let $\Err_2$ be a family of errors.
  If $\pgl[\mu]{\err}{\phi}$ and $\Exists x . \All a . 0 \leq \Err_2(a) \leq x$
  and $(\All a . \phi(a) \implies \pgl[f(a)]{\Err_2(a)}{\psi})$,
  then
  $\pgl[\mu \mbindi f]{\err_1 + \sum_{a \in A}\mu(a)\cdot\Err_2(a)}{\psi}$.
\end{lemma}
\vspace*{-0.3ex}
This lemma in turn is proven by carefully re-arranging the terms of the sums obtained from the definition of the bind of the probability monad.

Finally, the almost-sure termination theorem for \theaplogtotal (\cref{thm:twp-ast})
is proved by (1) proving the total adequacy theorem in much the same manner as the partial adequacy theorem (except that no later modalities are involved) and (2) by the completeness of the real numbers, in the sense that for any $x, y \in \real$, if $\All \err > 0 . x - \err \leq y$ then $x \leq y$.


\section{Related Work}
\label{sec:related}

\paragraph{Accuracy of probabilistic programs}

Our logic is inspired by aHL~\cite{ub}, which introduced the idea of using a
grading on Hoare triples that indicates the probability of the program failing
to satisfy the specification, and then adding those errors through the sequence
rule. This work considered an imperative probabilistic While language and used
their approach to reason about accuracy of differentially private mechanisms.
These ideas were then extended to the higher-order setting first
by~\citet{sato}, who consider a probabilistic lambda calculus with terminating
recursion, and then by~\citet{adversarial}, who add global first-order state via
a state monad. Compared to them, we consider full recursion and higher-order
state with dynamic allocation, and we validate new proof principles, including
expected error composition and value dependent error.

Expectation preserving composition of error can be related to
expectation-based logics, such as \citet{morganpgcl, kaminskiert,
quantitativesl}, where predicates are real-valued random variables. These logics are
presented via weakest-precondition-style predicate transformers, and
the weakest precondition of a sampling statement is precisely the
expected value of its postcondition, similar to how credits are transformed in our \rref{ht-rand-exp} rule. These
logics can also be used to reason about approximate correctness, but they
target first-order imperative languages. Recently, these techniques were
applied in \citet{amortizedert} to reason about amortized expected time complexity
of probabilistic programs. Various weakest pre-expectation-based logics also support techniques
for proving almost-sure termination~\citep[Chapter 6]{DBLP:phd/dnb/Kaminski19}. In a variant of one of these logics, \citet{newRuleAST} present
a powerful rule for proving almost-sure termination of probabilistic programs
that are out of scope of other techniques. While it is unclear if it is possible
to encode this general principle using error credits, we have used \theaplogtotal to prove
that their Example 5.4 is AST. The recently presented Caesar~\cite{caesar} provides
SMT-based support for verification in expectation-based logics.

Other approaches have tried to automate the computation of the probability that a program fails to satisfy a postcondition \cite{chakarovS13, traceabstraction, wang:quantitative:2021}.
These exploit different techniques of probability theory and programming language theory, such as martingales, concentration inequalities, approximants of fixed points, etc.

Approximate reasoning for probabilistic programs is also useful in the relational setting.
\citet{aprhl} introduce approximate couplings, which can be applied to prove different notions of approximate equivalence or differential privacy, in the setting of first-order imperative programs.
\citet{adversarial} also show that these techniques can be extended to the higher-order setting with global state.
Also in the relational setting, the line of work on Rely~\cite{rely} considers two kinds of approximate properties: the probability that a program executes correctly,
and the accuracy of the result itself.

\paragraph{Credit-based reasoning and resource analysis}

There is a long line of work on automated amortized resource analysis~\citep{DBLP:conf/popl/HofmannJ03, DBLP:journals/mscs/HoffmannJ22}, which uses a substructural type system that associates a \emph{potential} (a kind of stored credit) with a data structure.
Recent work has extended this approach to probabilistic programs~\citep{ngo18, DBLP:journals/pacmpl/WangKH20, DBLP:journals/pacmpl/DasWH23} to prove bounds on expected costs.
Analogously to our expected error rules, their typing rules for sampling
instructions allow to average the potential across all possible outcomes.

\citet{atkey} proposes a realization of ARAA-style potentials as a separation
logic resource, via a notion of \emph{credit}.  This idea was adapted to
$\lambda_{\text{ref}}$ and implemented in Coq by \citet{unionfind}, and later
brought to Iris \cite{tcred-iris, thunks}.  Error credits follow a parallel
story, namely realizing aHL-style error annotations on triples as a separation
logic resource and exploring the gains on expressive power we obtain. However,
there are no further similarities between the implementations of error credits
and time credits. In particular, the semantics of the languages,
the property being tracked and the approach to proving soundness are all
different.

\paragraph{Probabilities and separation logic}
A number of works in recent years have focused on the interactions between separation
logic and probabilities.
\citet{clutch} introduced Clutch, upon which we build.
They present a separation logic to reason about higher-order probabilistic programs,
focusing on relational properties and in particular contextual equivalence. \citet{quantitativesl}
present an expectation-based version of separation logic, which can be used to prove error
bounds for first-order pointer programs.

Polaris~\citep{polaris} is a concurrent program logic based on Iris for proving a coupling between a randomized program and a more abstract model.
The soundness theorem for Polaris allows bounds on probabilities and expectations in the model to be translated into bounds on the program across schedulers.

Other works focus on reinterpreting the notion of separating conjunction in
separation logic to represent probabilistic independence. This line of work originated
with \citet{psl}, and different variants have been developed (\citet{negativedep, lilac}).
These works also focus on first-order programs. In our work, the
separating conjunction has the standard meaning, it is only Hoare triples as
whole that have a probabilistic interpretation. Exploring a deeper connection between our
logic and separation logics for independence would be an interesting follow-up.

\section{Conclusions and Future Work}
\label{sec:conclusions}

In this paper we presented \theaplog, which develops the idea of
representing error as a resource to enable novel reasoning principles for
approximation bounds that lead to more modular and precise specifications
compared to prior work, including almost sure termination of probabilistic
algorithms.

There are multiple directions for future work. Firstly, it would be interesting to
extend \theaplog to a concurrent language, to support reasoning about
approximate randomized concurrent algorithms.
Secondly, the idea of expected
error composition should apply to other kinds of separation logic resources,
such as time credits, and could be used to reason about expected time complexity of
higher-order probabilistic programs. Thirdly, by integrating ideas
from separation logics for probabilistic independence, we could encode
concentration bounds that exploit this independence and thereby
obtain more precise error bounds. Finally, we believe our ideas should also
apply to the relational setting, where the error credits could be used to prove
approximate couplings, and have interesting applications to security and
differential privacy.


\begin{acks}
  This work was supported in part by
  the \grantsponsor{NSF}{National Science Foundation}{}, grant no.~\grantnum{NSF}{2225441},
  the \grantsponsor{Carlsberg Foundation}{Carlsberg Foundation}{}, grant no.~\grantnum{Carlsberg Foundation}{CF23-0791},
  a \grantsponsor{Villum}{Villum}{} Investigator grant, no. \grantnum{Villum}{25804}, Center for Basic Research in Program Verification (CPV), from the VILLUM Foundation,
  and the European Union (\grantsponsor{ERC}{ERC}{}, CHORDS, \grantnum{ERC}{101096090}).
  Views and opinions expressed are however those of the author(s) only and do not necessarily reflect those of the European Union or the European Research Council.
  Neither the European Union nor the granting authority can be held responsible for them.
\end{acks}


\bibliography{refs}


\begin{thebibliography}{47}


\ifx \showCODEN    \undefined \def \showCODEN     #1{\unskip}     \fi
\ifx \showDOI      \undefined \def \showDOI       #1{#1}\fi
\ifx \showISBNx    \undefined \def \showISBNx     #1{\unskip}     \fi
\ifx \showISBNxiii \undefined \def \showISBNxiii  #1{\unskip}     \fi
\ifx \showISSN     \undefined \def \showISSN      #1{\unskip}     \fi
\ifx \showLCCN     \undefined \def \showLCCN      #1{\unskip}     \fi
\ifx \shownote     \undefined \def \shownote      #1{#1}          \fi
\ifx \showarticletitle \undefined \def \showarticletitle #1{#1}   \fi
\ifx \showURL      \undefined \def \showURL       {\relax}        \fi
\providecommand\bibfield[2]{#2}
\providecommand\bibinfo[2]{#2}
\providecommand\natexlab[1]{#1}
\providecommand\showeprint[2][]{arXiv:#2}

\bibitem[Aguirre et~al\mbox{.}(2021)]%
        {adversarial}
\bibfield{author}{\bibinfo{person}{Alejandro Aguirre}, \bibinfo{person}{Gilles
  Barthe}, \bibinfo{person}{Marco Gaboardi}, \bibinfo{person}{Deepak Garg},
  \bibinfo{person}{Shin{-}ya Katsumata}, {and} \bibinfo{person}{Tetsuya Sato}.}
  \bibinfo{year}{2021}\natexlab{}.
\newblock \showarticletitle{Higher-order probabilistic adversarial
  computations: categorical semantics and program logics}.
\newblock \bibinfo{journal}{\emph{Proc. {ACM} Program. Lang.}}
  \bibinfo{volume}{5}, \bibinfo{number}{{ICFP}} (\bibinfo{year}{2021}),
  \bibinfo{pages}{1--30}.
\newblock
\urldef\tempurl%
\url{https://doi.org/10.1145/3473598}
\showDOI{\tempurl}


\bibitem[Aguirre et~al\mbox{.}(2024)]%
        {aguirre_2024_11489778}
\bibfield{author}{\bibinfo{person}{Alejandro Aguirre},
  \bibinfo{person}{Philipp~G. Haselwarter}, \bibinfo{person}{Markus de
  Medeiros}, \bibinfo{person}{Kwing~Hei Li}, \bibinfo{person}{Simon~Oddershede
  Gregersen}, \bibinfo{person}{Joseph Tassarotti}, {and} \bibinfo{person}{Lars
  Birkedal}.} \bibinfo{year}{2024}\natexlab{}.
\newblock \bibinfo{booktitle}{\emph{{Error Credits: Resourceful Reasoning about
  Error Bounds for Higher-Order Probabilistic Programs - Coq Artifact}}}.
\newblock
\urldef\tempurl%
\url{https://doi.org/10.5281/zenodo.11489778}
\showDOI{\tempurl}


\bibitem[Arons et~al\mbox{.}(2003)]%
        {planner}
\bibfield{author}{\bibinfo{person}{Tamarah Arons}, \bibinfo{person}{Amir
  Pnueli}, {and} \bibinfo{person}{Lenore~D. Zuck}.}
  \bibinfo{year}{2003}\natexlab{}.
\newblock \showarticletitle{Parameterized Verification by Probabilistic
  Abstraction}. In \bibinfo{booktitle}{\emph{Foundations of Software Science
  and Computational Structures, 6th International Conference, {FOSSACS} 2003
  Held as Part of the Joint European Conference on Theory and Practice of
  Software, {ETAPS} 2003, Warsaw, Poland, April 7-11, 2003, Proceedings}}
  \emph{(\bibinfo{series}{Lecture Notes in Computer Science},
  Vol.~\bibinfo{volume}{2620})}, \bibfield{editor}{\bibinfo{person}{Andrew~D.
  Gordon}} (Ed.). \bibinfo{publisher}{Springer}, \bibinfo{pages}{87--102}.
\newblock
\urldef\tempurl%
\url{https://doi.org/10.1007/3-540-36576-1\_6}
\showDOI{\tempurl}


\bibitem[Atkey(2011)]%
        {atkey}
\bibfield{author}{\bibinfo{person}{Robert Atkey}.}
  \bibinfo{year}{2011}\natexlab{}.
\newblock \showarticletitle{{Amortised Resource Analysis with Separation
  Logic}}.
\newblock \bibinfo{journal}{\emph{{Logical Methods in Computer Science}}}
  \bibinfo{volume}{{Volume 7, Issue 2}} (\bibinfo{date}{June}
  \bibinfo{year}{2011}).
\newblock
\urldef\tempurl%
\url{https://doi.org/10.2168/LMCS-7(2:17)2011}
\showDOI{\tempurl}


\bibitem[Bao et~al\mbox{.}(2022)]%
        {negativedep}
\bibfield{author}{\bibinfo{person}{Jialu Bao}, \bibinfo{person}{Marco
  Gaboardi}, \bibinfo{person}{Justin Hsu}, {and} \bibinfo{person}{Joseph
  Tassarotti}.} \bibinfo{year}{2022}\natexlab{}.
\newblock \showarticletitle{A separation logic for negative dependence}.
\newblock \bibinfo{journal}{\emph{Proc. {ACM} Program. Lang.}}
  \bibinfo{volume}{6}, \bibinfo{number}{{POPL}} (\bibinfo{year}{2022}),
  \bibinfo{pages}{1--29}.
\newblock
\urldef\tempurl%
\url{https://doi.org/10.1145/3498719}
\showDOI{\tempurl}


\bibitem[Barthe et~al\mbox{.}(2016a)]%
        {aprhl}
\bibfield{author}{\bibinfo{person}{Gilles Barthe},
  \bibinfo{person}{No{\'{e}}mie Fong}, \bibinfo{person}{Marco Gaboardi},
  \bibinfo{person}{Benjamin Gr{\'{e}}goire}, \bibinfo{person}{Justin Hsu},
  {and} \bibinfo{person}{Pierre{-}Yves Strub}.}
  \bibinfo{year}{2016}\natexlab{a}.
\newblock \showarticletitle{Advanced Probabilistic Couplings for Differential
  Privacy}. In \bibinfo{booktitle}{\emph{Proceedings of the 2016 {ACM} {SIGSAC}
  Conference on Computer and Communications Security, Vienna, Austria, October
  24-28, 2016}}, \bibfield{editor}{\bibinfo{person}{Edgar~R. Weippl},
  \bibinfo{person}{Stefan Katzenbeisser}, \bibinfo{person}{Christopher
  Kruegel}, \bibinfo{person}{Andrew~C. Myers}, {and} \bibinfo{person}{Shai
  Halevi}} (Eds.). \bibinfo{publisher}{{ACM}}, \bibinfo{pages}{55--67}.
\newblock
\urldef\tempurl%
\url{https://doi.org/10.1145/2976749.2978391}
\showDOI{\tempurl}


\bibitem[Barthe et~al\mbox{.}(2016b)]%
        {ub}
\bibfield{author}{\bibinfo{person}{Gilles Barthe}, \bibinfo{person}{Marco
  Gaboardi}, \bibinfo{person}{Benjamin Grégoire}, \bibinfo{person}{Justin
  Hsu}, {and} \bibinfo{person}{Pierre-Yves Strub}.}
  \bibinfo{year}{2016}\natexlab{b}.
\newblock \showarticletitle{A {{Program Logic}} for {{Union Bounds}}}. In
  \bibinfo{booktitle}{\emph{43rd {{International Colloquium}} on {{Automata}},
  {{Languages}}, and {{Programming}} ({{ICALP}} 2016)}}.
  \bibinfo{publisher}{{Schloss-Dagstuhl - Leibniz Zentrum für Informatik}}.
\newblock
\urldef\tempurl%
\url{https://doi.org/10.4230/LIPIcs.ICALP.2016.107}
\showDOI{\tempurl}


\bibitem[Barthe et~al\mbox{.}(2020)]%
        {psl}
\bibfield{author}{\bibinfo{person}{Gilles Barthe}, \bibinfo{person}{Justin
  Hsu}, {and} \bibinfo{person}{Kevin Liao}.} \bibinfo{year}{2020}\natexlab{}.
\newblock \showarticletitle{A probabilistic separation logic}.
\newblock \bibinfo{journal}{\emph{Proc. {ACM} Program. Lang.}}
  \bibinfo{volume}{4}, \bibinfo{number}{{POPL}} (\bibinfo{year}{2020}),
  \bibinfo{pages}{55:1--55:30}.
\newblock
\urldef\tempurl%
\url{https://doi.org/10.1145/3371123}
\showDOI{\tempurl}


\bibitem[Batz et~al\mbox{.}(2019)]%
        {quantitativesl}
\bibfield{author}{\bibinfo{person}{Kevin Batz},
  \bibinfo{person}{Benjamin~Lucien Kaminski}, \bibinfo{person}{Joost{-}Pieter
  Katoen}, \bibinfo{person}{Christoph Matheja}, {and} \bibinfo{person}{Thomas
  Noll}.} \bibinfo{year}{2019}\natexlab{}.
\newblock \showarticletitle{Quantitative separation logic: a logic for
  reasoning about probabilistic pointer programs}.
\newblock \bibinfo{journal}{\emph{Proc. {ACM} Program. Lang.}}
  \bibinfo{volume}{3}, \bibinfo{number}{{POPL}} (\bibinfo{year}{2019}),
  \bibinfo{pages}{34:1--34:29}.
\newblock
\urldef\tempurl%
\url{https://doi.org/10.1145/3290347}
\showDOI{\tempurl}


\bibitem[Batz et~al\mbox{.}(2023)]%
        {amortizedert}
\bibfield{author}{\bibinfo{person}{Kevin Batz},
  \bibinfo{person}{Benjamin~Lucien Kaminski}, \bibinfo{person}{Joost{-}Pieter
  Katoen}, \bibinfo{person}{Christoph Matheja}, {and} \bibinfo{person}{Lena
  Verscht}.} \bibinfo{year}{2023}\natexlab{}.
\newblock \showarticletitle{A Calculus for Amortized Expected Runtimes}.
\newblock \bibinfo{journal}{\emph{Proc. {ACM} Program. Lang.}}
  \bibinfo{volume}{7}, \bibinfo{number}{{POPL}} (\bibinfo{year}{2023}),
  \bibinfo{pages}{1957--1986}.
\newblock
\urldef\tempurl%
\url{https://doi.org/10.1145/3571260}
\showDOI{\tempurl}


\bibitem[Bellare and Rogaway(1993)]%
        {uniform-hash-assumption}
\bibfield{author}{\bibinfo{person}{Mihir Bellare} {and}
  \bibinfo{person}{Phillip Rogaway}.} \bibinfo{year}{1993}\natexlab{}.
\newblock \showarticletitle{Random Oracles are Practical: {A} Paradigm for
  Designing Efficient Protocols}. In \bibinfo{booktitle}{\emph{{CCS} '93,
  Proceedings of the 1st {ACM} Conference on Computer and Communications
  Security, Fairfax, Virginia, USA, November 3-5, 1993}},
  \bibfield{editor}{\bibinfo{person}{Dorothy~E. Denning},
  \bibinfo{person}{Raymond Pyle}, \bibinfo{person}{Ravi Ganesan},
  \bibinfo{person}{Ravi~S. Sandhu}, {and} \bibinfo{person}{Victoria Ashby}}
  (Eds.). \bibinfo{publisher}{{ACM}}, \bibinfo{pages}{62--73}.
\newblock
\urldef\tempurl%
\url{https://doi.org/10.1145/168588.168596}
\showDOI{\tempurl}


\bibitem[Benet(2014)]%
        {ipfs}
\bibfield{author}{\bibinfo{person}{Juan Benet}.}
  \bibinfo{year}{2014}\natexlab{}.
\newblock \showarticletitle{{IPFS} - Content Addressed, Versioned, {P2P} File
  System}.
\newblock \bibinfo{journal}{\emph{CoRR}}  \bibinfo{volume}{abs/1407.3561}
  (\bibinfo{year}{2014}).
\newblock
\showeprint[arXiv]{1407.3561}
\urldef\tempurl%
\url{http://arxiv.org/abs/1407.3561}
\showURL{%
\tempurl}


\bibitem[Boldo et~al\mbox{.}(2015)]%
        {coquelicot}
\bibfield{author}{\bibinfo{person}{Sylvie Boldo}, \bibinfo{person}{Catherine
  Lelay}, {and} \bibinfo{person}{Guillaume Melquiond}.}
  \bibinfo{year}{2015}\natexlab{}.
\newblock \showarticletitle{Coquelicot: {A} User-Friendly Library of Real
  Analysis for Coq}.
\newblock \bibinfo{journal}{\emph{Math. Comput. Sci.}} \bibinfo{volume}{9},
  \bibinfo{number}{1} (\bibinfo{year}{2015}), \bibinfo{pages}{41--62}.
\newblock
\urldef\tempurl%
\url{https://doi.org/10.1007/S11786-014-0181-1}
\showDOI{\tempurl}


\bibitem[Carbin et~al\mbox{.}(2013)]%
        {rely}
\bibfield{author}{\bibinfo{person}{Michael Carbin}, \bibinfo{person}{Sasa
  Misailovic}, {and} \bibinfo{person}{Martin~C. Rinard}.}
  \bibinfo{year}{2013}\natexlab{}.
\newblock \showarticletitle{Verifying quantitative reliability for programs
  that execute on unreliable hardware}. In
  \bibinfo{booktitle}{\emph{Proceedings of the 2013 ACM SIGPLAN International
  Conference on Object Oriented Programming Systems Languages \&amp;
  Applications}} (Indianapolis, Indiana, USA) \emph{(\bibinfo{series}{OOPSLA
  '13})}. \bibinfo{publisher}{Association for Computing Machinery},
  \bibinfo{address}{New York, NY, USA}, \bibinfo{pages}{33–52}.
\newblock
\showISBNx{9781450323741}
\urldef\tempurl%
\url{https://doi.org/10.1145/2509136.2509546}
\showDOI{\tempurl}


\bibitem[Chakarov and Sankaranarayanan(2013)]%
        {chakarovS13}
\bibfield{author}{\bibinfo{person}{Aleksandar Chakarov} {and}
  \bibinfo{person}{Sriram Sankaranarayanan}.} \bibinfo{year}{2013}\natexlab{}.
\newblock \showarticletitle{Probabilistic Program Analysis with Martingales}.
  In \bibinfo{booktitle}{\emph{Computer Aided Verification - 25th International
  Conference, {CAV} 2013, Saint Petersburg, Russia, July 13-19, 2013.
  Proceedings}} \emph{(\bibinfo{series}{Lecture Notes in Computer Science},
  Vol.~\bibinfo{volume}{8044})}, \bibfield{editor}{\bibinfo{person}{Natasha
  Sharygina} {and} \bibinfo{person}{Helmut Veith}} (Eds.).
  \bibinfo{publisher}{Springer}, \bibinfo{pages}{511--526}.
\newblock
\urldef\tempurl%
\url{https://doi.org/10.1007/978-3-642-39799-8\_34}
\showDOI{\tempurl}


\bibitem[Chargu{\'e}raud and Pottier(2019)]%
        {unionfind}
\bibfield{author}{\bibinfo{person}{Arthur Chargu{\'e}raud} {and}
  \bibinfo{person}{Fran{\c c}ois Pottier}.} \bibinfo{year}{2019}\natexlab{}.
\newblock \showarticletitle{Verifying the {{Correctness}} and {{Amortized
  Complexity}} of a {{Union-Find Implementation}} in {{Separation Logic}} with
  {{Time Credits}}}.
\newblock \bibinfo{journal}{\emph{Journal of Automated Reasoning}}
  \bibinfo{volume}{62}, \bibinfo{number}{3} (\bibinfo{date}{March}
  \bibinfo{year}{2019}), \bibinfo{pages}{331--365}.
\newblock
\showISSN{1573-0670}
\urldef\tempurl%
\url{https://doi.org/10.1007/s10817-017-9431-7}
\showDOI{\tempurl}


\bibitem[Cormen et~al\mbox{.}(2009)]%
        {clrs}
\bibfield{author}{\bibinfo{person}{Thomas~H. Cormen},
  \bibinfo{person}{Charles~E. Leiserson}, \bibinfo{person}{Ronald~L. Rivest},
  {and} \bibinfo{person}{Clifford Stein}.} \bibinfo{year}{2009}\natexlab{}.
\newblock \bibinfo{booktitle}{\emph{Introduction to Algorithms, 3rd Edition}}.
\newblock \bibinfo{publisher}{{MIT} Press}.
\newblock
\showISBNx{978-0-262-03384-8}
\urldef\tempurl%
\url{http://mitpress.mit.edu/books/introduction-algorithms}
\showURL{%
\tempurl}


\bibitem[Das et~al\mbox{.}(2023)]%
        {DBLP:journals/pacmpl/DasWH23}
\bibfield{author}{\bibinfo{person}{Ankush Das}, \bibinfo{person}{Di Wang},
  {and} \bibinfo{person}{Jan Hoffmann}.} \bibinfo{year}{2023}\natexlab{}.
\newblock \showarticletitle{Probabilistic Resource-Aware Session Types}.
\newblock \bibinfo{journal}{\emph{Proc. {ACM} Program. Lang.}}
  \bibinfo{volume}{7}, \bibinfo{number}{{POPL}} (\bibinfo{year}{2023}),
  \bibinfo{pages}{1925--1956}.
\newblock
\urldef\tempurl%
\url{https://doi.org/10.1145/3571259}
\showDOI{\tempurl}


\bibitem[DeCandia et~al\mbox{.}(2007)]%
        {dynamo}
\bibfield{author}{\bibinfo{person}{Giuseppe DeCandia}, \bibinfo{person}{Deniz
  Hastorun}, \bibinfo{person}{Madan Jampani}, \bibinfo{person}{Gunavardhan
  Kakulapati}, \bibinfo{person}{Avinash Lakshman}, \bibinfo{person}{Alex
  Pilchin}, \bibinfo{person}{Swaminathan Sivasubramanian},
  \bibinfo{person}{Peter Vosshall}, {and} \bibinfo{person}{Werner Vogels}.}
  \bibinfo{year}{2007}\natexlab{}.
\newblock \showarticletitle{Dynamo: amazon's highly available key-value store}.
  In \bibinfo{booktitle}{\emph{Proceedings of the 21st {ACM} Symposium on
  Operating Systems Principles 2007, {SOSP} 2007, Stevenson, Washington, USA,
  October 14-17, 2007}}, \bibfield{editor}{\bibinfo{person}{Thomas~C. Bressoud}
  {and} \bibinfo{person}{M.~Frans Kaashoek}} (Eds.).
  \bibinfo{publisher}{{ACM}}, \bibinfo{pages}{205--220}.
\newblock
\urldef\tempurl%
\url{https://doi.org/10.1145/1294261.1294281}
\showDOI{\tempurl}


\bibitem[Gregersen et~al\mbox{.}(2024)]%
        {clutch}
\bibfield{author}{\bibinfo{person}{Simon~Oddershede Gregersen},
  \bibinfo{person}{Alejandro Aguirre}, \bibinfo{person}{Philipp~G.
  Haselwarter}, \bibinfo{person}{Joseph Tassarotti}, {and}
  \bibinfo{person}{Lars Birkedal}.} \bibinfo{year}{2024}\natexlab{}.
\newblock \showarticletitle{Asynchronous Probabilistic Couplings in
  Higher-Order Separation Logic}.
\newblock \bibinfo{journal}{\emph{Proc. ACM Program. Lang.}}
  \bibinfo{volume}{8}, \bibinfo{number}{POPL}, Article \bibinfo{articleno}{26}
  (\bibinfo{year}{2024}).
\newblock
\urldef\tempurl%
\url{https://doi.org/10.1145/3632868}
\showDOI{\tempurl}


\bibitem[Hoffmann and Jost(2022)]%
        {DBLP:journals/mscs/HoffmannJ22}
\bibfield{author}{\bibinfo{person}{Jan Hoffmann} {and} \bibinfo{person}{Steffen
  Jost}.} \bibinfo{year}{2022}\natexlab{}.
\newblock \showarticletitle{Two decades of automatic amortized resource
  analysis}.
\newblock \bibinfo{journal}{\emph{Math. Struct. Comput. Sci.}}
  \bibinfo{volume}{32}, \bibinfo{number}{6} (\bibinfo{year}{2022}),
  \bibinfo{pages}{729--759}.
\newblock
\urldef\tempurl%
\url{https://doi.org/10.1017/S0960129521000487}
\showDOI{\tempurl}


\bibitem[Hofmann and Jost(2003)]%
        {DBLP:conf/popl/HofmannJ03}
\bibfield{author}{\bibinfo{person}{Martin Hofmann} {and}
  \bibinfo{person}{Steffen Jost}.} \bibinfo{year}{2003}\natexlab{}.
\newblock \showarticletitle{Static prediction of heap space usage for
  first-order functional programs}. In \bibinfo{booktitle}{\emph{Conference
  Record of {POPL} 2003: The 30th {SIGPLAN-SIGACT} Symposium on Principles of
  Programming Languages, New Orleans, Louisisana, USA, January 15-17, 2003}},
  \bibfield{editor}{\bibinfo{person}{Alex Aiken} {and} \bibinfo{person}{Greg
  Morrisett}} (Eds.). \bibinfo{publisher}{{ACM}}, \bibinfo{pages}{185--197}.
\newblock
\urldef\tempurl%
\url{https://doi.org/10.1145/604131.604148}
\showDOI{\tempurl}


\bibitem[Jung et~al\mbox{.}(2016)]%
        {irisho}
\bibfield{author}{\bibinfo{person}{Ralf Jung}, \bibinfo{person}{Robbert
  Krebbers}, \bibinfo{person}{Lars Birkedal}, {and} \bibinfo{person}{Derek
  Dreyer}.} \bibinfo{year}{2016}\natexlab{}.
\newblock \showarticletitle{Higher-order ghost state}.
\newblock \bibinfo{journal}{\emph{SIGPLAN Not.}} \bibinfo{volume}{51},
  \bibinfo{number}{9} (\bibinfo{date}{sep} \bibinfo{year}{2016}),
  \bibinfo{pages}{256–269}.
\newblock
\showISSN{0362-1340}
\urldef\tempurl%
\url{https://doi.org/10.1145/3022670.2951943}
\showDOI{\tempurl}


\bibitem[Jung et~al\mbox{.}(2018)]%
        {irisjournal}
\bibfield{author}{\bibinfo{person}{Ralf Jung}, \bibinfo{person}{Robbert
  Krebbers}, \bibinfo{person}{Jacques{-}Henri Jourdan}, \bibinfo{person}{Ales
  Bizjak}, \bibinfo{person}{Lars Birkedal}, {and} \bibinfo{person}{Derek
  Dreyer}.} \bibinfo{year}{2018}\natexlab{}.
\newblock \showarticletitle{Iris from the ground up: {A} modular foundation for
  higher-order concurrent separation logic}.
\newblock \bibinfo{journal}{\emph{J. Funct. Program.}}  \bibinfo{volume}{28}
  (\bibinfo{year}{2018}), \bibinfo{pages}{e20}.
\newblock
\urldef\tempurl%
\url{https://doi.org/10.1017/S0956796818000151}
\showDOI{\tempurl}


\bibitem[Jung et~al\mbox{.}(2015a)]%
        {irismonoid}
\bibfield{author}{\bibinfo{person}{Ralf Jung}, \bibinfo{person}{David Swasey},
  \bibinfo{person}{Filip Sieczkowski}, \bibinfo{person}{Kasper Svendsen},
  \bibinfo{person}{Aaron Turon}, \bibinfo{person}{Lars Birkedal}, {and}
  \bibinfo{person}{Derek Dreyer}.} \bibinfo{year}{2015}\natexlab{a}.
\newblock \showarticletitle{Iris: Monoids and Invariants as an Orthogonal Basis
  for Concurrent Reasoning}. In \bibinfo{booktitle}{\emph{ACM-SIGACT Symposium
  on Principles of Programming Languages}}.
\newblock
\urldef\tempurl%
\url{https://api.semanticscholar.org/CorpusID:1174404}
\showURL{%
\tempurl}


\bibitem[Jung et~al\mbox{.}(2015b)]%
        {iris}
\bibfield{author}{\bibinfo{person}{Ralf Jung}, \bibinfo{person}{David Swasey},
  \bibinfo{person}{Filip Sieczkowski}, \bibinfo{person}{Kasper Svendsen},
  \bibinfo{person}{Aaron Turon}, \bibinfo{person}{Lars Birkedal}, {and}
  \bibinfo{person}{Derek Dreyer}.} \bibinfo{year}{2015}\natexlab{b}.
\newblock \showarticletitle{Iris: Monoids and Invariants as an Orthogonal Basis
  for Concurrent Reasoning}. In \bibinfo{booktitle}{\emph{Proceedings of the
  42nd Annual {ACM} {SIGPLAN-SIGACT} Symposium on Principles of Programming
  Languages, {POPL} 2015, Mumbai, India, January 15-17, 2015}}.
  \bibinfo{pages}{637--650}.
\newblock
\urldef\tempurl%
\url{https://doi.org/10.1145/2676726.2676980}
\showDOI{\tempurl}


\bibitem[Kaminski(2019)]%
        {DBLP:phd/dnb/Kaminski19}
\bibfield{author}{\bibinfo{person}{Benjamin~Lucien Kaminski}.}
  \bibinfo{year}{2019}\natexlab{}.
\newblock \emph{\bibinfo{title}{Advanced weakest precondition calculi for
  probabilistic programs}}.
\newblock \bibinfo{thesistype}{Ph.\,D. Dissertation}. \bibinfo{school}{{RWTH}
  Aachen University, Germany}.
\newblock
\urldef\tempurl%
\url{http://publications.rwth-aachen.de/record/755408}
\showURL{%
\tempurl}


\bibitem[Kaminski et~al\mbox{.}(2016)]%
        {kaminskiert}
\bibfield{author}{\bibinfo{person}{Benjamin~Lucien Kaminski},
  \bibinfo{person}{Joost{-}Pieter Katoen}, \bibinfo{person}{Christoph Matheja},
  {and} \bibinfo{person}{Federico Olmedo}.} \bibinfo{year}{2016}\natexlab{}.
\newblock \showarticletitle{Weakest Precondition Reasoning for Expected
  Run-Times of Probabilistic Programs}. In
  \bibinfo{booktitle}{\emph{Programming Languages and Systems - 25th European
  Symposium on Programming, {ESOP} 2016, Held as Part of the European Joint
  Conferences on Theory and Practice of Software, {ETAPS} 2016, Eindhoven, The
  Netherlands, April 2-8, 2016, Proceedings}} \emph{(\bibinfo{series}{Lecture
  Notes in Computer Science}, Vol.~\bibinfo{volume}{9632})},
  \bibfield{editor}{\bibinfo{person}{Peter Thiemann}} (Ed.).
  \bibinfo{publisher}{Springer}, \bibinfo{pages}{364--389}.
\newblock
\urldef\tempurl%
\url{https://doi.org/10.1007/978-3-662-49498-1\_15}
\showDOI{\tempurl}


\bibitem[Krebbers et~al\mbox{.}(2017)]%
        {irisinteractive}
\bibfield{author}{\bibinfo{person}{Robbert Krebbers}, \bibinfo{person}{Amin
  Timany}, {and} \bibinfo{person}{Lars Birkedal}.}
  \bibinfo{year}{2017}\natexlab{}.
\newblock \showarticletitle{Interactive proofs in higher-order concurrent
  separation logic}. In \bibinfo{booktitle}{\emph{Proceedings of the 44th {ACM}
  {SIGPLAN} Symposium on Principles of Programming Languages, {POPL} 2017,
  Paris, France, January 18-20, 2017}},
  \bibfield{editor}{\bibinfo{person}{Giuseppe Castagna} {and}
  \bibinfo{person}{Andrew~D. Gordon}} (Eds.). \bibinfo{publisher}{{ACM}},
  \bibinfo{pages}{205--217}.
\newblock
\urldef\tempurl%
\url{https://doi.org/10.1145/3009837.3009855}
\showDOI{\tempurl}


\bibitem[Li et~al\mbox{.}(2023)]%
        {lilac}
\bibfield{author}{\bibinfo{person}{John~M. Li}, \bibinfo{person}{Amal Ahmed},
  {and} \bibinfo{person}{Steven Holtzen}.} \bibinfo{year}{2023}\natexlab{}.
\newblock \showarticletitle{Lilac: {A} Modal Separation Logic for Conditional
  Probability}.
\newblock \bibinfo{journal}{\emph{Proc. {ACM} Program. Lang.}}
  \bibinfo{volume}{7}, \bibinfo{number}{{PLDI}} (\bibinfo{year}{2023}),
  \bibinfo{pages}{148--171}.
\newblock
\urldef\tempurl%
\url{https://doi.org/10.1145/3591226}
\showDOI{\tempurl}


\bibitem[McIver et~al\mbox{.}(2018)]%
        {newRuleAST}
\bibfield{author}{\bibinfo{person}{Annabelle McIver}, \bibinfo{person}{Carroll
  Morgan}, \bibinfo{person}{Benjamin~Lucien Kaminski}, {and}
  \bibinfo{person}{Joost{-}Pieter Katoen}.} \bibinfo{year}{2018}\natexlab{}.
\newblock \showarticletitle{A new proof rule for almost-sure termination}.
\newblock \bibinfo{journal}{\emph{Proc. {ACM} Program. Lang.}}
  \bibinfo{volume}{2}, \bibinfo{number}{{POPL}} (\bibinfo{year}{2018}),
  \bibinfo{pages}{33:1--33:28}.
\newblock
\urldef\tempurl%
\url{https://doi.org/10.1145/3158121}
\showDOI{\tempurl}


\bibitem[Merkle(1987)]%
        {merkle-tree}
\bibfield{author}{\bibinfo{person}{Ralph~C. Merkle}.}
  \bibinfo{year}{1987}\natexlab{}.
\newblock \showarticletitle{A Digital Signature Based on a Conventional
  Encryption Function}. In \bibinfo{booktitle}{\emph{Advances in Cryptology -
  {CRYPTO} '87, {A} Conference on the Theory and Applications of Cryptographic
  Techniques, Santa Barbara, California, USA, August 16-20, 1987, Proceedings}}
  \emph{(\bibinfo{series}{Lecture Notes in Computer Science},
  Vol.~\bibinfo{volume}{293})}, \bibfield{editor}{\bibinfo{person}{Carl
  Pomerance}} (Ed.). \bibinfo{publisher}{Springer}, \bibinfo{pages}{369--378}.
\newblock
\urldef\tempurl%
\url{https://doi.org/10.1007/3-540-48184-2\_32}
\showDOI{\tempurl}


\bibitem[M{\'e}vel et~al\mbox{.}(2019)]%
        {tcred-iris}
\bibfield{author}{\bibinfo{person}{Glen M{\'e}vel},
  \bibinfo{person}{Jacques-Henri Jourdan}, {and} \bibinfo{person}{Fran{\c c}ois
  Pottier}.} \bibinfo{year}{2019}\natexlab{}.
\newblock \showarticletitle{Time {{Credits}} and {{Time Receipts}} in
  {{Iris}}}. In \bibinfo{booktitle}{\emph{Programming {{Languages}} and
  {{Systems}}}} \emph{(\bibinfo{series}{Lecture {{Notes}} in {{Computer
  Science}}})}, \bibfield{editor}{\bibinfo{person}{Lu{\'i}s Caires}} (Ed.).
  \bibinfo{publisher}{{Springer International Publishing}},
  \bibinfo{address}{{Cham}}, \bibinfo{pages}{3--29}.
\newblock
\showISBNx{978-3-030-17184-1}
\urldef\tempurl%
\url{https://doi.org/10.1007/978-3-030-17184-1_1}
\showDOI{\tempurl}


\bibitem[Miller(1975)]%
        {miller}
\bibfield{author}{\bibinfo{person}{Gary~L. Miller}.}
  \bibinfo{year}{1975}\natexlab{}.
\newblock \showarticletitle{Riemann's Hypothesis and tests for primality}. In
  \bibinfo{booktitle}{\emph{Proceedings of the Seventh Annual ACM Symposium on
  Theory of Computing}} (Albuquerque, New Mexico, USA)
  \emph{(\bibinfo{series}{STOC '75})}. \bibinfo{publisher}{Association for
  Computing Machinery}, \bibinfo{address}{New York, NY, USA},
  \bibinfo{pages}{234–239}.
\newblock
\showISBNx{9781450374194}
\urldef\tempurl%
\url{https://doi.org/10.1145/800116.803773}
\showDOI{\tempurl}


\bibitem[Morgan et~al\mbox{.}(1996)]%
        {morganpgcl}
\bibfield{author}{\bibinfo{person}{Carroll Morgan}, \bibinfo{person}{Annabelle
  McIver}, {and} \bibinfo{person}{Karen Seidel}.}
  \bibinfo{year}{1996}\natexlab{}.
\newblock \showarticletitle{Probabilistic Predicate Transformers}.
\newblock \bibinfo{journal}{\emph{{ACM} Trans. Program. Lang. Syst.}}
  \bibinfo{volume}{18}, \bibinfo{number}{3} (\bibinfo{year}{1996}),
  \bibinfo{pages}{325--353}.
\newblock
\urldef\tempurl%
\url{https://doi.org/10.1145/229542.229547}
\showDOI{\tempurl}


\bibitem[Ngo et~al\mbox{.}(2018)]%
        {ngo18}
\bibfield{author}{\bibinfo{person}{Van~Chan Ngo}, \bibinfo{person}{Quentin
  Carbonneaux}, {and} \bibinfo{person}{Jan Hoffmann}.}
  \bibinfo{year}{2018}\natexlab{}.
\newblock \showarticletitle{Bounded expectations: resource analysis for
  probabilistic programs}. In \bibinfo{booktitle}{\emph{Proceedings of the 39th
  {ACM} {SIGPLAN} Conference on Programming Language Design and Implementation,
  {PLDI} 2018, Philadelphia, PA, USA, June 18-22, 2018}},
  \bibfield{editor}{\bibinfo{person}{Jeffrey~S. Foster} {and}
  \bibinfo{person}{Dan Grossman}} (Eds.). \bibinfo{publisher}{{ACM}},
  \bibinfo{pages}{496--512}.
\newblock
\urldef\tempurl%
\url{https://doi.org/10.1145/3192366.3192394}
\showDOI{\tempurl}


\bibitem[Papadimitriou(1991)]%
        {walksat}
\bibfield{author}{\bibinfo{person}{C.H. Papadimitriou}.}
  \bibinfo{year}{1991}\natexlab{}.
\newblock \showarticletitle{On selecting a satisfying truth assignment}. In
  \bibinfo{booktitle}{\emph{[1991] Proceedings 32nd Annual Symposium of
  Foundations of Computer Science}}. \bibinfo{pages}{163--169}.
\newblock
\urldef\tempurl%
\url{https://doi.org/10.1109/SFCS.1991.185365}
\showDOI{\tempurl}


\bibitem[Pottier et~al\mbox{.}(2024)]%
        {thunks}
\bibfield{author}{\bibinfo{person}{Fran{\c{c}}ois Pottier},
  \bibinfo{person}{Arma{\"{e}}l Gu{\'{e}}neau},
  \bibinfo{person}{Jacques{-}Henri Jourdan}, {and} \bibinfo{person}{Glen
  M{\'{e}}vel}.} \bibinfo{year}{2024}\natexlab{}.
\newblock \showarticletitle{Thunks and Debits in Separation Logic with Time
  Credits}.
\newblock \bibinfo{journal}{\emph{Proc. {ACM} Program. Lang.}}
  \bibinfo{volume}{8}, \bibinfo{number}{{POPL}} (\bibinfo{year}{2024}),
  \bibinfo{pages}{1482--1508}.
\newblock
\urldef\tempurl%
\url{https://doi.org/10.1145/3632892}
\showDOI{\tempurl}


\bibitem[Rabin(1980)]%
        {rabin}
\bibfield{author}{\bibinfo{person}{Michael~O Rabin}.}
  \bibinfo{year}{1980}\natexlab{}.
\newblock \showarticletitle{Probabilistic algorithm for testing primality}.
\newblock \bibinfo{journal}{\emph{Journal of Number Theory}}
  \bibinfo{volume}{12}, \bibinfo{number}{1} (\bibinfo{date}{Feb.}
  \bibinfo{year}{1980}), \bibinfo{pages}{128–138}.
\newblock
\showISSN{0022-314X}
\urldef\tempurl%
\url{https://doi.org/10.1016/0022-314x(80)90084-0}
\showDOI{\tempurl}


\bibitem[Sato et~al\mbox{.}(2019)]%
        {sato}
\bibfield{author}{\bibinfo{person}{Tetsuya Sato}, \bibinfo{person}{Alejandro
  Aguirre}, \bibinfo{person}{Gilles Barthe}, \bibinfo{person}{Marco Gaboardi},
  \bibinfo{person}{Deepak Garg}, {and} \bibinfo{person}{Justin Hsu}.}
  \bibinfo{year}{2019}\natexlab{}.
\newblock \showarticletitle{Formal Verification of Higher-Order Probabilistic
  Programs: Reasoning about Approximation, Convergence, Bayesian Inference, and
  Optimization}.
\newblock \bibinfo{journal}{\emph{Proc. {ACM} Program. Lang.}}
  \bibinfo{volume}{3}, \bibinfo{number}{{POPL}} (\bibinfo{year}{2019}),
  \bibinfo{pages}{38:1--38:30}.
\newblock
\urldef\tempurl%
\url{https://doi.org/10.1145/3290351}
\showDOI{\tempurl}


\bibitem[Schr{\"{o}}er et~al\mbox{.}(2023)]%
        {caesar}
\bibfield{author}{\bibinfo{person}{Philipp Schr{\"{o}}er},
  \bibinfo{person}{Kevin Batz}, \bibinfo{person}{Benjamin~Lucien Kaminski},
  \bibinfo{person}{Joost{-}Pieter Katoen}, {and} \bibinfo{person}{Christoph
  Matheja}.} \bibinfo{year}{2023}\natexlab{}.
\newblock \showarticletitle{A Deductive Verification Infrastructure for
  Probabilistic Programs}.
\newblock \bibinfo{journal}{\emph{Proc. {ACM} Program. Lang.}}
  \bibinfo{volume}{7}, \bibinfo{number}{{OOPSLA2}} (\bibinfo{year}{2023}),
  \bibinfo{pages}{2052--2082}.
\newblock
\urldef\tempurl%
\url{https://doi.org/10.1145/3622870}
\showDOI{\tempurl}


\bibitem[Smith et~al\mbox{.}(2019)]%
        {traceabstraction}
\bibfield{author}{\bibinfo{person}{Calvin Smith}, \bibinfo{person}{Justin Hsu},
  {and} \bibinfo{person}{Aws Albarghouthi}.} \bibinfo{year}{2019}\natexlab{}.
\newblock \showarticletitle{Trace abstraction modulo probability}.
\newblock \bibinfo{journal}{\emph{Proc. {ACM} Program. Lang.}}
  \bibinfo{volume}{3}, \bibinfo{number}{{POPL}} (\bibinfo{year}{2019}),
  \bibinfo{pages}{39:1--39:31}.
\newblock
\urldef\tempurl%
\url{https://doi.org/10.1145/3290352}
\showDOI{\tempurl}


\bibitem[Solovay and Strassen(1977)]%
        {solovay-strassen}
\bibfield{author}{\bibinfo{person}{R. Solovay} {and} \bibinfo{person}{V.
  Strassen}.} \bibinfo{year}{1977}\natexlab{}.
\newblock \showarticletitle{A Fast Monte-Carlo Test for Primality}.
\newblock \bibinfo{journal}{\emph{SIAM J. Comput.}} \bibinfo{volume}{6},
  \bibinfo{number}{1} (\bibinfo{year}{1977}), \bibinfo{pages}{84--85}.
\newblock
\urldef\tempurl%
\url{https://doi.org/10.1137/0206006}
\showDOI{\tempurl}


\bibitem[Spies et~al\mbox{.}(2022)]%
        {DBLP:journals/pacmpl/SpiesGTJKBD22}
\bibfield{author}{\bibinfo{person}{Simon Spies}, \bibinfo{person}{Lennard
  G{\"{a}}her}, \bibinfo{person}{Joseph Tassarotti}, \bibinfo{person}{Ralf
  Jung}, \bibinfo{person}{Robbert Krebbers}, \bibinfo{person}{Lars Birkedal},
  {and} \bibinfo{person}{Derek Dreyer}.} \bibinfo{year}{2022}\natexlab{}.
\newblock \showarticletitle{Later credits: resourceful reasoning for the later
  modality}.
\newblock \bibinfo{journal}{\emph{Proc. {ACM} Program. Lang.}}
  \bibinfo{volume}{6}, \bibinfo{number}{{ICFP}} (\bibinfo{year}{2022}),
  \bibinfo{pages}{283--311}.
\newblock
\urldef\tempurl%
\url{https://doi.org/10.1145/3547631}
\showDOI{\tempurl}


\bibitem[Tassarotti and Harper(2019)]%
        {polaris}
\bibfield{author}{\bibinfo{person}{Joseph Tassarotti} {and}
  \bibinfo{person}{Robert Harper}.} \bibinfo{year}{2019}\natexlab{}.
\newblock \showarticletitle{A separation logic for concurrent randomized
  programs}.
\newblock \bibinfo{journal}{\emph{Proc. {ACM} Program. Lang.}}
  \bibinfo{volume}{3}, \bibinfo{number}{{POPL}} (\bibinfo{year}{2019}),
  \bibinfo{pages}{64:1--64:30}.
\newblock
\urldef\tempurl%
\url{https://doi.org/10.1145/3290377}
\showDOI{\tempurl}


\bibitem[Wang et~al\mbox{.}(2020)]%
        {DBLP:journals/pacmpl/WangKH20}
\bibfield{author}{\bibinfo{person}{Di Wang}, \bibinfo{person}{David~M. Kahn},
  {and} \bibinfo{person}{Jan Hoffmann}.} \bibinfo{year}{2020}\natexlab{}.
\newblock \showarticletitle{Raising expectations: automating expected cost
  analysis with types}.
\newblock \bibinfo{journal}{\emph{Proc. {ACM} Program. Lang.}}
  \bibinfo{volume}{4}, \bibinfo{number}{{ICFP}} (\bibinfo{year}{2020}),
  \bibinfo{pages}{110:1--110:31}.
\newblock
\urldef\tempurl%
\url{https://doi.org/10.1145/3408992}
\showDOI{\tempurl}


\bibitem[Wang et~al\mbox{.}(2021)]%
        {wang:quantitative:2021}
\bibfield{author}{\bibinfo{person}{Jinyi Wang}, \bibinfo{person}{Yican Sun},
  \bibinfo{person}{Hongfei Fu}, \bibinfo{person}{Krishnendu Chatterjee}, {and}
  \bibinfo{person}{Amir~Kafshdar Goharshady}.} \bibinfo{year}{2021}\natexlab{}.
\newblock \showarticletitle{Quantitative Analysis of Assertion Violations in
  Probabilistic Programs}. In \bibinfo{booktitle}{\emph{Proceedings of the 42nd
  {{ACM SIGPLAN International Conference}} on {{Programming Language Design}}
  and {{Implementation}}}} \emph{(\bibinfo{series}{{{PLDI}} 2021})}.
  \bibinfo{publisher}{{Association for Computing Machinery}},
  \bibinfo{address}{{New York, NY, USA}}, \bibinfo{pages}{1171--1186}.
\newblock
\showISBNx{978-1-4503-8391-2}
\urldef\tempurl%
\url{https://doi.org/10.1145/3453483.3454102}
\showDOI{\tempurl}


\end{thebibliography}

 \pagebreak

 \appendix

 \section{Appendix: Full Definition of the Weakest Precondition and Graded Lifting Modality}

The definitions of the weakest precondition and of the graded lifting modality as presented in \cref{sec:soundness} contain some simplifications for the sake of pedagogy.
We now restate the definitions in full detail.

\subsection{The Weakest Precondition in Detail}
\label{sec:wp-full}

The full definition of the weakest precondition differs from the one presented in \cref{sec:soundness} in that it also contains the invariant mask annotation and fancy update modality of Iris.
\begin{align*}
  \label{def:wp-full}
  \wpre{\expr_1}[\mask]{\Phi} \eqdef{}
   & (\expr_1 \in \Val \mathrel{\land} \pvs[\mask] \Phi(\expr_1)) \lor{}                                                                       \\
   & (\expr_1 \not\in \Val \mathrel{\land} \All \state_1, \err_1 . \stateinterp(\state_1) \sep \uptoauth{\err_1} \wand \pvs[\mask][\emptyset] \\
   & \quad
  \execub \
  ( \expr_1, \state_1,
     \err_1,
  \ (\Lam \, \expr_2, \state_2, \err_2
     \ . \later \pvs[\emptyset][\mask]
     (\stateinterp(\state_2) \sep \uptoauth{\err_2} \sep \wpre{\expr_2}[\mask]{\Phi})))
\end{align*}
Just as before, the full definition of the total weakest precondition is obtained by omitting the later modality on the last line, and by taking the least instead of the guarded fixed point of the recursive definition.

\subsection{The Graded Lifting Modality in Detail}
\label{sec:modality-full}

In \cref{sec:modality}, we presented a simplified version of the graded lifting modality which does not support presampling tapes. The full definition of $\execub$ contains two clauses: \rref{step-exp} for expected error lifting of program steps and \rref{statestep-exp} for the presampling analog.

The rules in this section should be read as defining $\execub$ as an inductive predicate, \ie as the least fixed point of the closure system associated to the rules.

\paragraph{Adding presampling tapes}
The presampling ghost operations on tapes are realized through an auxiliary \defemph{state steps} relation $\statestepdistr : \Lbl \times \State \ra \Distr \State $. If $\lbl$ is the label associated to an (allocated) tape with bound $\tapebound$, then $\statestepdistr_\lbl(\state_1)$ denotes the distribution on states obtained by appending a uniformly randomly sampled value between $0$ and $\tapebound$ to the end of tape $\lbl$:
\begin{align*}
  \statestepdistr_\lbl(\state_1)(\state_2) =
  \begin{cases}
    \textstyle\frac 1 {\tapebound + 1}
    & \text{if }
      \state_2 = \state_1[\lbl \to (\tapebound, \tape \cdot n)] \text{ and } \state_1(\lbl) = (\tapebound, \tape) \text{ and } n \leq \tapebound, \\
    0 & \text{otherwise.}
  \end{cases}
\end{align*}
We can now extend the graded lifting modality to allow taking state steps. Just as we did with \rref{step-simple}, we will first discuss a simplified rule that does not support reasoning about errors in expectation. In practice, this rule is derivable from the rule \rref{statestep-exp} below.
\begin{mathpar}
  \infrule{statestep-simple}{
    \err_1 + \err_2 \leq \err
    \and
    \pgl[\statestepdistr_\lbl(\state_1)] {\err_1} R
    \and
    \All \state_2 .  R(\state_2) \wand \execub(\expr_1, \state_2, \err_2, Z)
  }{
    \execub (\expr_1, \state_1, \err, Z)
  }
\end{mathpar}
The user of the rule can, once again, split the error budget $\err$ between a first step and the remainder of the program. In \rref{statestep-simple}, however, the first step is a state step, \ie a purely logical step in the ghost state.
The recursive occurrence of $\execub$ in the last premise of \rref{statestep-simple} allows the user of the modality to perform a number of state steps before eventually proving the base case of the modality, namely \rref{step-exp}.

Finally, we can combine the idea of reasoning of expected-error reasoning with state steps via the following rule.
\begin{mathpar}
  \infrule{statestep-exp}{
    \red (\expr_1, \state_1)
    \and
    \pgl[\statestepdistr_\lbl(\state_1)] {\err_1} R
    \and
    \Exists r . \All \cfg_2 . \Err_2(\cfg_2) \leq r
    \\ \rule{0pt}{1.3\baselineskip}
    \err_1 + \sum_{\state_2 \in \State} \statestepdistr(\expr_1, \state_1)(\expr_1, \state_2) \cdot \Err_2(\expr_1, \state_2) \leq \err
    \\ \rule{0pt}{1.1\baselineskip}
    \All \state_2 .  R(\state_2) \wand
    \Err_2(\expr_1, \state_2) \geq 1 \lor  \execub(\expr_1, \state_2, \Err_2(\expr_1, \state_2))
  }{ \execub (\expr_1, \state_1, \err, Z)}
\end{mathpar}
This rule is used, for instance, to derive the program logic rule \rref{presample-exp}.

\newcommand{\mapprog}{\langv{map}}
\newcommand{\listmap}{\langv{list.map}}

\newcommand{\ecAmp}{\logv{ecAmp}}
\newcommand{\ecRem}{\logv{ecRem}}
\newcommand{\ecExc}{\logv{ecExc}}

\section{A Specification for Map}
\label{appendix:map}



Consider the higher-order $\mapexpr$ function, which takes a function $\textit{fv}$ and a list $lv$ as input, and returns a list where $\textit{fv}$ is applied to each element of the original list. Ignoring all error bounds reasoning, a reasonable specification for $\mapexpr$ could look something like the following (here, $\mapfun$ is a Coq-level map):
\begin{equation}
\small
\hoare {
  \begin{array}{c}
     \islist\ xs\ l \sep \\
        \All \var . \hoare{P (\var)}{e\ x}{x' .\ x' = f(x) \sep Q(x')}\sep \\
        \textstyle\Sep_{x\in xs} P(x)
        \end{array}
  }{\mapexpr\ \expr\ l}
  {l' . \begin{array}{c}
      \islist\ (\mapfun\ f\ xs)\ l' \sep \\
      \Sep_{x' \in (\mapfun\ f\ xs)} Q(x')
    \end{array}
  }
\label{eqn:precise-map}
\end{equation}

Now consider the case where each application of $\textit{fv}$ incurs $\Err(x)$ error credit, varying for each $x$ stored in the list. It follows that we would want the term $\mapexpr\ \textit{fv}\ lv$ to incur exactly $\sum_{x \in l}, \Err(x)$ error credits, however if we attempt to prove the error-incurring specification by annotating Hoare triples with an ``error budget'' similar to the style of aHL~\citet{ub} we come across a few difficulties:
\begin{enumerate}
  \item By default, the logic itself does not allow the error to \emph{depend on program terms}, i.e.\ in our example, the term $l$. We could circumvent this issue by forcibly adding a parameter representing the elements of $l$, but this involves a quantification in the metalogic which can make our specification harder to reuse.
  \item Even though the proof of an approximated specification follows a similar structure to the original, we cannot \emph{reuse} the original specification to prove the approximate one as they live in different logics. One could retroactively reimplement the non-approximate specification in aHL using an error annotation of $0$, but this is not a good strategy in general as it creates extra overhead and exposes unnecessary details to the proof engineer.
\end{enumerate}

Instead, using the fact that error in \theaplog is an ordinary separation logic resource, we can easily verify an approximated specification by instantiating the predicates in \ref{eqn:precise-map} to include error credits:
\begin{equation}
\small
\hoare {
  \begin{array}{c}
     \islist\ xs\ l \sep \\
        \All \var . \hoare{P (\var) \sep \mhl{\upto{\Err(\var)}}}{e\ x}{x' .\ x' = f(x) \sep Q(x')}\sep \\
        \textstyle\Sep_{x\in xs}  \left(P(x)  \sep \mhl{\upto{\Err(x)}}\right)
        \end{array}
  }{\mapexpr\ \expr\ l}
  {l' . \begin{array}{c}
      \islist\ (\mapfun\ f\ xs)\ l' \sep \\
      \Sep_{x' \in (\mapfun\ f\ xs)} Q(x')
    \end{array}
  }
  \label{eqn:approx-map}
\end{equation}

We do not run into the same issues as our earlier example.
By representing error as a resource, \theaplog can define flexible, value-dependent
credits $\upto{\Err(x)}$ without appealing to the metalogic.
Moreover this error-incurring specification is a corrolary of the original non-approximate
specification, facilitating more modular proofs which avoid proving redundant lemmas.

We mention that for a more coarse grained application, one can define the weaker specification where the total error credit for the map function is $\upto{\text{length}(l) * \text{max}_{x\in l} \Err(x)}$.
This specification is typical of approximate logics where value-dependent error is difficult to express, and
follows from \ref{eqn:approx-map} by credit weakening.

\section{Amortized Hash Functions and Merkle Trees}
\label{appendix:merkle}

In this case study, we highlight three striking aspects of \theaplog: the support for higher-order specifications, modularity reasoning, and amortized error reasoning.
The example builds up through three levels of abstraction:
\begin{enumerate}
  \item we implement an amortized non-resizing hash function from a non-amortized one,
  \item we implement a Merkle tree library with the amortized hash function, and
  \item we apply the Merkle tree library to validate data stored in some unreliable storage.
\end{enumerate}

\subsection{Non-amortized Non-resizing Hash Function}
We first implement a hash function under the \textit{uniform hash assumption}~\cite{uniform-hash-assumption}, i.e.\ a hash function $h$ from a set of keys $K$ to values $V$ behaves as if for each key $k$, the hash $h(K)$ is randomly sampled uniformly and independently over $V$. We can implement such a hash function using mutable map $m$: to hash each key $k$ we first check the map to determine whether it has been hashed before and if so, we return the hash value stored in $m(k)$. Otherwise, we randomly sample a value from $V=\{0, \dots, n\}$, store the value in $m(k)$, and return it.

\begin{align*}
  \computehash\ m\ v \eqdef{}
   & \MatchML \mapget\ m\ v with
  \Some(b) => b
  | \None =>
  {\begin{array}[t]{l}
     \Let b = \Rand n  in \\
     \mapset\ m\ v\ b;    \\
     b
   \end{array}}
  end {}
\end{align*}

In the analysis of data structrures which use hash maps, one often assumes that a hash function is \textit{collision-free}, in the sense that for a finite number of queries to the hash function with unique inputs, the output will never repeat. As written, this is \textit{impossible}; the probability of a hash collision increases with the number of keys hashed, and in the extreme case where we query the hash function more than $n+1$ times the pigeonhole principle ensures a collision must occur.

Nevertheless, this is not an issue in practice because the size of $V$ is usually many magnitudes larger than the number of queries to the hash function. Therefore, it is the case that the hash function remains collision-free up to some small error.
To be more concrete, consider the state where we have queried the hash function $f \eqdef{} \computehash\ m$ a total of $s$ times each with a distinct input (so the map is now of size $s$). Suppose also that no query has produced a hash collision.
Now to hash a fresh key without a hash collision, the hash function must ``avoid'' sampling from any of its prior $s$ outputs. We can specify this in \theaplog by requiring that a query to $f$ pay $\upto{\frac{s}{\totalsize}}$:

\begin{mathpar}
  \hoare {
    \begin{array}{c}
      n\notin \dom\ m \sep \\
      \cfhf\ f\ m \sep     \\
      \upto{\frac{\text{size}(m)}{\totalsize}}
    \end{array}
  }{f\ n}
  {v . \cfhf\ f\ (\mapinsert{n}{v}{m})}
\end{mathpar}

\subsection{Amortized Non-resizing Hash Function}
One limitation of the above specification is that the error for each hash operation is proportional to the size of the map. This can complicate the error credit reasoning, particularly when many different clients perform a sequence of consecutive hash operations. Ideally we would want to amortize the error credits over a fixed number of hash queries $\maxsize$ so that for each query, the requisite error is a fixed constant that does not depend on the state of inner map. We will realize this using the \theaplog interpretation of errors credits.

Starting from an empty map, if we bound the number of queries to be $\maxsize$ the total number of errors used is $\upto {\sum_{i=0}^{\maxsize-1} \frac{i}{\totalsize}} = \upto{\frac{(\maxsize-1)*\maxsize}{2(\totalsize)}}$. We can evenly ammortize these errors across the sequence of queries so that each query pays $\upto {\frac{(\maxsize-1)*\maxsize}{2(\totalsize)*\maxsize}}=\upto {\frac{(\maxsize-1)}{2(\totalsize)}} \eqdef \upto{\amortizederr}$. Using this strategy, we can prove the following improved specification for our hash function:
\begin{mathpar}
  \hoare {
    \begin{array}{c}
      \mhl{\text{size}(m)<\maxsize} \sep \\
      n\notin \dom\ m \sep           \\
      \acfhf\ f\ m \sep              \\
      \mhl{\upto{\amortizederr}}
    \end{array}
  }{f n}
  {v . \acfhf\ f\ \mapinsert{n}{v}{m}}
\end{mathpar}

Our proof reuses the original non-amortized specification by using the ``piggy bank'' method; the abstract predicate $\acfhf$ not only contains the $\cfhf$ resource, but also any the error credits paid in excess during the first $\maxsize/2$ hash operations.
That is, the first half of the queries store their extra credit into the ``piggy bank'', which the latter operations can use to pay for their more expensive queries.

\section{Validating Data from an Unreliable Source}

\newcommand{\leaflist}{\textit{lis}}

We tie up this case study by presenting an application of the Merkle tree library to
validating of data stored in an unreliable source.
By \textit{unreliable}, we mean that we do not have any guarantees on the correctness
of the ``read'' and ``write'' operations, meaning that after reading a client must
also validate that their data has been stored faithfully.

Let $\leaflist$ be a list of naturals that we want to store in the unreliable storage.
We will represent $\leaflist$ using a Merkle tree, where the elements of the list constitute the leaves.
To implement a ``write'' operation we compute the hashes for the Merkle tree top to bottom
and then store the entire tree into the unreliable storage,
returning a reference to the merkle tree and a closure which remembers the root hash of the tree.
This closure is essentially the $\checker$ from the aforementioned Merkle tree library, and we will use it to
validate that the tree has correctly stored.
%
To ``read'' an element from the list, we recurisvely walk down the Merkle tree in the unreliable storage, obtaining the final leaf node and a potential proof of its correctness.
This information is then passed to the checker to determine whether the computed root hash coincides with the real root hash value.
In total, applying our Merkle tree library each read operation requires an error credit of $\upto{\amortizederr * \mtreeheight}$.

\newcommand{\treevalidwithleaflist}{\textit{tree\_valid\_with\_leaf\_list}}
\newcommand{\idx}{\textit{idx}}
\newcommand{\leaflookup}{\textit{leaf\_lookup}}
\newcommand{\some}{\textlog{Some}}
\newcommand{\checkerspec}{\textit{checker\_spec}}
\newcommand{\height}{\textit{height}}
\newcommand{\readtree}{\textit{read\_tree}}
\newcommand{\boundscheck}{\textit{bounds\_check}}
\newcommand{\location}{\textit{loc}}

\begin{align*}
  &\leaflookup\ \location\ \height\ \idx\ \checker \eqdef{}\\
  &{\begin{array}[t]{l}
       \Let (\lproof, \lleaf) = \readtree\ \location\ \height\ \idx in\\
       \If \boundscheck\ \lproof\ \lleaf\ then\\
       \hspace{1em}  \If \checker\ \lproof\ \lleaf \\
       \hspace{2em} then \Some \lleaf \\
       \hspace{2em}\Else \None\\
       \Else \None
   \end{array}}
\end{align*}

Here the $\readtree$ program takes the location of the root of the tree, the height, and the index of the leaf node we are reading, and returns a tuple containing the proof and the leaf value. The function $\boundscheck$ verifies that the proof and the leaf value lie within expected bounds. It is worth noting that the implementation and correctness guarantees of $\readtree$ and $\boundscheck$ are relatively unimportant to the specification of $\leaflookup$, as our specification guarantees that we return $\Some\ x$ only if $x$ is the true value of the $\idx$-th leaf value of the Merkle tree regardless. 

Altogether, to correctly execute $\leaflookup\ l\ (\mtreeheight)\ \idx\ \checker$ for some location $l$ in unreliable storage we require the following preconditions:

\begin{enumerate}
  \item The number $\idx$ is smaller than $2^{\mtreeheight}$.
  \item The function $\checker$ from the merkle tree library is specialized to both $\mtree$ and $m$.
  \item The merkle tree $\mtree$ is built correctly from hash map $m$ and leaves $\leaflist$.
  \item The function $f$ encodes the amortized hash function under the map $m$.
  \item The size of $m$ plus the height of the tree is smaller or equals to $\maxsize$.
  \item At least enough error credit for the read.
\end{enumerate}

Expressed in \theaplog,
\begin{mathpar}
  \hoareV {
    \begin{array}{c}
      \idx < 2^{\mtreeheight}      \sep                 \\
      \checkerspec\ \checker\ \mtree\ m \sep            \\
      \treevalidwithleaflist\ \mtree\ \leaflist\ m \sep \\
      \acfhf\ f\ m \sep                                 \\
      \text{size}(m)+\mtreeheight <= \maxsize \sep    \\
      \upto{\amortizederr * \mtreeheight}
    \end{array}
  }
  {\leaflookup\ l\ (\mtreeheight)\ \idx\ \checker}
  {v . v = \some\ x \implies \leaflist[\idx]=x
  }
\end{mathpar}

For a similar reason as to how we do not impose many guarantees on the functions $\readtree$ and $\boundscheck$,  nowhere in the specification do we impose any restrictions on the location $l$. In particular, we do not even know that $l$ is a reference to the correct Merkle tree in the unreliable storage as we do not assume any behavior of the unreliable storage.

\section{Appendix: Additional Details for Collision-Free Hash Map}
\label{app:cf-hash-map}

Hash maps are one of the most ubiquitous data structures in programming, since they
can represent large sets with efficient insertion, deletion and lookup operations.
This efficiency relies on having a low number of collisions, so that each location
on the table contains a small number of values. As the number of collisions increases,
and thus the performance of the hash map worsens, it is often beneficial to resize the
table, thus redistributing the hashed values and freeing up space for new insertions.

In order to be able to reason about the efficiency of hash maps, we need to compute
the probability of a collision happening. However, computing this probability over a
sequence of insertions is cumbersome, due to its dependence on the current size of
the hash table and the number of elements it contains. Moreover, it also leads to less
modular specifications for programs that use hash maps as their components.


We will use the dynamically-resizing hash function defined in the main body of
the paper to implement a collision-free dynamically-resizing hash map, with
amortized cost of insertion.  We will have an array of size $v$, in which $s$
entries are filled with a hashed value and the rest are uninitialized. Once we
fill in $r$ elements, we resize the table to have size $2*v$ and we set $r$ to
$2*r$. The code is shown below:

We include the code for the insertion function for a collision-free, resizing
hash map, as studied in~\cref{sec:cf-hash-map}. A hash map is represented as a tuple
$(l, hf, v, s, r)$, where $l$ is an array containing the physical representation
of the map, $hf$ is a (collision-free) hash function as shown in~\cref{sec:cf-hash}, 
$v$ is the current size of the array containing the hash map, $s$ is the number of initialized
entries and $r$ is the size threshold before resizing:
\begin{align*}
  \mathit{insert}\ \mathit{hm}\ w\ \triangleq \quad
  &\smash{\Let (l, hf, v, s, r) = \mathit{hm} in}\\
  &\smash{\Let (b, hf') = \computehashrs\ hf\ w\ in}\\
  &\smash{\Let w' = \deref l[b]\ in}\\
  &\smash{\If w' = () then }\\
  &\smash{\quad l[b] \gets w}\\
  &\smash{\quad \If s + 1 = r then }\\
  &\smash{\quad\quad \Let l' = \mathit{resize}\ l\ v\ v\ in}\\
  &\smash{\quad\quad (l', hf', 2*v, s+1, 2*r)}\\
  &\smash{\quad\Else (l, hf', v, s+1, r)}\\
  &\smash{\Else (l, hf', v, s, r)}
\end{align*}
Note in particular that if we try to insert an element $w$ and there is another
element $w'$ with the same hash, then $w$ will not get inserted into the table.
However, the specification of $\computehashrs$ ensures that this will not happen
if we have ownership of $\upto{(3 \cdot R_0) / (4 \cdot V_0)}$. 

The core part of the representation predicate for the hash map is shown below:
\begin{align*}
	\mathsf{isHashmap}\ hm\ ns\ \triangleq \quad
	& \exists l, hf, v, s, r, m, tbl. ( hm = (l, hf, v, s, r) ) \ast \\
	& \quad l \mapsto^* tbl \ast
    ( (\mathit{filterUnits}\ tbl) \equiv ns ) \ast \\
	  & \quad \acfhfrs\ hf\ m\ v\ s\ r \ast \\
	  & \quad ( \forall (i, w : \nat). m[w] = i \leftrightarrow tbl[i] = w ) \ast \\
	  & \quad ( \forall i < v, i \not\in \mathsf{img}\ m \rightarrow tbl[i] = () ) \ast (\dots)
\end{align*}
This should be read as ``$hm$ is a hash map representing the set (of natural numbers) $ns$''.
The hash map contains a table $tbl$ whose contents are either natural numbers or unit,
and the set of natural numbers it contains is exactly $ns$. The index at which every element
is located is controlled by a collision-free hash function $hf$, that tracks a partial map
$m$. Thus the table will contain an element $w$ at index $i$ if and only if $m$ maps $w$ to $i$.

Crucially, this predicate keeps no track of error credits, all of the error accounting is done
through the $\acfhfrs$ predicate, which is used as a client.
With this representation predicate, we can prove the following specification for $\mathit{insert}$.
\[
  \hoare{ \mathsf{isHashmap}\ hm\ ns\ \ast \upto{(3 \cdot R_0) / (4 \cdot V_0)} }
	{  \mathit{insert}\ hm\ w }     
 { hm', \mathsf{isHashmap}\ hm' (ns \cup \{ w \})  }
\]
This specification states that if we own $\upto{(3 \cdot R_0) / (4 \cdot V_0)}$,
then insertion of an element $w$ will always succeed. There are two ways in
which this can be validated. Either $w$ was already in the hash map (and
therefore $ns = ns \cup \{ w \}$) or it is a new element. If it is a new
element, we also have to case on whether we have to resize or not. In either of
those cases, we can use $\upto{(3 \cdot R_0) / (4 \cdot V_0)}$ to sample a
fresh value from the hash map, following the specifications proven for the
resizing hash function. This ensures that the location in the table corresponding to
that index is uninitialized. Furthermore, by ensuring that the hash map resizes
at the same time as the hash table does, this specification will be valid no
matter how many insertions have been performed before.

Notice here how the
modularity of \theaplog enables us to have a relatively simple specification,
which besides the presence of error credits, coincides with the specification
one would expect for a hash map.

\section{Appendix: Expectation-Preserving Composition on Words}
\label{appendix:iter-adv-comp}

While \rref{ht-rand-exp} and \rref{presample-exp} can move error credits between the outcomes of a single
random event, in order to prove the planner rule
we need to move error credits out of the sequence of events which sample an entire target word.
Defining a suitable sequence of expectation-preserving composition steps to
accomplish this can be subtle: sampling any prefix of our target word should
\textit{decrease} our error credit, but sampling a
prefix of the target followed by an erroneous sample should yield an
amplification on our \textit{initial} credit amount.

Let $\vec{w}$ be a word of length $L > 0$ in the alphabet $[0, N]$.
For $0 \leq i \leq L$, define the constants
\begin{equation*}
  \ecAmp_{N, L} \eqdef 1 + \frac{1}{(N+1)^{L} - 1} \quad\quad\quad\quad
  \ecRem_{N, L}(i) \eqdef 1 - \frac{(N+1)^{i} - 1}{(N+1)^{L}-1}
\end{equation*}
so that $0 \leq \ecRem_{N, L}(i) < 1 < \ecAmp_{N, L}$.
Suppose we want to amplify some positive amount of credit $\err$
against $\vec{w}$; that is we seek to either sample all of $\vec{w}$, or obtain extra error credits.
For $0 \le i < L$, define the error distribution functions
\begin{equation*}
  D_{N, L}^{\err}(i, c) \eqdef
  \begin{cases}
    \ecRem_{N, L}(i + 1) \cdot \err & c = \vec{w}[i]  \\
    \ecAmp_{N, L}\cdot\err & \textrm{otherwise}
  \end{cases}
\end{equation*}

Starting with $\upto{\ecRem_{N, L}(i) \cdot \err}$ the function $D_{N, L}^{\err}(i, \_)$ is mean-preserving, since
\begin{align*}
  \sum_{c = 0}^{N}D_{N, L}^{\err}(i, c)
  & = \ecRem_{N, L}(i+1) \cdot \epsilon + N \cdot \ecAmp_{N, L} \cdot \err \\
  & = \left( \frac{(N+1)(N+1)^{L}-(N+1)^{i+1}}{(N+1)^{L}-1}\right) \cdot \err \\
  & = (N + 1) \cdot \ecRem_{N, L}(i) \cdot \err
\end{align*}

Now we can redistribute the error
credit out of the event where we sample $\vec{w}$ and distribute it evenly into all other cases,
using $L-1$ steps of advanced composition.
Starting with $i=0$, at the beginning of the $i^{\textrm{th}}$ sample we
have will have correctly sampled the first $i$ characters of $\vec{w}$ and own
$\upto{\ecRem_{N, L}(i) \cdot \err}$.
At step $i$, perform expectation preserving composition using the
error function $D_{N, L}^{\err}(i, \_)$.
Each composition either correctly samples the next character of $\vec{w}$ and decreases
the error credit supply to $\upto{\ecRem_{N, L}(i + 1) \cdot \err}$, or increases it to $\upto{\ecAmp_{N, L} \cdot \err}$.
Note that $\upto{\ecRem_{N, L}(0) \cdot \err} = \upto{\err}$ for the initial case,
and $\upto{\ecRem_{N, L}(L) \cdot \err} = \upto{0}$ once $\vec{w}$ is completely sampled.
In aggregate, this sequence of proof steps will either result in sampling $\vec{w}$
or increasing our error credit by a factor of $\ecAmp_{N, L}$.

Implemented using \rref{presample-exp}, this procedure proves the amplification lemma from \cref{sec:planner}:
\begin{equation*}
  \infrule[Right]{}
  {
    \thoare{\Exists \vec{j} . \progtape{\lbl}{\tapebound}{\tape \dplus \vec{j}} \sep \upto{\ecAmp_{N, L} \cdot \err}}{e}{\phi}
    \\
    \thoare{\progtape{\lbl}{\tapebound}{\tape \dplus \vec{w}}}{e}{\phi}
  }
  { \thoare{\progtape{\lbl}{\tapebound}{\tape} \sep \upto \err}{e}{\phi} }
\end{equation*}

Finally, it will be convenient to define a lower bound on the amount of extra credit generated each time our chain of advanced composition fails to sample $\vec{w}$: $\ecExc_{N, L} \eqdef \ecAmp_{N, L} - 1$.
Since $\ecRem_{N, L}(i) < 1$ for all $0 \leq i \leq L$, we can prove that
\begin{equation}
  \upto{\ecAmp_{N, L}} \wand \upto{\ecRem_{N, L}(i)} \sep \upto{\ecExc_{N, L}}
  \label{eqn:ec-retry}
\end{equation}
In other words, when we fail to sample $\vec{w}$ using this technique we have at least enough credit to try again, plus an additional $\upto{\ecExc_{N, L} \cdot \err}$.

\section{Randomized SAT Solving}
\label{appendix:walk-sat}

Our \textit{induction by error amplification} technique it not limited to proving termination
for simple rejection samplers.
To demonstrate the versatility of this approach, we develop a higher order specification
which can prove termination properties of ``check and retry'' algorithms which have
complex dependencies on state.
%

\subsection{A Higher Order Specification for Rejection Samplers}
\label{appendix:higher-order-rejection-samplers}

Let $s$ and $c$ be a sampler/checker pair, and $\Theta : \Val \to \iProp$ be a property of samples.
We seek to establish a relationship between $s$ and $c$ such that $(S\ s\ c)$ almost certainly
terminates with a value satisfying $\Theta$.

Recall that our formulation of the planner rule allows us to almost surely sample some
target word onto a tape, and that the target word is free to depend on the state of the
tape beforehand.
Explicitly providing such a target word as a function of a sequence of sampling events
may be unnecessarily cumbersome, and may not even be sufficient to execute the program,
since the planner rule makes no assertions over the \textit{junk} section sampled
before the target word.
Rather than explicitly specifying a target sampling event,
it may be more straightforward to leave the target event implicit and
specify how $c$ behaves in terms of $s$ directly.

\newcommand{\hospec}{\mathit{sc}}

\begin{equation}
\begin{aligned}
  \hospec(s, c, k, \Theta) \eqdef
  & \; \ulcorner k > 1 \urcorner \sep \forall \err > 0.\,  \\
  &\quad\quad\quad \thoare{\upto{\epsilon}}{s\ ()}{v.\,
    \begin{array}{l}
      (\thoare{\TRUE}{c\ v}{r : \bool .\, \ulcorner r = \True \urcorner \sep \Theta\ v}) \ \vee \\
      (\thoare{\TRUE}{c\ v}{r : \bool .\, \ulcorner r = \True \urcorner \wand \Theta\ v} \sep\upto{k\ \err})) \\
    \end{array}}
\end{aligned}
\label{eqn:hospec}
\end{equation}

Figure \ref{eqn:hospec} leverages a higher-order specification in order to establish
such a relationship.\footnote{In this section, we abbreviate
  $\thoare{\Phi}{e}{v.\ \ulcorner v = \True \vee v = \False \urcorner \sep \Psi}$ as $\thoare{\Phi}{e}{v : \bool.\, \Psi}$.}
The specification reads as follows: for a fixed constant $k$ and any positive
amount of error credit $\err$, we can execute the sampler to obtain one of two outcomes:
\begin{enumerate}
  \item a \textit{productive sample}, which ensures the checker will step to $\True$, or
  \item an \textit{erroneous sample}, which does not guarantee the sampler will step to $\True$, but amplifies our error credit by $k$.
\end{enumerate}
Figure \ref{eqn:hospec} additionally specifies that $\Theta$ holds whenever $c$ accepts a
sample, even if that sample is erroneous.
Since we do not need an exact error bound to prove almost sure termination, this allows
a prover to underapproximate the cases where $c$ will accept.

We can perform induction by error amplification for any sampler and checker pair satisyfing
this specification.
A sequence of $d = \lceil \log_{k}(1/\epsilon) \rceil$ erroneous samples guarantees that we accumulate $\upto{1}$,
so by induction on $d$ we can obtain
\begin{equation}
  (\exists k,\hospec(s, c, k, \Theta)) \wand \forall \err > 0, \ \thoare{\upto{\err}}{S\ s\ c }{r .\, \Theta \, r}.
  \label{formula:sampler-int}
\end{equation}
When $\Theta$ is pure\footnote{To prove termination alone, one can use $\Theta = (\Lam \_ . \top)$, which is pure. }
our continuity result yields
\begin{equation*}
  (\exists k,\ \hospec(s, c, k, \Theta)) \to \forall \sigma.\, \prex{(S\ s\ c, \sigma)}{\Theta} = 1.
  \label{formula:sampler-lim}
\end{equation*}

\subsection{An Incrementalized Specification}

\newcommand{\specisc}{\mathit{isc}}

Next, we generalize specification \ref{eqn:hospec} to allow the sampler to have effects which
persist between attempts.
This is necessary in order to support samplers whose ``productive'' guesses may not
immediately cause the checker to accept, but rather make incremental progress
towards an accepting state.

\begin{equation}
  \begin{aligned}
  & \specisc(s, c, P, E, L_{p}, L_{e}, \Theta) \eqdef \\
  & \hspace{40pt} \thoare{\upto{E(0)} \lor P(0)}{s\ ()}{v.\, \thoare{\TRUE}{c\ v}{r : \bool .\, \ulcorner r = \True \urcorner \sep \Theta\ v}} \sep \\
  & \hspace{40pt} \forall i < L_{e},\, j < L_{p}.\, \\
  & \hspace{55pt} \thoare{
    \begin{array}{l}
      \upto{E(i+1)} \\
      \sep P(j + 1)
    \end{array}}
    {s\ ()}
    {r.\, \begin{array}{l}
            \thoare{\TRUE}{c\ v}{r: \bool.\, \ulcorner r = \True \urcorner \sep \Theta\ v}\ \vee \vspace{10pt} \\
            \thoare{\TRUE}{c\ v}{r: \bool.\, \begin{array}{l}
                                           \upto{E(i + 1)} \sep P(j) \sep \\
                                           (\ulcorner r = \True \urcorner \wand \Theta\ v)
                                         \end{array}}\ \vee \vspace{10pt} \\
            \thoare{\TRUE}{c\ v}{r: \bool.\, \exists j' \leq L_{p}.\,
                                        \begin{array}{l}
                                           \upto{E(i)} \sep P(j') \sep \\
                                           (\ulcorner r = \True \urcorner \wand \Theta\ v)
                                         \end{array}}\\
          \end{array}
      }
    \end{aligned}
    \label{eqn:incrhospec}
\end{equation}

This specification generalizes \ref{eqn:hospec} by paramaterizing two key components.
The first is $E : \nat \to \real_{\geq 0}$, which quantifies the \textit{error} in
terms of the number of failed samples remaining before we can apply a credit spending argument.
The specification also generalizes over the remaining \textit{progress}
towards an accepted sample via the term $P : \nat \to \iProp$.
As a regular Iris proposition, $P$ can contain invariants relating to program state
and the resources required to execute $s$ or $c$.
Both $E$ and $P$ have fixed bounds on their argument $L_{e}$ and $L_{p}$.
typical instansiation sets $E(L_{e}) = 0$ so that we can start with $\upto{E(L_{e})}$,
and $E(0) = 1$ so $\upto{E(0)} \wand \bot$.

We will now break down \ref{eqn:incrhospec} in more detail.
The first conjunct is a base case, which states that either complete error
$\upto{E(0)}$ or complete progresss $P(0)$ are enough to step the sampler
and checker to the desired result.
Like the non-incremental specification, the second conjunct of \ref{eqn:incrhospec} specifies the
behavior of the checker in terms of the sampler's result.
Namely, executing $s$ yields one of three cases:
\begin{enumerate}
  \item a \textit{lucky sample}, which allows the checker to terminate immediately,
  \item an \textit{improvement sample}, which makes progress by decreasing the argument to $P$, or
  \item an \textit{erroneous sample}, which loses progress but generates enough credit to decrease $E$.
\end{enumerate}

We can prove a total specification for samplers and checkers which satisfy \ref{eqn:incrhospec}:
\begin{equation}
  \specisc(s, c, P, E, L_{p}, L_{e}, \Theta) \wand \forall i \leq L_{e},\, j \leq L_{p}.\, \ \thoare{E(i) \sep P(j)}{S\ s\ c }{r .\, \Theta \, r}
  \label{formula:inc-sampler-int}
\end{equation}

The proof of \ref{formula:inc-sampler-int} proceeds by a nested induction, firstly over the remaining
erroneous samples, and secondly over the remaining progress samples.
That is, the proof attempts to sample a sufficiently long sequence of improvement samples or improve
the error at least once.
Even through it is possible for the sampler to lose progress on $P$, it never loses the error credits in
$E$, so this induction is well-founded.




\subsection{Almost Sure Recognition in Randomized SAT Solving}

\newcommand{\evalasn}{\langv{eval\_asn}}
\newcommand{\initasn}{\langv{init\_asn}}
\newcommand{\updasn}{\langv{upd\_asn}}
\newcommand{\evalvar}{\mathit{eval\_var}}
\newcommand{\evalclause}{\mathit{eval\_clause}}
\newcommand{\resample}{\mathit{resample}}
\newcommand{\wssample}{\mathit{sampler}}
\newcommand{\wscheck}{\mathit{checker}}
\newcommand{\invasn}{\mathit{inv\_asn}}
\newcommand{\SATpos}{\mathit{Pos}}
\newcommand{\SATneg}{\mathit{Neg}}
\newcommand{\SAT}{\mathit{SAT}}
\textit{WalkSAT} \cite{walksat} is a randomized algorithm for recognizing satisfiable boolean formulas.
Given a boolean formula in conjunctive normal form, WalkSAT searches for a solution by iteratively
flipping random variables inside unsatisfied clauses.
Importantly, the possible variables which WalkSAT might flip depends both on
both the current assignment and also the formula under test.
This means that, unlike the case for a simple random walk on $\bool^{N}$, it is not immediate
that WalkSAT almost-surely terminates even when the formula is satisfiable.\footnote{For
  example, a buggy implementation of WalkSAT which never resamples from the first clause
  will fail to recognize the satisfiable formula $(X_{1} \vee X_{1} \vee X_{1}) \wedge (X_{2} \vee X_{2} \vee X_{2})$
  if $X_{1}$ is $\textrm{False}$ in its inital assignment. }
By proving an instance of \ref{eqn:incrhospec} and applying a continuity argument,
we will show that WalkSAT almost surely recognizes satisfiable boolean formulas.

Figure \ref{fig:walksat} depicts our implementation of WalkSAT as a
sampler/checker pair.
Our \thelang{} program is paramaterized by the number of varibles $N$
and a 3SAT formula $f$ at the Coq-level.
We represent a formula $f$ as a list of \textit{clauses}, each consisting
of a triple of \textit{variables}, comprised of a \textit{variable index} $n < N$
and a \textit{polarity}.
As an example, we represent the 3SAT clause $(x_{1} \vee x_{2} \vee \overline{x_{3}})$ as the triple $((1, \SATpos), (2, \SATpos), (3, \SATneg))$.

We represent an \textit{assignment} of variable indices to boolean values
in \thelang{} as a length $N$ linked list, and in Coq as a length $N$ list of booleans.
The Coq-level predicate $\SAT_{f}(a)$ means that $a$ is a length $N$ boolean list,
and it satisfies the formula $f$.
The proposition $\invasn \ v \ s$ asserts that the ProbLang value $v$
represents the same assignment as the list $s$.
The standard programs $\evalasn$ and $\updasn$ look up and update
values in a (\thelang{} level) assignment by index, and the program $\initasn$ allocates a
new assignment with random values.

\begin{figure}
  \begin{minipage}[t]{0.33\linewidth}
    \begin{align*}
      \evalvar((n, p)) \eqdef
      &\smash{\;\Lam a.\; \Let b = \evalasn \ a \ n in} \\
      & \quad\quad\quad{\MatchML p with
        \SATpos => b
        | \SATneg => {\sim} b
        end {}} \\
      \\
      \evalclause((x_{0}, x_{1}, x_{2})) \eqdef
      & \smash{\;\Lam a .\, \evalvar \ x_{0} \ a \ \And \evalvar \ x_{1} \ a \And \evalvar \ x_{2} \ a } \\
      \\
      \resample((n_{0}, \_), (n_{1}, \_), (n_{2}, \_)) \eqdef
      &\smash{\;\Lam l . \Let a = \ \deref l in}\\
      &\smash{\quad\quad\Let i = \Rand \ \#2 in } \\
      &\smash{\quad\quad\Let b = \evalasn \ a \ n_{i} in } \\
      &\smash{\quad\quad l \leftarrow \updasn \ a \ n_{i} \ (\sim b) } \\
      \\
      \wssample(f) \eqdef
      &\; \Lam l .\,
          \MatchML f with
            [] => ()
          | (c : cs) => {\begin{array}[t]{l}
                           \If \evalclause \ c \ (\deref l) \\
                           \quad then \wssample \ cs \ l \\
                        \quad \Else \resample \ c \ l \\
                       \end{array}}
        end {} \\
      \\
      \wscheck(f) \eqdef
      &\;\Lam l .\,
        \MatchML f with
          [] => \True
        | (c : cs) => \evalclause \ c \ (\deref l) \ \And \ \wscheck \ cs \ l
        end{} \\
      \\
      W_{f} & \eqdef \Lam l .\, S\ (\wssample\ f\ l)\ (\wscheck\ f \ l) \\
    \end{align*}
  \end{minipage}
  \caption{A sampler and checker implementing a simplified version of WalkSAT. }
  \label{fig:walksat}
\end{figure}

The functions $\evalvar$, $\evalclause$ and $\wscheck$ determine if the current
\thelang{} assignemnt satisfies a variable, clause, and formula respectively.
Given a clause, the function $\resample$ chooses a variable index referenced by
that clause uniformly at random, and negates it in the current assignment.
Finally, the function $\wssample$ resamples from the first unsatisfied clause,
doing no such clauses exist.
Combined, the program $W_{f}$ implements a simplified\footnote{Implementations of
  WalkSAT typically choose a \textit{random} unsatisfied clause, rather than the first
  one. This detail is irrelevant to our analysis. } version of the WalkSAT algorithm:
the program repeatedly resamples the first unsatisfied clause and terminates when
the current assignemnt satisfies $f$.

Let us assume there is some solution $t$ to a 3SAT formula $f$, that we own a
positive amount of error credit $\upto{\err}$, and that we own an assignment $l$.
Denote the Hamming distance between $t$ and a boolean list of the same length as $d^{(0)}(a)$.
Now, we can define the key components for an instance of \ref{eqn:incrhospec}:
\begin{align*}
  L_{p} & \eqdef N \\
  L_{e} & \eqdef \lceil 1 / \ecExc_{3, N} \rceil  \\
  P_{W}(n) & \eqdef \exists a,\; m. \; (l \mapsto a) \sep \upto{\ecRem_{3, N}(N - n) \, \err} \sep \ulcorner\invasn \ a \ m \urcorner \sep \ulcorner d^{(0)}(asn) \le n \urcorner\\
  E_{W}(n) & \eqdef \max(0, 1 - n \, \ecExc_{3, N} \, \err) \\
  \Theta_{W} & \eqdef \exists a,\; m .\; (l \mapsto a) \sep \ulcorner \invasn \ a \ m \urcorner \sep \ulcorner \SAT_{f} \ m \urcorner
\end{align*}

Our error measure $E_W$ is a linear function of the minimum amplification of
length $N$ words in 3 characters.
Our progress measure $P_{W}$ asserts both ownership over a \thelang{} level assignment,
and bounds the Hamming distance between that assignment and $t$.
Interestingly, the progress measure also contains an error term.
Because correcting an assignment can take up to $N$ lucky resamples, our amplification strategy
requires some nonzero source of credit which is free to decrease as \textit{improvement} samples
are drawn.
As a resouce, separating error credits to isolate such dependencies is no issue in \theaplog.
Finally, the resulting proposition $\Theta_{W}$ asserts that the \thelang{} level value
corresponds to a satisfying assignment for $f$.
Note that we are not proving that the final assignemnt matches $t$; for formulas
with multiples solutions this is in fact not true.


Now we can outline the proof of $\textrm{isc}(\wssample\ f\ l, \wscheck\ f\ l, P_{W}, E_{W}, L_{p}, L_{e}, \Theta_{W})$.
The first conjunct is trivial, we must show that either $E_{W}(0)$ or $P_{W}(0)$ is enough
for $W_{f}$ to terminate and satsify $\Theta$.
The final error case is a straightforward credit spending argument, since $\upto{E_{W}(0)} = \upto(1)$.
For the final progress case, the Hamming Distance critereon in $P_{W}$ ensures
that the current state is $t$, which we know to be a solution to $f$ as required.

The second conjuct pertains to the behavior of the sampler given some
initial $\upto{E_{W}(i+1)}$ and $P_{W}(j+1)$.
If the intial state is some solution to $f$, the sampler will not resample
any clause and we can finish by the \textit{lucky sample} case.
Otherwise, the current assignment is not a solution to $f$, meaning it must
necessarily differ  from $t$ at some variable $0 \leq z \leq 2$ in its first unsatsified
clause (or else that clause would be satisfied).
The resources in $P_{W}(j+1)$ are enough to step us up to the point where
$\resample$ chooses which variable to flip, at which point we perform an
expectation-preserving composition using the function $D_{3, N}(N-n, z)$
and the credits $\upto{\ecRem_{3, N}(N - j) \, \err}$.
The case where $\resample$ flips the $z^{th}$ variable decreases the Hamming distance
between the current assignment and $t$ by 1, meaning we can establish
the \textit{improvement sample} case.
Otherwise, the program flips some other variable, and we will establish the
\textit{erroneous sample} case.
This goal has two main parts: we must find some $j' \leq L_{p}$ such that prove $P_{W}(j')$,
and we must improve our error measure from $\upto{E(i+1)}$ to $\upto{E(i)}$.
As remarked in section~\ref{appendix:iter-adv-comp}, we can split our newly
generated error credit $\upto{\ecAmp_{3, N}\,\err}$ into $\upto{\ecExc_{3, N}\,\err}$
and $\upto{\ecRem_{3, N}(L_{p})}$.
The excess $\upto{\ecExc_{3, N}\,\err}$ credit is exactly the difference between $E(i+1)$
and $E(i)$, allowing us to improve the error measure.
The $\upto{\ecRem_{3, N}(L_{p})}$ term is enough to re-establish $P_{W}(L_{p})$ regardless of the
effect the incorrect flip had on the assignment, since the Hamming distance between
any assignment and $t$ is at most $L_{p}$.
Therefore, our implementation WalkSAT satisfies our general incremental specification
$\textrm{isc}(\wssample\ f\ l, \wscheck\ f\ l, P_{W}, E_{W}, L_{p}, L_{e}, \Theta_{W})$.

To conclude this section, we will show that WalkSAT almost surely terminates when
the formula is satisfiable.
We have shown that we can construct a $\specisc$ instance for any initial
postitive error credit $\err$, so by \ref{formula:inc-sampler-int} we have
\begin{equation*}
  \forall i \leq L_{e},\, j \leq L_{p}.\, \ \thoare{E(i) \sep P(j)}{S\ s\ c }{r .\, \Theta \, r}
\end{equation*}

Furthermore $\ecRem_{3, N}(L_{p}) = E_{W}(L_{e}) = 0$, and $d^{(0)}(a) \le L_{p}$ for
all assignments $a$.
This means that given $\upto{\err}$ we can establish our intial
conditions $\upto{E_{W}(L_{e})}$ and $P_{W}(L_{p})$ for any starting assignment:
\begin{equation}
  \forall \epsilon > 0, \thoare{\upto{\epsilon} \sep \ulcorner \exists t.\, SAT_{f} \ t\urcorner}{\Let l = \initasn in (W_{f} \ l)}{\exists v, a.\, l \mapsto v \ast \ulcorner \invasn \ v \ a \urcorner \ast \ulcorner SAT_{f}\ a \urcorner}
  \label{eqn:walksat-err}
\end{equation}

Finally, since \ref{eqn:walksat-err} is quantified over $\err$, \cref{thm:twp-ast}
allows us to conclude that WalkSAT almost surely terminates for a satisfiable formula.


\end{document}